\documentstyle[12pt,pstricks,psfig]{article}
\newcommand{\doe}{This work was supported by the Director, Office of Energy 
                  Research, Division of Nuclear Physics of the Office of High 
                  Energy and Nuclear Physics of the U.S. Department of Energy 
                  under Contract No. DE-AC03-76SF00098 and U.S. - Hungary 
                  Science and Technology Joint Fund J. F. No. 378}

\newcommand{\lton}
        {\mathrel{\lower.9ex\hbox{$\stackrel{\displaystyle <}{\sim}$}}}
\newcommand{\lsim}
{\ \raisebox{2.75pt}{$<$}\hspace{-9.0pt}\raisebox{-2.75pt}{$\sim$}\ }
\newcommand{\gsim}
{\ \raisebox{2.75pt}{$>$}\hspace{-9.0pt}\raisebox{-2.75pt}{$\sim$}\ }

\begin{document}
\baselineskip=16pt
\parindent=0.25in
\abovedisplayskip=14pt
\belowdisplayskip=14pt
\parindent=0.5in

\begin{flushright}
{\large LBL-38145}
  \end{flushright}

\vskip 2\baselineskip
\renewcommand{\thefootnote}{\fnsymbol{footnote}}
\setcounter{footnote}{0}
\begin{center}
\mbox{}\\[5ex]
{\Large\bf 
A pQCD-based Approach to Parton Production and
Equilibration in High-Energy Nuclear Collisions\footnote{\doe}
}\\
\vspace{1.5cm}
{\bf XIN-NIAN WANG}\\
{\it Nuclear Science Division, Mailstop 70A-3307}\\
{\it Lawrence Berkeley National Laboratory}\\
{\it University of California, Berkeley, CA 94720 USA}\\
\end{center}
\vskip \baselineskip

\tableofcontents

\vskip 6\baselineskip

\centerline{\bf ABSTRACT}
\centerline{\parbox[t]{5in}{
\hspace{0.25in}A pQCD-based model for parton production  
and equilibration in ultrarelativistic heavy-ion collisions 
is reviewed. The model combines pQCD processes including initial and
final state radiations together with string phenomenology for 
nonperturbative soft processes. Nuclear effects on the initial parton
production, such as multiple parton scattering and nuclear shadowing of
parton distribution functions are considered. Comparisons with existing 
data are made and further tests of the model to constrain model 
parameters are proposed. With the obtained space-time history of the 
parton production, evolution of the minijet gas toward a fully 
equilibrated parton plasma is studied. Direct probes of the
early parton dynamics, such as pre-equilibrium photon and 
dilepton production, open charm production, $J/\psi$ suppression
and jet quenching are also reviewed.
}}

\vskip 4\baselineskip
\newpage

\section{Introduction}

According to the Big Bang theory, during the early stage of the evolution 
of our universe, there existed a state of matter consisting mainly of 
unbound quarks and gluons at a temperature of about 200 MeV. 
As the universe expanded and cooled down, it went through a phase
transition and eventually evolved to what we see today. Such a phase 
transition was predicted \cite{PRQCD,REVIEW} shortly after
Quantum Chromodynamics (QCD) was established as the fundamental 
theory for strong interactions. In this theory, quarks, which have
three types of color charges, interact via non-Abelian gauge bosons,
i.e., gluons. Because of the non-Abelian nature of strong interactions,
quarks and gluons are confined to hadrons under normal circumstances.
However, hadronic matter under extremely dense and hot conditions
will go through a phase transition to form a quark-gluon plasma in which
quarks and gluons are no longer confined to the size of a hadron. 
Lattice gauge studies of QCD at finite temperatures have indeed found 
such a phase transition, though the nature of the transition has not yet
been clarified \cite{LATTICE}.

To recreate the early universe and experimentally verify the QCD phase 
transition in the laboratory, it was proposed that nuclei be accelerated to
extremely high energies and then allowed to collide with each other. 
During the collisions, both high baryon densities and high temperatures 
might be reached. Thus, a quark-gluon plasma could be formed which will
subsequently go through the phase transition and hadronize into
particles in the final states. During the last decade, heavy-ion 
experiments have been done at the AGS of BNL and the SPS of CERN with 
$E_{\rm lab} \approx 14$ and 200 AGeV, respectively.
The results from those experiments \cite{QM95} have demonstrated 
extremely rich physics which cannot be explained
by simple extrapolation of $pp$ collisions. However, up to now, 
there is no unambiguous evidence for the existence of a
quark-gluon plasma over a significantly large space-time region.
While experiments at AGS and SPS and their analyses continue,
new experiments \cite{STAR} have been planned at 
RHIC of BNL and LHC of CERN with $E_{\rm cm}=200$ AGeV and 5.5 ATeV, 
respectively.  What I would like to emphasize in this review is 
that heavy-ion collisions at these ultrarelativistic energies 
will demonstrate a completely new dynamics which is not accessible 
at the present energies.

The key issue here is the nuclear structure at different scales.
When the transverse momentum transfer involved in each 
nucleon-nucleon collision is small 
($p_T {\ \raisebox{2.75pt}{$<$}\hspace{-9.0pt}
\raisebox{-2.75pt}{$\sim$}\ }\Lambda_{\rm QCD}$), effective models
based on, {\it e.g.}, meson-exchange and resonance formation
are sufficient to describe multiple interactions between hadrons 
in which the parton structure of the hadrons cannot  yet be resolved.
These coherent (with respect to the partons inside the hadron) 
interactions lead to collective behavior in low-energy heavy-ion 
collisions as were first observed in the Bevalac experiments \cite{BEVL}
and recently at the AGS energies \cite{STACHEL}. However, when $p_T$ 
becomes large enough to resolve individual partons inside a nucleon,
the dynamics is best described on the  parton level
via perturbative QCD (pQCD). Though hard  parton interactions 
occur at CERN-SPS energies ($\sqrt{s}\lsim$ 20 AGeV), they play a 
negligible role in the global features of heavy-ion collisions. 
However, at collider energies ($\sqrt{s} \gsim 100$ AGeV) the importance 
of hard or semihard parton  scatterings is clearly seen in 
high-energy $pp$ and $p\bar{p}$ collisions \cite{WANGPP}. 
They are therefore also expected to be dominant in heavy-ion 
collisions at  RHIC and LHC energies \cite{JBAM,KLL}.
These hard or semihard interactions happen on a very short
time scale and they generally break color coherence inside the
individual nucleons. After the fast partons pass through
each other and leave the central region, a dense partonic 
system will be left behind which is not immediately in thermal 
and chemical equilibrium. The partons inside such a system 
will then further interact with each other and equilibration 
will eventually be established if the interactions are frequent
enough among a sufficiently large number of initially produced
partons. Due to the asymptotic behavior of QCD, production 
rates of hard and semihard partons are calculable via perturbative 
QCD during the initial stage of heavy-ion collisions. The 
color screening mechanism in the initially produced dense partonic
system makes it also possible to use pQCD to investigate
thermal and chemical equilibration of the system. 

Currently, there are many models to describe the coherent hadronic
interaction in which parton degrees of freedom are not important yet.
In these models, hadrons remain as individual entities and particle
production is mainly through resonance formation. Indeed, most 
of the experimental data at AGS energies can be explained
by RQMD \cite{RQMD} and ARC \cite{ARC} Monte Carlo simulations.
As the colliding energy increases, partons inside nucleons
become more relevant as the basic constituents of the interaction.
However, nonperturbative soft interactions still dominate
the collision dynamics in the energy range $\sqrt{s}\lsim 100$ AGeV.
In order to describe particle production in this energy range,
phenomenological string models\cite{FRIT,DPM,ATTILA,VENUS}
have been developed and they can explain well the global 
properties of particle production at CERN SPS energies. The
physics behind these string models is that multiple partons
produced in soft interactions are not independent of each other.
Coherence among them can be modeled by strings
or flux tubes between leading quarks and diquarks (or antiquarks). Particle
production comes from the fragmentation of the string through
$q\bar{q}$ creation inside the flux tube. At collider energies,
$\sqrt{s}>100$  AGeV, hard and semihard scatterings, which can 
be calculated via pQCD, become increasingly important.  
Since nonperturbative physics is always present in any 
processes involving strong interactions, pQCD has to be combined
with models of nonperturbative interactions, {\em e.g.},
string models. Such pQCD-inspired models, like PYTHIA \cite{PYTHIA}, 
ISAJET \cite{ISAJET}, describe well the high $p_T$ phenomena
in high-energy hadronic collisions. Along the same line,
M.~Gyulassy and I extended this approach to high-energy
heavy-ion collisions and developed the HIJING (Heavy Ion Jet INtercation 
Generator) Monte Carlo model \cite{HIJING} to take into account 
the physics of hard and semihard parton production.
Nearly at the same time, similar models, PCM \cite{PCM} and 
DTUNUC \cite{DTUNUC}, also emerged. In particular, PCM
has explicitly modeled the space-time evolution of the
parton production processes and rescatterings along the lines
of Boal \cite{BOAL}. Though hard and semihard processes
become dominant at high energies, they are always
accompanied by soft interactions. These soft interactions
in general produce many soft final partons and take longer 
time ($\sim$ 1 fm/$c$) to complete. Therefore, they cannot be
modeled by simple elastic parton-parton scatterings.
However, these nonperturbative interactions might be modified 
inside a dense partonic medium, so that a perturbative method might 
be sufficient to describe the evolution of the initially produced 
dense partonic system towards an equilibrated quark-gluon plasma.

The crux of parton dynamics in heavy-ion collisions rests in
the complexity of multiple parton scatterings and
the associated bremsstrahlung. An exact quantum
field treatment of these multiple parton interactions
must include all possible matrix elements and the
interference between them. Although quantum transport 
theory \cite{HEHE,KLUGER,BLIA} can in principle include 
all these interference effects and developments in this
subject are impressive in the last few years, its 
applicability to heavy-ion collisions is still far from 
being realistic. An alternative and more realistic 
solution to this problem is to find ways to incorporate
correctly the subtle interference phenomena in a {\em classical}
parton cascade model. There are two important interference 
phenomena in multiple parton interactions.
(1) The interference among different amplitudes of
multiple parton scatterings, especially in the initial
parton scatterings when the two beams of fast partons pass
through each other, leads to the Glauber formula of multiple
interactions \cite{GLAUBER,GWLPM1}. Thus, a semiclassical cascade
cannot be used to treat the initial parton production during
the overlapping period of the two colliding nuclei.
The resultant interference is
responsible, as I will demonstrate later, for the disappearance
of the nuclear enhancement of jet production in $pA$ collisions
at large $p_T$ and high energies. Therefore, we can neglect
processes in which a parton suffers multiple hard scatterings in
the initial parton scatterings. Brodsky and Lu have shown \cite{BRLU} 
that a Glauber analysis of multiple antiquark-nucleon scatterings 
also leads to nonperturbative shadowing of the quark distributions
inside heavy  nuclei. (2) The spectrum of induced gluon
bremsstrahlung in multiple scatterings is also modified
by the destructive interference among different radiation
amplitudes, the so-called Landau-Pomeranchuk-Migdal (LPM) 
effect \cite{LPM,SHUL}.
In a detailed analysis \cite{GWLPM1,GWLPM2}, one finds
that the interference actually happens between two
amplitudes in which the beam parton has completely different
virtualities, {\em i.e.}, time-like in the final state radiation 
of one scattering and space-like in the initial state radiation of 
the previous one. Therefore, a propagating parton can no longer
be considered as {\em always} time-like with a decreasing 
virtuality in a parton cascade simulation, {\em e.g.}, PCM \cite{PCM}.

Another aspect of parton production and interaction
in ultrarelativistic heavy-ion collisions is that the
same processes responsible for parton equilibration
can also provide direct probes of the early parton
dynamics and the properties of the parton gas
during its evolution toward a fully equilibrated
quark-gluon plasma. For example, parton scatterings in
an equilibrating system also lead to dilepton, photon
and charmed quark production. Since the production rates
of these three processes are proportional to the product of 
quark-quark, quark-gluon and gluon-gluon densities, respectively,
measurements of the thermal enhancement of dileptons,
photons and charmed quarks can provide us with information
of the quark and gluon densities in the plasma, and thus on
the thermal and chemical equilibration. Interactions of an initially 
produced high-$p_T$ jet with partons inside the equilibrating 
plasma, in the meantime, will cause the jet losing energy.
Studying of jet quenching due to energy loss can thus
tell us whether and how the the parton gas is thermalized.
Because of the same mechanism, $J/\psi$ is also suppressed
due to its interaction with the parton gas. Since the
$J/\psi$ dissociation cross section inside a deconfined partons
is very different from that inside a hadronic gas, the study
of $J/\psi$ suppression can then provide us with evidence for 
color deconfinement in the parton plasma and possibly with
information on the QCD phase transition.

This review will be structured as follows: In the 
next section, I will review a pQCD-inspired model
for multiple parton production in $pp$, $pA$ and $AA$
collisions. Special emphasis will be devoted to the role
of hard and semihard parton collisions and their connection
to Pomeron structure and soft interactions. I will also discuss
the interpolation between hard and soft physics and
the possible treatment of nonperturbative interactions.
In Section 3, nuclear effects such as multiple
initial and final state interactions and nuclear shadowing
of parton distribution functions will be discussed. The
Monte Carlo implementation of parton production will also be
reviewed. In Section 4, I will discuss the evolution of a 
partonic system via pQCD, including the Landau-Pomeranchuk-Migdal (LPM) 
effect on induced gluon radiation. In Section 5, I will 
discuss the possibility of using hard processes as probes of
the early parton dynamics, the formation of quark-gluon plasma 
formation and the QCD phase transition. In particular,
photon and dilepton production, charm production, pre-equilibrium 
$J/\psi$ suppression, jet quenching, and monojet production will
be discussed. Finally I will give a summary
and discussion in Section 6.

\section{Parton Scatterings in Hadron-Hadron Collisions}

High-energy nuclear collisions are expected to be dominated 
by semihard parton collisions with transverse momentum transfers 
$p_T \gsim p_0 \sim 1-2$ GeV/$c$. The produced
partons carrying such transverse momenta are often referred
to as minijets. Multiple parton production has been estimated\cite{KLL} to 
produce up to 50\% (80\%) of the transverse energy per unit rapidity 
in the collisions of heavy nuclei at RHIC (LHC) energies.
While not resolvable as distinct jets, minijets are expected to lead to a 
wide variety of correlations among observables, such as transverse 
momentum, strangeness and fluctuation enhancements, that compete 
with expected signatures of a
QGP. Therefore, it is especially important to calculate these background
processes as reliably as possible. In addition, it has been shown that
multiple mini-jet production is important in $p\bar{p}$ interactions to
account for the increase of the total cross section\cite{GAISSER} with energy, 
the increase of the average transverse momentum with charged 
multiplicity\cite{XWRH}, and the violation of Koba-Nielsen-Olesen (KNO) 
scaling of the charged multiplicity distributions\cite{PYTHIA,WANG91}.

It has long been recognized \cite{GAISSER} that the inclusive jet
cross section $\sigma_{\rm jet}$ in $pp$ or $p\bar{p}$ collisions 
increases very rapidly with energy and eventually will be larger 
than the total inelastic cross section $\sigma_{\rm in}$. This
is because the inclusive jet cross section also contains
the average number of jets per inelastic event. Thus,
\begin{equation}
  \sigma_{\rm jet}=\langle n_{\rm jet}\rangle\sigma_{\rm in}, \label{eq:jet1}
\end{equation}
where $\langle n_{\rm jet}\rangle$ is the average number of hard
or semihard scatterings per inelastic event. One way to include
hard or semihard processes in the unitarized cross sections of
$pp$ and $p\bar{p}$ is to introduce the idea of Pomeron exchanges
in an eikonal formalism. I will argue that hard or semihard 
scatterings can be considered as the hard loops inside a Pomeron 
exchange, and the interpolation between hard and soft components
will become clear. This will also shed some light on the modeling
of the soft component.

\subsection{Pomeron Exchange and Minijet Production}

The Pomeron was introduced to describe the effective 
interaction between quarks and gluons \cite{POM,GLR,LANDSHOFF1}. 
It has been used to model hadronic cross sections, diffractive
interactions and multiple particle production. In general,
it is considered to be a color singlet object with the quantum numbers
of a photon. It can couple both to quarks and gluons. The imaginary
part of a Pomeron exchange amplitude is usually related to multiple
particle production in hadronic interactions. Let us first
consider the eikonal formalism for quark scatterings
via multiple  Pomeron exchange.

Denote the matrix element for a single Pomeron exchange as
\begin{equation}
{\cal M}_{\rm el}^{(1)}=2\pi i\delta(E_i-E_f)2\pi\sqrt{s}f_1(s,t),
\label{eq:ap1}
\end{equation}
\begin{eqnarray}
f_1(s,t)&=&\frac{-i}{2\pi}\frac{g_p}{i\sqrt{s}}\bar u_{\sigma_f}(p_f)
\not\!\!A({\bf q})u_{\sigma_i}(p_i) \nonumber\\
        &\equiv&\frac{-i}{2\pi}\frac{1}{i\sqrt{s}}T(s,{\bf q}), \label{eq:ap2}
\end{eqnarray}
where $\sqrt{s}/2=E_i=E_f$ and $g_p$ is a coupling constant between a quark
and a Pomeron. The amplitude $f(s,t)$ is defined such that the 
differential cross section for elastic scatterings is given by
\begin{equation}
  \frac{d\sigma}{dt}=\pi |f(s,t)|^2. \label{eq:ap3}
\end{equation}
The initial and final polarizations, $\sigma_i$, $\sigma_f$,
should be averaged and summed over in the calculation of the cross 
sections. One can check that if $A({\bf q})$ is replaced by
a Debye screened Coulomb potential, $A({\bf q})=g/({\bf q}^2+\mu^2)$,
and $g$ by the strong coupling constant, the above formula 
leads to $d\sigma/dt=4\pi\alpha_s^2/({\bf q}^2+\mu^2)^2$.

One can similarly write down the amplitude for double Pomeron exchange,
\begin{eqnarray}
{\cal M}_{\rm el}^{(2)}=2\pi i\delta(E_i-E_f)(-g^2_p)&\int\frac{d^3\ell}{(2\pi)^3}
\bar u_{\sigma_f}(p_f)\not\!\!A({\bf p}_f-{\bf \ell})
\frac{\not\,\ell}{\ell^2+i\epsilon}\not\!\!A({\bf \ell}-{\bf p}_i)
 u_{\sigma_i}(p_i) \nonumber \\
&e^{-i({\bf \ell}-{\bf p}_i)
\cdot{\bf x}_1-i({\bf p}_f-{\bf \ell})\cdot{\bf x}_2},\label{eq:ap4}
\end{eqnarray}
where energy conservation at each coupling vertex sets the energy
of the internal line to $\ell^0=\sqrt{s}/2=E_f$. 
Amplitudes involving backscattering are suppressed 
at high energies, because of the limited momentum transfer
that each the Pomeron can impart. If we assume that Pomerons do not overlap 
in space, the singularity in $A({\bf q})$ can be neglected, and the 
integration over $\ell_z$ (with respect to the $\hat{z}$ direction of 
${\bf x}_{21}={\bf x}_2-{\bf x}_1=L\hat{z}+{\bf r}_{\perp}$) gives us
\begin{eqnarray}
{\cal M}_{\rm el}^{(2)}&=&2\pi i\delta(E_i-E_f)
\int\frac{d^2\ell_{\perp}}{(2\pi)^2}
\bar u_{\sigma_f}(p_f)(-g^2_p)\Gamma_{(2)} u_{\sigma_i}(p_i),\nonumber\\
\Gamma_{(2)}&=&\not\!\!A({\bf p}_f-{\bf p})
\frac{\not\!p}{2ip_z}\not\!\!A({\bf p}-{\bf p}_i)
e^{-i({\bf p}-{\bf p}_i)
\cdot{\bf x}_1-i({\bf p}_f-{\bf p})\cdot{\bf x}_2},\label{eq:ap5}
\end{eqnarray}
where $p=(E_f,\sqrt{E_f^2-\ell_{\perp}^2},{\bf \ell}_{\perp})$ is the
four-momentum of the internal line. One can derive the classical
Glauber multiple collision cross section from this amplitude by 
averaging and summing over the initial and final state ensemble of the
target\cite{GWLPM1}. In the limit of high-energy and small angle scattering,
one can neglect the phase factor in the above equation and obtain
the amplitude [as defined in Eq.~(\ref{eq:ap1})],
\begin{equation}
f_2(s,t)=\frac{-i}{2\pi}\int d^2b\frac{1}{2!}[-\chi({\bf b},s)]^2
e^{i{\bf q}_{\perp}\cdot{\bf b}}, \label{eq:ap6}
\end{equation}
where $1/2!$ comes from the different orderings of the target potentials
and ${\bf q}_{\perp}={\bf p}_{f\perp}-{\bf p}_{i\perp}$ is the total
transverse momentum transfer from the multiple scatterings. The
eikonal function $\chi_{\sigma_1,\sigma_2}({\bf b},s)$ is defined as the
Fourier transform of the single scattering amplitude (besides a 
factor $i/2\pi$),
\begin{eqnarray}
\chi({\bf b},s)&\equiv&\frac{i}{\sqrt{s}}\int\frac{d^2q_{\perp}}{(2\pi)^2}
e^{-i{\bf q}_{\perp}\cdot{\bf b}}T(s,{\bf q}_{\perp}) \nonumber \\
   &=&\frac{i}{4\pi \sqrt{s}}\int_{-\infty}^0dt J_0(b\sqrt{-t})T(s,t), 
\label{eq:ap7}
\end{eqnarray}
where $t=-{\bf q}_{\perp}^2$ and $J_0$ is the zeroth order Bessel 
function. In the definition of the product of eikonal functions, 
summation over the polarizations of the intermediate lines is implied. 
If we are only interested in unpolarized collisions, we
can regard $T(s,{\bf q}_{\perp})$ as a scalar in the following.
One can generalize the double scattering amplitude to multiple
scatterings and sum them together to get the total amplitude,
\begin{eqnarray}
f(s,t)=\sum_{n=1}^{\infty}f_n(s,t)&=&\frac{-i}{2\pi}
\int d^2b \sum_n \frac{1}{n!}[-\chi({\bf b},s)]^n 
e^{i{\bf q}_{\perp}\cdot{\bf b}} \nonumber \\
&=&\frac{i}{2\pi}\int d^2b [1-e^{-\chi({\bf b},s)}]
e^{i{\bf q}_{\perp}\cdot{\bf b}}. \label{eq:ap9}
\end{eqnarray}
This is the total elastic quark scattering amplitude in the 
eikonal formalism. This amplitude can also be derived from the 
scattering theory in quantum mechanics \cite{SCHIFF}, where the 
eikonal function is related to the potential experienced by the
scattering particle.

One can consider a $pp$ interaction as multiple quark-quark scatterings
and derive a similar formula for the total amplitude of an elastic
$pp$ collision. The effective eikonal function for $pp$ collisions
now should be given by
\begin{equation}
\chi({\bf b},s)=\frac{i}{\sqrt{s}}\int\frac{d^2q_{\perp}}{(2\pi)^2}
e^{-i{\bf q}_{\perp}\cdot{\bf b}}t({\bf q}_{\perp},s)t(-{\bf q}_{\perp},s)
T(s,{\bf q}_{\perp}), \label{eq:eikpp}
\end{equation}
where $t({\bf q}_{\perp},s)$ is the Fourier transform of the
quark density distribution inside a nucleon and $T(s,{\bf q}_{\perp})$
is the amplitude of a quark-quark scattering with one Pomeron exchange.
Here we neglected the contribution from Reggeon exchanges, so that
the amplitude $T(s,{\bf q}_{\perp})$ is purely 
imaginary  \cite{BRLU} (or the eikonal function is real); 
thus the elastic amplitude of $pp$
collisions, Eq.(~\ref{eq:ap9}), is also purely imaginary. Experimental
measurements of high-energy $pp$ or $p\bar{p}$ collisions indeed 
find the real part of the elastic amplitude to be small~\cite{ELST}.

With the above elastic amplitude, the differential elastic cross 
section  for $pp$ collisions is given by Eq.~(\ref{eq:ap3}). Using 
the identity
\begin{equation}
\int^0_{-\infty}dt J_0(b\sqrt{-t})J_0(b^{\prime}\sqrt{-t})
       =\frac{2}{b}\delta(b-b^{\prime}),
\end{equation}
one can obtain the elastic cross section,
\begin{equation}
\sigma_{\rm el}=\int{d^2b}\left[1-e^{-\chi(b,s)}\right]^2, \label{eq:elcr}
\end{equation}
where we assumed that the imaginary part of the eikonal function 
is negligible. Using the optical theorem, one can also get the 
total and inelastic cross sections of $pp$ collisions\cite{WANG91,WANGTH},
\begin{eqnarray}
\sigma_{\rm tot}&=&4\pi{\rm Im}f(s,t=0)
           =2\int{d^2b}\left[1-e^{-\chi(b,s)}\right], \label{eq:totcr} \\
\sigma_{\rm in}&=&\sigma_{\rm tot}-\sigma_{\rm el}
           =\int{d^2b}\left[1-e^{-2\chi(b,s)}\right]. \label{eq:incr}
\end{eqnarray}

The approximation of a small imaginary (real) part of the eikonal 
function (elastic amplitude) in Eq.~(\ref{eq:eikpp}) also leads 
us to assume that the quark-quark scattering amplitude 
$T(s,{\bf q}_{\perp})$ with one Pomeron exchange is also 
purely imaginary [cf. Eqs.~(\ref{eq:ap9}) and (\ref{eq:eikpp})]. 
According to Eq.~(\ref{eq:ap2}) and the optical theorem ,
$T(s,{\bf q}_{\perp})$ should be related to the total
inclusive cross section of quark-quark scatterings,
\begin{equation}
T(s,{\bf 0})=-i\frac{\sqrt{s}}{2}\sigma_{\rm incl}(s). \label{eq:incl}
\end{equation}
Assuming the dominance of small angle contributions in Eq.~(\ref{eq:eikpp}),
we have then the eikonal function for $pp$ collisions,
\begin{equation}
\chi(b,s)\approx \frac{1}{2}\sigma_{\rm incl}(s)T_N(b,s), \label{eq:eikp2}
\end{equation}
where $T_N(b,s)$ is the overlap function of two nucleons,
\begin{equation}
  T_N(b,s)=\int{d^2b^{\prime}}t({\bf b}^{\prime},s)
                          t({\bf b}-{\bf b}^{\prime},s),
\end{equation}
which is normalized to one, $\int{d^2b}T_N(b,s)=1$.

\subsection{Cross Sections: Soft vs. Hard}

\begin{figure}
\centerline{\psfig{figure=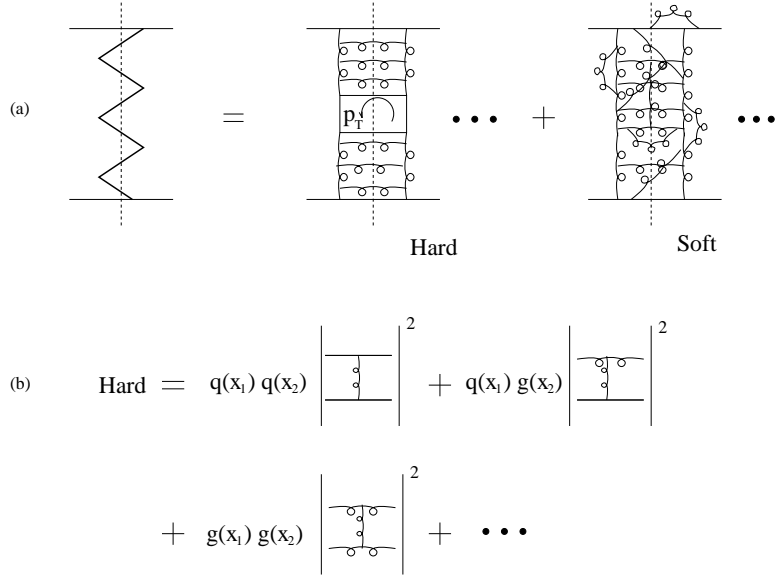,width=4in,height=3in}}
\caption{
(a) A Pomeron can be divided into the ladder diagrams of a hard
  part,  in which at least one loop momentum $p_T$ is large, and a 
  corresponding soft part. (b) The hard part of a cut-Pomeron can be 
  factorized into parton distributions and hard scattering matrix elements.}
\label{fig1}
\end{figure}

To calculate the total inclusive cross section of quark-quark
scatterings, $\sigma_{\rm incl}(s)$, one needs to know the Pomeron
structure and its coupling to quarks and gluons. In a simple model,
one can regard a Pomeron as a double gluon exchange with a ladder
structure as illustrated in Fig.~\ref{fig1}(a). In general, most of the
loops inside the ladder have soft momenta and thus cannot be
calculated via pQCD. We consider this nonperturbative contribution
to the inclusive cross section as $\sigma_{\rm soft}(s)$. As
the colliding energy $\sqrt{s}$ increases, contributions from
large momentum, $p_T>p_0$, to the loop integral also increase.
As illustrated in Fig.~\ref{fig1}(b), the cutting diagram of such a
Pomeron exchange which contains at least one hard loop corresponds
to hard scatterings in $pp$ collisions. If one assumes that the
loop momenta are ordered so that they increase toward the
hard loop, then the rest of the cut-Pomeron beside the hard
loop can be resummed to give the quark or gluon (parton) distributions
of a nucleon and their Gribov-Lipatov-Altarelli-Parisi 
(GLAP) \cite{GLAP} evolution. The corresponding contribution to 
the inclusive cross section can then be identified as the inclusive
jet cross section \cite{EHLQ},
\begin{eqnarray}
        \sigma_{\rm jet}&=&\int_{p_0^2}^{s/4}dp_T^2dy_1dy_2\frac{1}{2}
                \frac{d\sigma_{\rm jet}}{dp_T^2dy_1dy_2}, \label{eq:sjet2}\\
        \frac{d\sigma_{\rm jet}}{dp_T^2dy_1dy_2} &=& K
        \sum_{a,b} x_1  f_a(x_1,p_T^2) x_2 f_b(x_2,p_T^2)
        \frac{d\sigma^{ab}(\hat{s},\hat{t},\hat{u})}{d\hat{t}},\label{eq:sjet1}
\end{eqnarray}
where the summation runs over all parton species, $y_1$ and $y_2$ 
are the rapidities of the scattered partons,  and $x_1$ and $x_2$ 
are the light-cone momentum fractions
carried by the initial partons. These variables are related by
$x_1=x_T(e^{y_1}+e^{y_2})/2$, $x_2=x_T(e^{-y_1}+e^{-y_2})/2$,
$x_T=2p_T/\sqrt{s}$. The $f_a(x,Q^2)$ in Eq.~(\ref{eq:sjet1}) 
are the parton distribution
functions. The pQCD cross sections, $d\sigma_{ab}$, depend on 
the subprocess variables
$\hat{s}=x_1 x_2 s$, $\hat{t}=-p_T^2(1+ \exp(y_2-y_1))$,
and $\hat{u}=-p_T^2 (1+\exp(y_1-y_2))$.
A factor $K\approx 2$ is included to correct the lowest order 
pQCD rates for next to leading order effects \cite{EKS,EWPROB}.

At small $x\sim p_T/\sqrt{s}$, the momentum in the ladder
diagram might not be ordered anymore and there might be
more than one hard loop inside the ladder. However, 
calculations \cite{LANDSHOFF2} have shown that such contributions 
are still small at present and future collider energies. If we
neglect contributions from multiple hard loops , the total inclusive
quark-quark cross section can be written as
\begin{equation}
  \sigma_{\rm incl}=\sigma_{\rm soft}(s)+\sigma_{\rm jet}(s).
                        \label{eq:incl2}
\end{equation}
where $\sigma_{\rm soft}$ represents the contribution from
the cut-Pomeron in which the transverse momenta of the loops
are limited to $p_T<p_0$, and therefore is not calculable via
pQCD. However, as we can see from its definition,
$\sigma_{\rm soft}$ represents processes involving many
soft partons in the final state. It cannot be modeled by
a simple two parton elastic scattering. Since there is no
ordering  in $p_T$, interference becomes important in soft 
particle production, which can be modeled by, {\em e.g.}, 
a string model.

Given the above decomposition of the total inclusive cross section
and the eikonal function in Eq.~(\ref{eq:eikp2}), we can rewrite the
inelastic cross section for $pp$ collisions, Eq.~(\ref{eq:incr}), as
\begin{eqnarray}
  \sigma_{\rm in}&=&\int{d^2b}[1
      -e^{-(\sigma_{\rm soft}+\sigma_{\rm jet})T_N(b,s)}]
      \equiv\int{d^2b}\sum_{j=0}^{\infty}g_j(b); \label{eq:cin} \\
      g_j(b)&=&\frac{[\sigma_{\rm jet}T_N(b,s)]^j}{j!}
      e^{-\sigma_{\rm jet}T_N(b,s)},\;\; j\geq 1; \label{eq:sjet3} \\
    g_0(b)&=&[1-e^{-\sigma_{\rm soft}T_N(b,s)}]e^{-\sigma_{\rm jet}T_N(b,s)}.
                \label{eq:sjet4}
\end{eqnarray}
where $g_j(b)$ can be considered as the probability for $j$
hard scatterings \cite{WANG91,DURAND} among inelastic collisions
at fixed impact paramenter $b$. Note that a jet in our terminology 
refers to a large $p_T$ hard scattering. The average number of jets 
with $p_T\geq p_0$ in $pp$ collisions is thus 
$\langle n_{\rm jet}\rangle=\sigma_{\rm jet}/\sigma_{\rm in}$.
The above probabilistic interpretation of multiple hard 
scatterings depends on our assumption that multiple 
cut-Pomerons are independent from each other.
This holds as long as the average number of hard scatterings
is not too large. Given an interaction transverse area,
$\sim \pi/p_0^2$, for processes with $p_T\sim p_0$, independence
requires that the total interaction area is less than $\pi R_N^2$,
where $R_N\approx 0.85$ fm is the nucleon radius; {\it i.e.},
\begin{equation}
\sigma_{\rm jet}\lsim (p_0 R_N)^2\sigma_{\rm in}\equiv\sigma_{\rm max}\;\;.
         \label{indep}
\end{equation}
For $p_0\gsim 2 $ GeV/$c$, and $\sigma_{\rm in}\approx 40$ mb, the 
right hand side is $\sigma_{\rm max} \approx 3$ barns, 
and thus the independent approximation should hold up
to the highest energies foreseen. For nuclear collisions the total 
number of jets is given by
\begin{equation}
N^{AA}_{\rm jet}=T_{AA}(b) \sigma_{\rm jet}\; \; , \label{naa}
\end{equation}
where $T_{AA}(b)$ is the nuclear overlap function
at an impact parameter $b$. For $b=0$, $T_{AA}\approx A^2/\pi R_A^2$,
and multiple mini-jets may be independent as long as
\begin{equation}
\sigma_{\rm jet} \lsim (p_0 R_A)^2 \pi R_A^2/A^2 \approx 2 
\sigma_{\rm max}/A^{2/3} 
                \; \; . \label{indepaa}
\end{equation}
For $A=197$ the right hand side is 180 mb, and thus independence
should apply up to LHC energies (see Fig.~\ref{fig2} below). 
On the other hand,
nuclear shadowing of the initial structure functions will reduce the
jet cross section so that independence should be valid beyond LHC
energies.

       For $p_0 > 1$ GeV, $\sigma_{\rm jet}(s)$ is found
to be very small when $\sqrt{s}$\lsim~20~GeV and only the
soft component is important. The low-energy data of diffractive 
nucleon-nucleon scatterings exhibit a number of geometrical 
scaling properties\cite{GSC1} in the range $10<\sqrt{s}<100$~GeV, 
{\em e.g.},  $\sigma_{\rm el}/\sigma_{\rm tot}\cong 0.175$, and 
$B/\sigma_{\rm tot}\cong 0.3$, where $B$ is the slope of the 
diffractive peak of the differential elastic cross section. 
This suggests a geometrical scaling form \cite{GSC2}
for the eikonal function at low energies; {\em i.e.}, 
it is only  a function of $\xi=b/b_0(s)$, 
with $\pi b_0^2(s)\equiv\sigma_{\rm soft}(s)/2$ providing 
a measure of the geometrical size of the nucleon. We further 
assume that the nucleon overlap function is given by the 
{}Fourier transform of a dipole form factor so that,
\begin{eqnarray}
        T_N(b,s)&=&\frac{\chi_0(\xi)}{\sigma_{\rm soft}(s)};\label{eq:over1}\\
        \chi_0(\xi)&=&\frac{\mu_0^2}{48}(\mu_0 \xi)^3 K_3(\mu_0 \xi),
                \;\; \xi=b/b_0(s),\label{eq:over2}
\end{eqnarray}
where $\mu_0=3.9$.  With this assumption, the eikonal function 
can be written as,
\begin{equation}
        \chi(b,s)\equiv\chi(\xi,s)
        =\chi_0(\xi)[1 +\sigma_{\rm jet}(s)/\sigma_{\rm soft}(s)]\;.
\end{equation}
This form ensures that geometrical scaling\cite{GSC2}
is recovered at low energies when $\sigma_{\rm jet} \ll \sigma_{\rm soft}$.

        Choosing $p_0\simeq 2$ GeV/$c$ and assuming a constant 
value of $\sigma_{\rm soft}=57$ mb at high energies, 
the calculated cross sections and the multiplicity 
distributions in $pp$ and $p\bar{p}$ collisions agree well with 
experiments\cite{WANG91}. This is the model adopted in 
HIJING \cite{HIJING} to simulate multiple jet
production at the level of nucleon-nucleon collisions.
In Fig.~\ref{fig2}, the calculated total, inelastic and elastic 
cross sections of $pp$ or $p\bar{p}$ collisions are shown as 
functions of $\sqrt{s}$ (solid line) together with experimental 
data\cite{GSC1,CRS2,CRS3,CRS4,CRS5,CRS6}. The dashed line
corresponds to the inclusive jet cross section. The calculated
cross sections agree well with experiments from ISR to Tevatron 
and cosmic-ray energies. We note that the total inclusive
jet cross section increases much faster than the inelastic
cross section as a function of $\sqrt{s}$, leading to an 
increase of the average number of minijets, 
$\sigma_{\rm jet}/\sigma_{\rm in}$, with energy.

\begin{figure}
\centerline{\psfig{figure=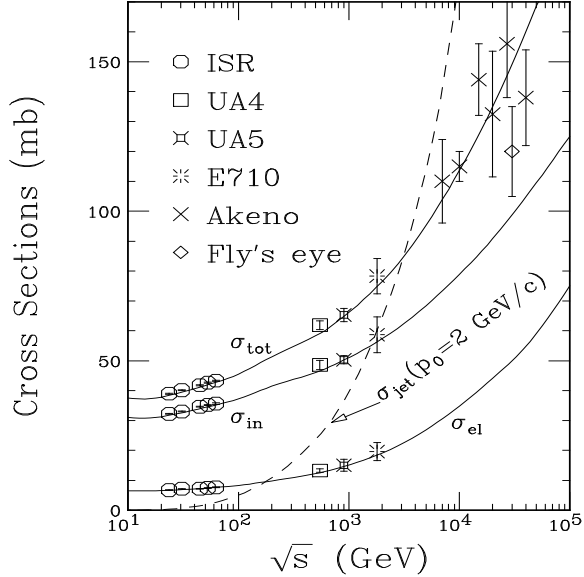,width=3in,height=3in}}
\caption{
The total, inelastic, and elastic cross sections of
        $pp$ and $p\bar{p}$ collisions calculated by HIJING (solid
        lines) as compared to the data 
        \protect\cite{GSC1,CRS2,CRS3,CRS4,CRS5,CRS6}.
        The dashed line is the total inclusive jet cross section
        with $p_T\geq p_0=2$ GeV/$c$.}
\label{fig2}
\end{figure}

        We emphasize that the value of $p_0=2$ GeV used in the above
calculation is only a phenomenological parameter. In order for the 
model to have some predictive power, $p_0$ should not depend 
on $\sqrt{s}$.  However, its value is subject to considerable 
controversy\cite{GAISSER}. The problem arises from the fact that
there is not a clear boundary specified by $p_0$ between soft and 
hard processes, as shown in Eqs.~(\ref{eq:sjet2}) and (\ref{eq:incl2}).
If we believe that pQCD can be reliably applied to calculate
hard contributions to the total inclusive cross section up to $p_0$,
then the rest must be modeled by a phenomenological $\sigma_{\rm soft}$.
Clearly, $\sigma_{\rm soft}(s)$ must depend on the choice of $p_0$.
With a smaller (larger) $p_0$, more (less) contributions are included 
as hard collisions. Hence $\sigma_{\rm soft}(s)$ would be smaller (larger).  
Obviously  many choices of $p_0$ and $\sigma_{\rm soft}(s)$ 
can give the same total cross section $\sigma_{\rm tot}(s)$.
The only restriction is that the sum $\sigma_{\rm soft}(s)
+\sigma_{\rm jet}(s)$ must have the right value to give the
right energy dependence of the total cross section $\sigma_{\rm tot}(s)$.
Since $\sigma_{\rm jet}(s)$ increases with decreasing $p_0$ 
and $\sigma_{\rm soft}(s)$ is non-negative, $p_0$ must be 
bounded from below by the experimental data on the total 
cross section $\sigma_{\rm tot}(s)$. We find that this lower
limit in our model is $p_0=1.2$ GeV with the given parton 
distribution functions. For $p_0$ smaller than $1.2$~GeV, 
the inclusive jet cross section at high energies is overestimated
and the resultant $\sigma_{\rm tot}(s)$ can never fit the data. 
In addition, one must keep in mind that the lowest $p_0$ is also 
bounded by the relevant $Q_0=p_0$ in the evolution of the parton 
distribution functions.

In HIJING, the parton distribution functions are taken to be the 
Duke-Owens\cite{DUKE} parametrization set 1 with $Q_0=2$ GeV.
The latest version of this parameterization (DO1.1) for $f_a(x,Q^2)$ 
is adequate through RHIC energies. However, at higher energies, 
more updated distributions, such as GRV \cite{GRV} or 
MRSD$-^{\prime}$ \cite{MRS} should be used. These distributions,
constrained by the most recent HERA data \cite{HERA} in deeply 
inelastic $ep$ collisions, are more divergent at small $x$ and give
rise to larger minijet cross sections, especially at the LHC
energy \cite{EKV94}. Using these distributions, 
one cannot fit $\sigma_{\rm tot}(s)$ at 
high energies with fixed $p_0$. As has been discussed in 
Ref.~\cite{LANDSHOFF2}, this is due to multiple
hard loops in one Pomeron exchange. The divergent behavior
of the distribution functions is usually related to the
fact that the hard loops are not ordered in $p_T$, which
gives rise to multiple jet pair production per cut-Pomeron.
Thus the inclusive jet cross section in the eikonal function
must be normalized by the average number of such jet pairs \cite{KEXW1}.

\subsection{Modeling the Soft Interactions}

From Fig.~\ref{fig1}(a), we can see that a cut-Pomeron always 
produces many soft partons with small transverse momenta no matter
whether there is a hard process present or not. Unlike
partons from initial and final state radiations associated with
a hard scattering, these 
soft partons are not ordered in transverse momentum and
coherence is extremely important, which could virtually produce
a coherent color field. Therefore, soft interactions
may not be modeled simply by regularized parton-parton
elastic scatterings with small transverse momentum transfers.
On the other hand, the color field could also be screened
significantly later by the interaction with the hard or semihard 
partons from  the hard collisions \cite{KEMG} in
heavy-ion collisions. The screening effect will  decrease
the final particle production from the color field.

As demonstrated in three-jet events in $e^+e^-$ annihilations, 
the color interference effects can be approximated fairly well
by a string model \cite{LUND}. In HIJING we adopted a 
variant of the multiple string phenomenological  model 
for the soft interaction as developed in 
Refs.~\cite{FRIT,DPM,VENUS}. Those soft interactions 
must naturally involve small $p_T$ transfer to the constituent 
quarks, as well as induced soft gluon radiation which can be 
modeled by introducing kinks in the strings. The produced
partons are treated either as hard kinks (for gluons) on 
the string or the end points (for quarks) of another string.
The strings are assumed to decay independently via quark-antiquark 
creation using, in our case, the Lund JETSET7.2 \cite{JETSET} 
fragmentation routine to describe the hadronization. 

The string excitation is achieved by a collective momentum
transfer $P=(P^+,P^-,{\bf p}_T)$ between the hadrons. Given 
initial light-cone momenta 
\begin{equation}
        p_1=(p_1^+,\frac{m_1^2}{p_1^+},{\bf 0}_T),\;\;\;\; 
        p_2=(\frac{m_2^2}{p_2^-}, p_2^-,{\bf 0}_T), 
\end{equation} 
with $(p_1^+ +m_2^2/p_2^-)(p_2^- +m_1^2/p_1^+)\equiv s$, the final momenta 
of the strings are assumed to be 
\begin{equation}
        p'_1=(p_1^+ -P^+,\frac{m_1^2}{p_1^+}+P^-, {\bf p}_T),\;\;\;\;
        p'_2=(\frac{m_2^2}{p_2^-}+P^+, p_2^- -P^-,-{\bf p}_T).
\end{equation}
The remarkable feature of soft interactions is that low
transverse momentum exchange processes with $p_T\lsim 1$ GeV/$c$ 
can result in large effective light-cone momentum exchanges\cite{FRIT}, 
giving rise to two excited strings with large invariant mass.  Defining
\begin{equation}
        P^+=x_+\sqrt{s}-\frac{m_2^2}{p_2^-},\;\;\;\;
        P^-=x_-\sqrt{s}-\frac{m_1^2}{p_1^+}\;, 
\end{equation}
the excited masses of the two strings will be 
\begin{equation}
        M_1^2=x_-(1-x_+)s-p_T^2, \;\;\;\; M_2^2=x_+(1-x_-)s-p_T^2,
                        \label{eq:strnms} 
\end{equation} 
respectively. In HIJING, we require that the excited 
string mass must exceed a minimum value $M_{\rm cut}=1.5$
GeV, and therefore the kinematically allowed region of $x^{\pm}$ 
is restricted to 
\begin{equation}
  x_-(1-x_+)\geq M_{Tcut1}^2/s,\;\;\;\; x_+(1-x_-)\geq M_{Tcut2}^2/s,
                        \label{eq:xregn} 
\end{equation} 
where $M_{T{\rm cut}1}^2=M_{\rm cut}^2+p_T^2$, 
$M_{T{\rm cut}2}^2=M_{\rm cut}^2+p_T^2$.
Only collisions with 
\begin{equation}
        s\geq s_{min}=(M_{T{\rm cut}1}+M_{T{\rm cut}2})^2 \label{eq:smin} 
\end{equation}
are allowed to form excited strings. 
Eq.~(\ref{eq:smin}) also determines the maximum $p_T$ that the
strings can obtain from the soft interactions.
In events with both hard and soft processes,
two strings are still assumed to form but with a kinetic boundary
reduced by the hard scatterings.

In HIJING, the probability for light-cone momentum transfer 
is assumed to be 
\begin{equation}
        P(x_{\pm})=\frac{(1.0-x_{\pm})^{1.5}}
                {(x_{\pm}^2+c^2/s)^{1/4}}     \label{eq:xdistr1}
\end{equation} 
for nucleons and 
\begin{equation}
       P(x_{\pm})=\frac{1}{(x_{\pm}^2+c^2/s)^{1/4}
                [(1-x_{\pm})^2+c^2/s]^{1/4}}   \label{eq:xdistr2}
\end{equation} 
for mesons, with $c=0.1$ GeV, along the lines of the DPM model\cite{DPM}. 
Soft gluon bremsstrahlung processes with $p_T<p_0$ are also introduced
as kinks along the excited string.
In addition, HIJING also includes an extra low $p_T<p_0$ transfer
to the constituent quarks and diquarks at the string end points
in soft interactions. This effect is important at low energies,
$E_{\rm lab} \sim 20 $ GeV, to account for the high $p_T$ tails of 
the pion and proton distributions \cite{HIJING}.

I emphasize that the low $p_T$ algorithm used in HIJING
is a phenomenological model needed to incorporate non-perturbative
aspects of beam jet physics. Many variants of soft dynamics can be 
envisioned, but none can be rigorously defended from fundamental QCD. 
One of the attractive aspects of going to the highest possible 
collider energies is that the theoretical uncertainties due to 
soft dynamics are reduced, as more and more of the dynamics becomes 
dominated by calculable semi-hard and hard QCD processes. 

\subsection{Minijets and Transverse Flow}

\begin{figure}
\centerline{\rotateright{\psfig{figure=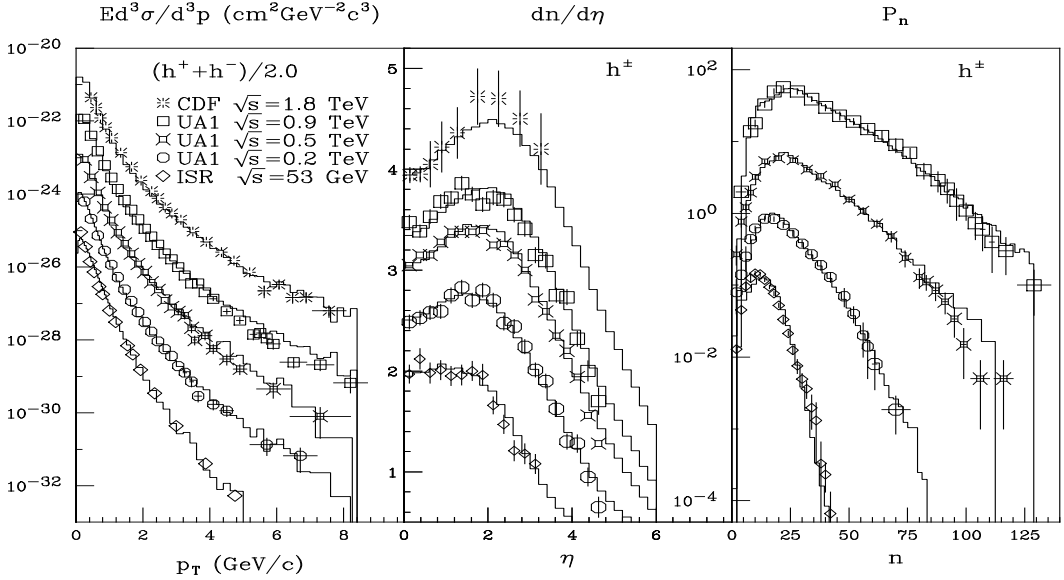,width=3in,height=5.5in}}}
\caption{
Data on charged particle inclusive $p_T$ 
        spectra $Ed^3\sigma/d^3p$ \protect\cite{bspt,ua1pt,cdfpt}, 
        pseudorapidity distributions $dn/d\eta$ 
        \protect\cite{dndyua5,dndycdf}, and multiplicity 
        distributions \protect\cite{pnisr,pnua5,pn2ua5} $P_n$ in $pp$ 
        and $p\bar{p}$ collisions compared to HIJING calculations 
        at different energies $\protect\sqrt{s}$.
        The transverse momentum spectra and multiplicity
        distributions have been displaced for clarity by extra factors of 10
        relative to the absolutely normalized data at 
        $\protect\sqrt{s}=53 $ GeV.}
\label{fig3}
\end{figure}

\begin{figure}
\centerline{\rotateright{\psfig{figure=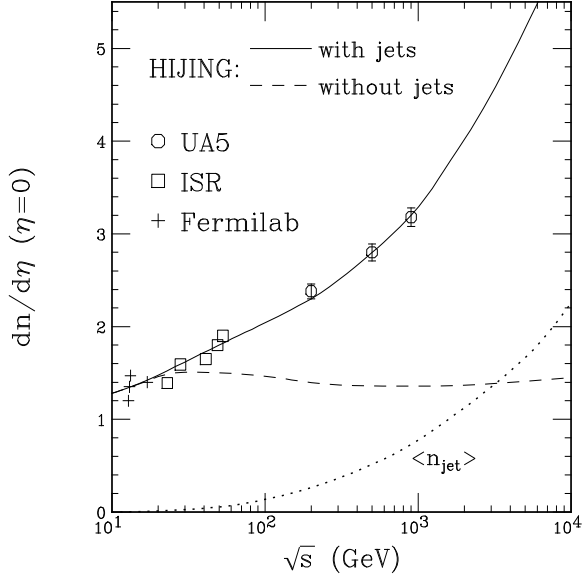,width=3in,height=3in}}}
\caption{
$dn_{\rm ch}/d\eta(\eta=0)$ in inelastic $pp$ and 
        $p\bar{p}$ collisions as a function of $\protect\sqrt{s}$. 
        The solid line is the HIJING calculation compared to
        the data \protect\cite{dndyua5,dndyisr,dndyfnal}. The dashed line 
        is for events without jet production in HIJING simulations. 
        The dotted line is the calculated average number
        of jets,$\langle n_{jet}\rangle
        =\sigma_{\rm jet}/\sigma_{\rm in}$.}
\label{fig4}
\end{figure}

        To summarize the consistency of HIJING calculations with
the available experimental data, I show in Fig.~\ref{fig3} the 
inclusive spectra of charged particles in $pp$ and $p\bar{p}$ over a wide
energy range, $\sqrt{s}=$50--1800 GeV. We see that the model accounts
well for the energy dependence of not only the transverse momentum
distribution, but also the rapidity distribution, as well as the
multiparticle fluctuations. This overall quantitative understanding
of multiparticle observables in hadronic interactions, especially
the magnitude and energy dependence of the conspicuous power-law
tail of the $p_T$ spectrum characteristic of pQCD, strongly supports
the importance of minijets physics at collider energies. To
demonstrate the onset of particle production with minijet
production, I plot in Fig.~\ref{fig4} the energy dependence of the
central rapidity density of produced charged particles.
I also show the contribution from purely soft interactions.
Since the central rapidity density from soft string
fragmentations is almost constant as a function of the
colliding energy, the increased $dn_{\rm ch}/d\eta$ mainly
comes from the hadronization of jets at high energies.
The correlation between the central rapidity density and 
minijets is very clear when $dn_{\rm ch}/d\eta$ is compared 
with the average number of minijets, $\langle n_{jet}\rangle
=\sigma_{\rm jet}/\sigma_{\rm in}$, as functions of $\sqrt{s}$.

        Beside providing overall agreement with experimental data 
on multiparticle distribution and spectra \cite{PYTHIA,HIJINGPP}, 
minijet production will also influence correlations between 
observed quantities, some of which could be mistakenly 
attributed to QGP and other collective phenomena.
One typical example is the multiplicity and mass dependence
of the average transverse momentum observed in $p\bar{p}$
collisions at Tevatron Collider energy \cite{TEVAFLOW}.
In hydrodynamic models, transverse
collective flow is usually generated from the expansion of 
the thermalized dense system \cite{HYDRO1}. Since all hadrons 
have the same flow velocity, heavy particles tend to have 
larger transverse momentum when they finally freeze out.  
If the average transverse momentum is plotted against 
the total multiplicity, it is anticipated that 
for heavy particles it will be larger and the increase 
with the multiplicity will be faster than for light ones. 
Quite surprisingly, experiments on $p\bar{p}$ collisions 
at Tevatron Collider energy have recently reported
observation of just this effect \cite{TEVAFLOW}.
L\'{e}vai and M\"{u}ller \cite{BMPL}  have studied this 
reaction in a linearized transport theory. They found that there 
is no time for the baryons  to equilibrate with the  pions 
during the expansion of a hadronic fireball. They therefore
suggested that a more novel explanation could be
required to account for the apparent similarity of the 
flow velocities of the mesons and baryons. They
noted  that these observations could be understood
if an equilibrated quark-gluon plasma (QGP) were formed in these 
collisions\cite{VHOVE}.

\begin{figure}
\centerline{\rotateright{\psfig{figure=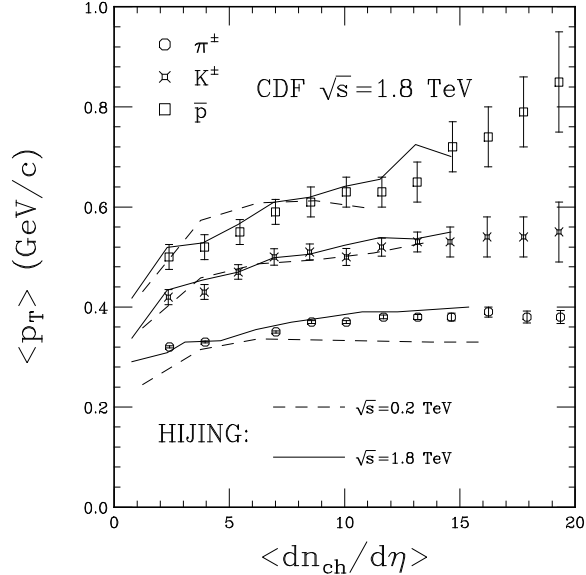,width=3in,height=3in}}}
\caption{$\langle p_T\rangle$ of 
        pions, kaons, and anti-protons
        (from bottom to top) in $-0.36<\eta<1.0$ versus the average
        charged multiplicity density $dn_{\rm ch}/d\eta$ in $|\eta|<3.25$.
        The histograms are HIJING results and points are 
        the data \protect\cite{TEVAFLOW}.
        The dashed lines are for $pp$ at $\protect\sqrt{s}=200$ GeV.}
\label{fig5}
\end{figure}

        However, as also noted by L\'{e}vai and M\"{u}ller, 
the common transverse flow of  hadrons may also
arise accidentally from the fragmentation
of minijets.  This is indeed what has been found with
HIJING calculations \cite{HIFLOW}. Shown in Fig.~\ref{fig5} is the 
HIJING result of the correlation between $\langle p_T\rangle$ and 
the total charged multiplicity $n_{\rm ch}$ for
pions, kaons, and antiprotons (from bottom to top) as solid lines
together with the data at Tevatron energy. The average multiplicity 
density $\langle dn_{\rm ch}/d\eta\rangle$ is calculated as 
$n_{\rm ch}(|\eta|<\Delta\eta)$ divided by $2\Delta\eta$.
The average $p_T$ is obtained by applying the same procedure 
as used in the experiment \cite{TEVAFLOW} in which
the $p_T$ distributions are first fitted with 
parametrizations [power law $a(p_T+b)^{-n}$ for pions and 
exponential $\beta\exp(-\alpha p_T)$ for kaons and antiprotons] 
and then the fitted parameters are used to calculate 
$\langle p_T\rangle$ in the restricted range $0<p_T<1.5$ GeV/$c$.
Apparently, the data are well accounted for by our calculation.
Shown as dashed lines in Fig.~\ref{fig5} are the calculated results
for $pp$ collisions at RHIC energy, $\sqrt{s}=200$ GeV. They are
similar to the results at $\sqrt{s}=1.8$ TeV, except that
pions have a lower saturated value of $\langle p_T\rangle$ at
RHIC energy. Since pions are the dominant produced particles,
the high multiplicity $\langle p_T\rangle$ for all charged hadrons
at $\sqrt{s}=200$ GeV is smaller than at $\sqrt{s}=1.8$ TeV.

\begin{figure}
\centerline{\psfig{figure=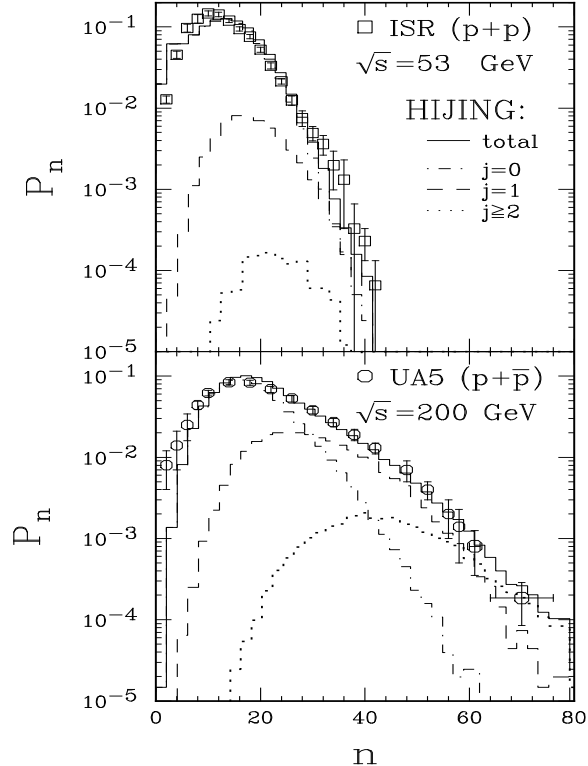,width=3in,height=4in}}
\caption{Charged multiplicity distributions in NSD $pp$ at 
        $\protect\sqrt{s}=53$ GeV and $p\bar{p}$ 
        collisions at $\protect\sqrt{s}=200$ GeV. The data are
        from Refs.~\protect\cite{pnisr,pnua5}. The solid histograms are from
        the HIJING calculation with contributions from events with 
        $j=0$ (dot-dashed histograms), $j=1$ (dashed histograms), 
        and $j\geq 2$ (dotted histograms) jet production.}
\label{fig6}
\end{figure}

        In a model with multiple parton production, it is easy
to understand why $\langle p_T\rangle$ increases with
$n_{\rm ch}$. If we decompose the multiplicity distribution into
different contributions from events with different number
of minijets as shown in Fig.~\ref{fig6}, we find that large multiplicity
events are dominated by multiple minijet production while low
multiplicity events are dominated by those of no jet production.
The average transverse momentum in events with multiple minijets
is certainly larger than those without, thus leading to the
increase of $\langle p_T\rangle$ with $n_{\rm ch}$. 
In order to understand the different behavior of the 
correlations between $\langle p_T\rangle$ and $n_{\rm ch}$ 
for different particles, we  recall that
the jet fragmentation functions for heavy hadrons
tend to be harder than for light hadrons, as measured in $e^+e^-$ 
annihilation experiments\cite{SLWU}. Therefore, heavy hadrons from
jet fragmentation carry larger transverse momenta than
light hadrons in $pp$ or $p\bar{p}$ collisions. 
This leads naturally to larger slopes of
the $\langle p_T\rangle$ vs $n_{\rm ch}$ correlation for heavier 
particles. In other words, the fragmentation of minijets
can mimic the transverse ``flow'' effect giving the resultant 
appearance of collective behavior.

Now one may ask: if totally different models \cite{BMPL,STRING} 
can both describe the data, which one has the true underlying
dynamics and how can they be distinguished from each other?
To answer these questions, one has to turn to the properties
of jet fragmentation.

\subsection{Resolving Minijets in Hadronic Interactions}

     Jets in hadronic interactions, as defined earlier, are 
produced by large $p_T$ parton scatterings in pQCD.
Experimentally, jets are identified as hadronic clusters whose
transverse energy $E_T$ can be reconstructed from a calorimetrical
study of the events \cite{UA1JET}. However, this cluster-finding method 
becomes questionable for small and even intermediate $E_T$ values due 
to the background of fluctuations \cite{BYON} from the underlying
soft interactions. It has therefore been very difficult to resolve 
minijets with $p_T\gsim$ 2 GeV/$c$ from the underlying soft background.

        Although minijets with $p_T\gsim$ 2 GeV/$c$ are difficult to 
resolve as distinct jets from the background, their effects even 
in minimum biased events, as we have demonstrated, can explain
many aspects of hadronic collisions and the associated multiparticle 
production \cite{GAISSER}. I have proposed \cite{XWCORR} 
that the $p_T$ dependence of the two-particle correlation function 
can be utilized to study
the minijet content in the minimum biased events of hadronic
interactions. Because particles from jet fragmentation
tend to cluster in phase space, two-particle correlations
must be enhanced due to minijet  production. Especially for
two-particle correlations in the azimuthal angle $\phi$, contributions
from back-to-back minijets should be strongly peaked in both
forward ($\Delta\phi=0$) and backward ($\Delta\phi=\pi$) directions.
If we calculate the same correlations, but for some selected
particles whose transverse momenta are larger than a certain
$p_T$ cut, the two peaks should be more prominent because
these particles are more likely to come from minijets.
On the other hand,  particles from soft production or an expanding
quark gluon plasma are isotropical in the transverse plane and 
would only have some nominal correlation in the backward direction 
due to momentum conservation. Therefore, the experimental
measurement of two-particle correlation functions and their
$p_T$ dependence, especially in $p\bar{p}$ collisions at the
Fermilab Tevatron energy where multiplicity and mass dependence
of $\langle p_T\rangle$ were first observed, is essential to
end the present controversy over whether the phenomenon is due
to minijets, or string interaction \cite{STRING} or the formation of a
quark gluon plasma \cite{BMPL}.

        It is well known that particles from high $p_T$ jets are
very concentrated in both directions of the back-to-back jets.
The widths of these high $p_T$ jet profiles are about 1 in both
pseudorapidity $\eta$ and azimuthal angle $\phi$ \cite{AKESSON}. 
Minijets, though with smaller $p_T$, should have similar properties.
Since particles with $p_T>p_{T{\rm cut}}$
are more likely to come from jet fragmentation, we can 
expect that the two-particle correlation functions 
are more characteristic of jet profiles when 
$p_{T{\rm cut}}$ is larger. This method of two-particle
correlations is unique because it can determine
contributions to particle production from minijets
which are intangible under the traditional cluster-finding
algorithm \cite{UA1JET} due to their small $p_T \sim $ 2 GeV/$c$.

\begin{figure}
\centerline{\psfig{figure=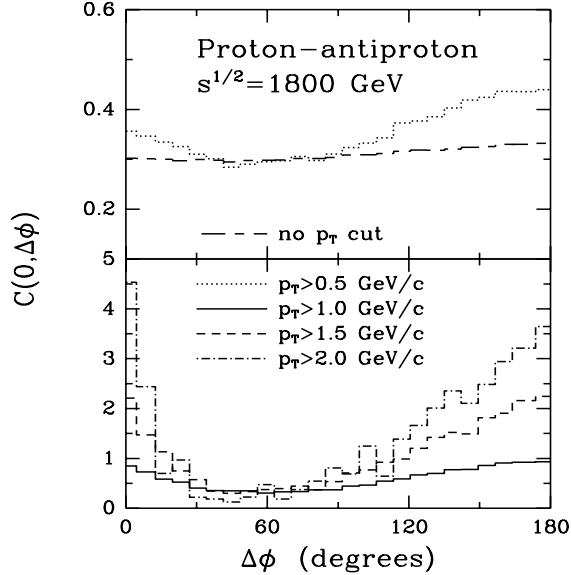,width=3in,height=3in}}
\caption{The correlation functions $c(0,\Delta\phi)$ vs $\Delta\phi$ 
        between two charged particles in the full rapidity range in
        $p\bar{p}$ collisions at $\protect\sqrt{s}=1.8$ TeV. The dashed line
        is for all charged particles, dotted line for particles with
        $p_T>0.5$ GeV/$c$, dash-dotted line for $p_T>1.0$ GeV/$c$, and
        solid line for $p_T>1.5$ GeV/$c$. }
\label{fig7}
\end{figure}

The normalized two-particle correlation functions in
the azimuthal angle $\phi$ are defined as
\begin{equation}
        c(\phi_1,\phi_2)=\frac{\rho(\phi_1,\phi_2)}
                {\rho(\phi_1)\rho(\phi_2)} -1,
\end{equation}
where $\rho(\phi)$ is the averaged particle density
in $\phi$ and $\rho(\phi_1,\phi_2)$ is the two-particle 
density which is proportional to the probability of
joint particle production at $\phi_1$ and $\phi_2$.

        Shown in Fig.~\ref{fig7} are our calculated results 
on two-particle correlation functions in $p\bar{p}$ collisions 
at $\sqrt{s}=1.8$ TeV for selected particles with 
different $p_{T{\rm cut}}$. The calculation includes all charged
particles in the full rapidity range. As we have expected, 
due to minijets, there is strong two-particle
correlation at both $\Delta\phi=0$ and $\pi$, forming a valley
at $\Delta\phi\sim \pi/3$. For large $p_{T{\rm cut}}$, the correlation
functions are very similar to large $p_T$ jet profiles as functions
of $\phi$ relative to the triggered jet axis \cite{AKESSON}. 
These features are, however, absent in low-energy hadronic collisions 
where minijet production is negligible \cite{XWCORR}. Since there 
are still many particles from soft production which can contribute 
only to the backward correlation due to momentum conservation,
the study of energy dependence of the relative heights of the two
peaks at $\Delta\phi=0$ and $\pi$ could provide us information
about the energy dependence of minijet production. The background
at $\Delta\phi\sim\pi/3$ also depends on the average number
of minijets produced\cite{XWCORR}.

\begin{figure}
\centerline{\psfig{figure=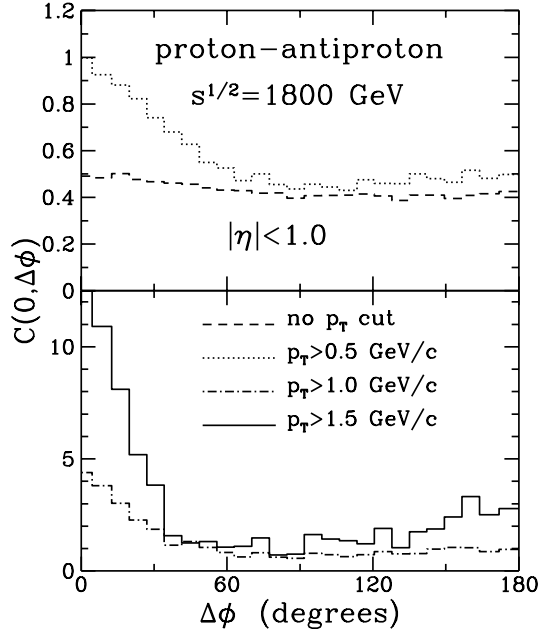,width=3.5in,height=3.5in}}
\caption{The same as Fig.~\protect\ref{fig7}, except for charged particles 
        in a limited rapidity range of $|\eta|<1$.}
\label{fig8}
\end{figure}

        Unlike high $p_T$ back-to-back jets which are both kinematically 
bounded to the central rapidity region, a pair of minijets can be easily 
produced with a large rapidity gap between them. When we trigger one minijet
in a limited rapidity window, the  other one which is produced in the same
parton scattering often falls outside the fixed rapidity window.
Therefore, if we calculate two-particle correlations for particles 
in a limited rapidity range, the minijet contribution to the backward
correlation ($\Delta\phi=\pi$) will mostly drop out while the
contribution to forward correlation ($\Delta\phi=0$) still remains. 
Indeed, as shown in Fig.~\ref{fig8}, the forward correlation 
at $\Delta\phi=0$ for particles in $|\eta|<1$ is very strong,
but the backward correlation at $\Delta\phi=\pi$ is drastically 
reduced as compared to the correlation pattern in the full
rapidity range in Fig.~\ref{fig7}. Furthermore, 
due to strong short range
two-particle correlations in rapidity \cite{HIJINGPP}, the forward 
correlation at $\Delta\phi=0$ is also enhanced by restricting
particles to $|\eta|<1$. At $\sqrt{s}=1.8$ TeV, we find that
the enhancement of backward correlation at $\Delta\phi=\pi$
due to minijets becomes important only when the rapidity window
is $|\eta|\gsim 2$.

\section{Parton Production in Nucleus-Nucleus Collisions}

        When we extrapolate the model to hadron-nucleus and 
nucleus-nucleus collisions, nuclear effects due to multiple
parton scatterings and the interference have to be considered.
Those nuclear effects can be generally divided into the two categories 
of initial and final state interactions. Initial state interactions
and the interference lead to an apparent depletion of the effective
parton density inside a nucleus, the so-called nuclear shadowing.
They also lead to a modification of the momentum spectra
of produced partons, the Cronin effect \cite{CRON1}, at intermediate 
energies. However, as I will demonstrate below, the Cronin effect will
disappear at large transverse momentum and at high energies,
like other high-twist processes. Both initial
and final state scatterings can cause a fast parton
to lose its energy. Due to color interference, this energy loss is 
not directly proportional \cite{GWLPM1,GWLPM2,DOKS} to the parton's 
energy. Thus for extremely energetic initial
partons, this energy loss will be negligible. However, for
produced partons which travel in the transverse direction,
the energy loss becomes important relative to their finite
transverse momentum. This energy loss will essentially
modify the parton fragmentation, as well as accelerate
parton equilibration.

\subsection{Nuclear Shadowing of Parton Distribution Functions}

Let us start first with the effect of nuclear modification
of the parton distributions, or nuclear shadowing due to 
initial state interactions. What we are mostly concerned about 
here is the depletion of the effective parton density which
will reduce the initial produced parton density and the 
transverse energy density\cite{HIJING,KE91,WG92}. This Section
is based mainly on the work by Eskola, Qiu and myself in Ref.~\cite{EQW94}.

``Nuclear shadowing'' in the context of the deeply inelastic 
lepton-nucleus scattering refers to the measured depletion of the 
nuclear structure function $F_2^A$ at small $x_{\rm Bj}$, as
compared to $F_2$ of unbound nucleons \cite{EMC90}. The same kind
of depletion at small $x$ is expected to happen also in the nuclear 
gluon distributions. During the recent years there have been many 
efforts to explain the measured nuclear shadowing of quarks and 
antiquarks \cite{BRLU,NZ75,FS88,CQR89,CD89,JK90},
but for  gluons the situation is still inconclusive.
In these models, shadowing at small $x$ can be attributed to 
parton fusions before the hard scattering which probes the
parton distributions. In terms of parton fusions, shadowing 
is also predicted to happen in protons. In this case,
``shadowing'' refers to the saturation of the actual parton 
distributions caused  by fusions of overcrowding gluons 
at very small $x$. This mechanism proceeds through 
perturbative QCD-evolution  as formulated in 
\cite{GLR,MQ86}. It has been  shown by Collins and 
Kwieci{\'n}ski that the singular gluon distributions actually 
saturate due to gluon fusions \cite{CK90}.  
I will first review QCD-evolution of parton distributions
and then demonstrate how parton shadowing arises from the
inclusion of a fusion term in the evolution equation.

\begin{figure}
\centerline{\rotateright{\psfig{figure=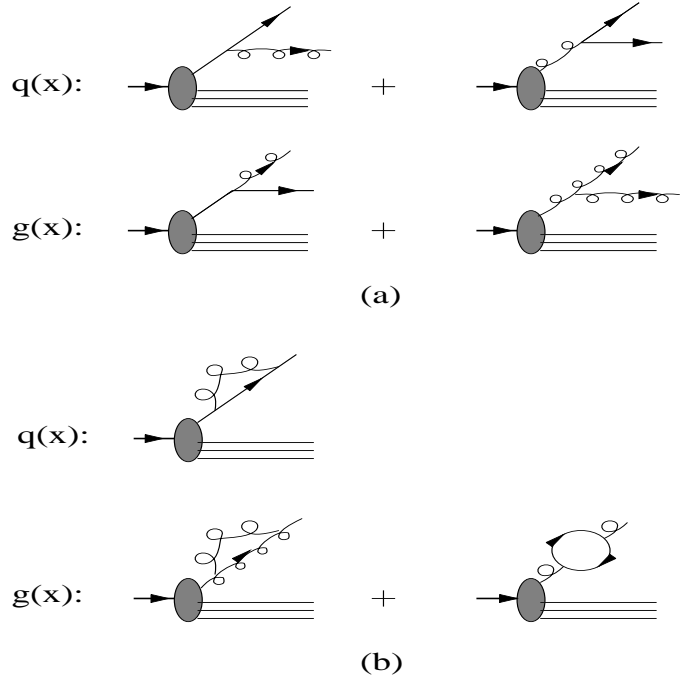,width=3.5in,height=3.5in}}}
\caption{(a) Radiative and (b) virtual corrections to parton distributions.}
\label{fig9}
\end{figure}

Parton distribution functions inside a nucleon or nucleus are 
closely related to the Pomeron substructure and its coupling to quarks
and gluons \cite{GLR,LANDSHOFF2}. If one assumes that the ladder
diagram inside a Pomeron is ordered in $p_T$, then one
can derive a set of evolution equations for the
parton distribution functions with respect to the momentum
scale of the hard scattering \cite{GLAP,FIELD}:
\begin{eqnarray}
Q^2\frac{\partial q_i(x,Q^2)}{\partial Q^2}&=& \frac{\alpha_s(Q^2)}{2\pi}
\int^1_x\frac{dy}{y}\left[P_{q\rightarrow qg}(y)q_i(\frac{x}{y},Q^2)
+P_{g\rightarrow q\bar{q}}(y)g(\frac{x}{y},Q^2)\right], \label{eq:evq}\\
Q^2\frac{\partial g(x,Q^2)}{\partial Q^2}&=& \frac{\alpha_s(Q^2)}{2\pi}
\int^1_x\frac{dy}{y}\left[\sum_{i=1}^{2N_f}
P_{q\rightarrow gq}(y)q_i(\frac{x}{y},Q^2)
+P_{g\rightarrow gg}(y)g(\frac{x}{y},Q^2)\right], \label{eq:evg}
\end{eqnarray}
where the splitting functions, $P_{a\rightarrow bc}(y)$, are the
probability distribution functions for the respective radiative processes 
as illustrated in Fig.~\ref{fig9}(a), $x$ is the
fractional momentum of the specified partons, and $N_f$ is the number
of quark flavors.  In addition to
these splitting processes, one should also take into account
the virtual corrections as shown in Fig.~\ref{fig9}(b).
Their contributions to the evolution equations are \cite{JCJQ}
\begin{eqnarray}
q_i(x,Q^2):&\;\;\;\;& -\frac{\alpha_s(Q^2)}{2\pi}q_i(x,Q^2)\int^1_0 dy
  P_{q\rightarrow qg}(y),\label{eq:virq}\\
g(x,Q^2):&\;\;\;\;& -\frac{\alpha_s(Q^2)}{2\pi}g(x,Q^2)\int^1_0 dy
  \frac{1}{2}\left[P_{g\rightarrow gg}(y)
  +N_fP_{g\rightarrow q\bar{q}}(y)\right]. \label{eq:virlg}
\end{eqnarray}
These virtual corrections are important to regularize the
soft divergences in the splitting functions and guarantee flavor
and momentum conservation. If one wants a probabilistic
interpretation of the evolution and especially a numerical simulation,
the above virtual corrections will give rise to the Sudakov
form factors which are necessary to guarantee unitarity.
As one can see, the above virtual corrections can be incorporated
into the evolution equations by introducing a $\delta$-function and
``+functions'' (see \cite{FIELD} for their definitions) 
in the splitting functions
\begin{eqnarray}
  P_{g\rightarrow q\bar{q}}(y)&=&\frac{1}{2}[y^2+(1-y)^2]; \label{eq:split1}\\
  P_{q\rightarrow gq}(y)&=&\frac{4}{3}\frac{1+(1-y)^2}{y};\label{eq:split2}\\
  P_{q\rightarrow qg}(y)&=&\frac{4}{3}\left[\frac{1+y^2}{(1-y)_+}
                         +\frac{3}{2}\delta(1-y)\right];\label{eq:split3}\\
  P_{g\rightarrow gg}(y)&=&6\left[\frac{y}{(1-y)_+}+\frac{1-y}{y}
  +y(1-y)+\frac{1}{12}(11-\frac{2}{3}N_f)\delta(1-y)\right].\label{eq:split4}
\end{eqnarray}
With the above splitting functions one can check that flavor and
momentum are conserved, {\it i.e.},
\begin{eqnarray}
\frac{\partial}{\partial Q^2}\int_0^1 dx  
                  \left[q_i(x.Q^2)-\bar{q}_i(x,Q^2)\right] &=&0,\\
\frac{\partial}{\partial Q^2}\int_0^1 dx  x \left[g(x,Q^2)
                  +\sum_{i=1}^{2N_f}q_i(x,Q^2)\right]&=&0.
\end{eqnarray}

In the parton model, a nucleus consists of many valence quarks.
Those valence quarks, however, are dressed with clouds of
sea quarks and gluons due to vacuum fluctuations (which
corresponds to the loops inside a Pomeron ladder). The
life time of the virtual partons is of the order of the
hadron size, $R_N\sim 1$ fm. If those valence quarks are 
probed by a hard scattering with resolution $Q\gg 1/R_N$, 
the coherence of the parton clouds will be broken and numerous
partons will be released. The parton density inside the clouds 
will then depend on the resolution of the scattering $Q$.
The larger the $Q$ the denser the parton clouds, as predicted
by the QCD evolution equations. These evolution equations
are based on perturbative QCD which can only be applied down to some
scale $Q_0$. Given the initial values of parton distributions
at $Q_0$, the evolution in $Q>Q_0$ can be predicted by pQCD.
However, the parton distributions below $Q_0$ cannot be obtained
from pQCD and so far can only be determined from experiments.

At small values of $x$, the leading order QCD evolution equations 
predict that the number of gluons becomes extremely large.
It has been known \cite{GLR,MQ86} that for 
sufficiently small values of $x$ and/or of $Q^2$, the total
transverse area occupied by the gluons will be larger than 
the transverse area of a hadron, so that the interactions among
gluons can no longer be neglected.  Such gluon recombination results 
in a modification of the QCD evolution equations, 
Eqs.~(\ref{eq:evq}) and (\ref{eq:evg}).
In the limit of small-$x$ and neglecting quark distributions,
the modified QCD evolution equation for the gluon
distribution can be cast in the form \cite{GLR,MQ86}
\begin{equation} \hskip-0.3truecm
\partial_y\partial_t G(y,t) = cG(y,t) - \gamma\exp(-t-{\rm e}^t)[G(y,t)]^2,
\label{APMQ}
\end{equation} 
where $y = \ln(1/x)$, $t = \ln[\ln(Q^2/\Lambda^2_{\rm QCD})]$, $G(y,t) =
xg(x,Q^2)$ and $c = 12/(11-2N_f/3)$ with $N_f$ the number of quark
flavors.

The seond term in Eq.~(\ref{APMQ}) corresponds to the change of
the distribution due to gluon recombination. The strength of the 
gluon recombination is controlled by the factor $\gamma$, 
originating from two possible sources. One can consider
gluon fusion in terms of one Pomeron to two-Pomeron coupling.
The two fusing gluon ladders (two Pomerons), which couple 4 gluons 
to 2 gluons, can arise either from independent constituents 
of the proton/nucleus or from the same one, 
as discussed in  \cite{MQ86,CK90,KMSR90}. 
We will refer to the former case as ``independent'' and to the latter as
``non-independent'' fusion. Since  recombinations from  both 
sources happen simultaneously, we divide the
parameter $\gamma$ into two parts: 
\begin{equation} 
\gamma = \gamma_{\rm I}+\gamma_{\rm II},\label{lambda}  
\end{equation}
where $\gamma_{\rm I}$ corresponds to the independent recombination and
$\gamma_{\rm II}$ to the non-independent one. 

To understand the form of the fusion term in Eq.~(\ref{APMQ}),
one can consider $xg(x,Q^2)$ as the gluon number $n_g$ per
unit rapidity ($dy=dx/x$). If we note that the gluon-gluon
cross section $\hat{\sigma}\sim \alpha_s/\Lambda_{\rm QCD}^2$,
the independent gluon fusion probability inside a nucleon with 
transverse area $\pi R_N^2$ is then,
\begin{equation}
  W=\frac{n_g^2\hat{\sigma}}{\pi R_N^2}\sim 
  \frac{\alpha_s}{R_N^2\Lambda_{\rm QCD}^2} [xg(x,Q^2)]^2.
\end{equation}
This corresponds to the gluon fusion term in Eq.~(\ref{APMQ})
(note that $e^{-t}\sim \alpha_s$). One can calculate
this fusion term in a QCD recombination model \cite{MQ86}. 
In a proton, the strength of the independent fusion then takes the form 
\begin{equation}
\gamma_{\rm I} = \frac{2}{3}\frac{1}{\pi R_N^2}\cdot
\frac{\pi^3c^2}{2\Lambda_{\rm QCD}^2}, \label{lambdaIp} 
\end{equation}
where $R_N\sim 1$ fm is the radius of a proton.

The non-independent fusion happens inside the same valence quark
which is assumed to have a transverse size $\sim 1/Q_i$.
The magnitude of the  non-independent fusion of the gluon ladders 
can be estimated as \cite{MQ86}
\begin{equation}
\gamma_{\rm II} \approx \frac{16}{81}\frac{1}{\pi (2/Q_i)^2}\cdot 
\frac{\pi^3c^2}{2\Lambda_{\rm QCD}^2}, \label{lambdaIIp} 
\end{equation}
where a simplification is made by fixing the initial $x$ of the valence
quark to $x_i\sim 1$. We also approximate the scale of the initial
valence quark by $Q_i\sim 2$ GeV.

Let us then consider a large loosely bound nucleus.
Naturally, both types of fusions are still there but only for 
the independent one an $A^{1/3}$-scaling arises. In this case
\begin{equation}
\gamma_{\rm I}^A = \frac{9}{8}\frac{A}{\pi R_A^2}\cdot
\frac{\pi^3c^2}{2\Lambda_{\rm QCD}^2},\label{lambdaIA}
\end{equation}
where the nucleus is taken to be a sphere with a 
sharp surface at $R_A = 1.12A^{1/3}$ fm. The strength 
of  the non-independent fusion remains the same as in the 
case of a free proton: $\gamma_{\rm II}^A = \gamma_{\rm II}$.

It is interesting to notice how the relative contributions of 
the two types of recombination will change when going from a 
proton to a nucleus of $A\sim 200$:
$\gamma_{\rm II}/\gamma_{\rm I}\approx 7.6$ and 
$\gamma_{\rm II}^A/\gamma_{\rm I}^A\approx 1.0$.  
Thus the non-independent
fusion is clearly dominant in a free proton while in a large nucleus the
contributions from both types are of the same order.  
As a result, parton recombination is strongly enhanced in a heavy nucleus.

In order to solve Eq.~(\ref{APMQ}) exactly by integration, 
one would need the initial distribution either at fixed $y_0$ 
or $t_0$ {\it and} the derivatives along a boundary line $(y,t_0)$ 
or $(y_0,t)$, respectively.
However, since the expression for the non-linear term in
Eq.~(\ref{APMQ}) is not valid for the regions where $x$ is large,
or where both $x$ and $Q$ are very small,
the natural boundary condition at $x=1$ (or $y_0=0$)
is not suitable here. In addition,
since we do not have sufficient information on other boundary lines,
we cannot solve Eq.~(\ref{APMQ}) by direct integration.
Instead, with the semiclassical approximation \cite{GLR}, we 
use the method of characteristics \cite{CK90,EQW94}, 
so that we can avoid the region of both small $x$ and small $Q^2$.

The semiclassical approximation corresponds to
neglecting the second order derivative term, 
$\partial_y\partial_t\ln(G)$, which leaves us with the evolution equation as
\begin{equation}
\partial_yz(y,t)\partial_tz(y,t) = c-\gamma\exp[-t-{\rm e}^t+z(y,t)],
\label{semicl}
\end{equation}
where $z(y,t) = \ln[G(y,t)]$. The reason why this approximation
is called semiclassical is that it corresponds to using a saddle-point
approximation to the integration in the integral form of the
evolution equation. The above equation can 
then be cast and solved in the form of a set of characteristic 
equations as shown in detail in \cite{CK90}. What is needed
here is the gluon distribution at the boundary $y_0[=-\ln(x_0)]$
at all $t$, which can be obtained by evolving the distribution
in the region $x>x_0$ according to the original GLAP evolution
equations, Eqs.~(\ref{eq:evq}) and (\ref{eq:evg}). 

For a proton, we assume that the recombination becomes effective 
at $x\sim x_0\sim 0.01$, which is consistent with
\cite{CK90,KMSR90}. We can use the results from 
a global fitting to the parton distributions, like CTEQ \cite{CTEQ93},
to constrain $x_0$. 
In fact, we will see that with $x_0=0.01$ the shadowed gluons 
deviate considerably from the CTEQ gluons only after $x<0.001$,
so our choice for $x_0$ seems to be reasonable, and we do not 
expect  the results to be very sensitive to small changes of $x_0$.

As explained above, the gluon recombination is strongly 
enhanced in heavy nuclei and it starts at somewhat
larger values of $x$ than in protons. The corresponding boundary line $x_0^A$
for a nucleus is approximately determined by the relative magnitude of the
evolution terms in Eq.~(\ref{APMQ}): for 
$G_A(x_0^A)\sim G(x_0)\gamma_A/\gamma$, the relative contribution 
from the gluon fusion in a nucleus is about the same as in a nucleon.
This gives $x_0^A\sim$ 0.05--0.1. This range of $x_0^A$ is also supported by
other studies \cite{KE93}. 

Let the total gluonic fractional momentum in the non-shadowed
parton distributions be $f_0=\int_0^1 dx \,xg_{\rm CTEQ}(x,Q_0^2)$. 
In the case of a proton, 
shadowing changes the gluonic momentum typically by less than a per
cent, which we can clearly neglect as a small overall change.

Perturbative shadowing at small $x$ reduces the gluonic momentum more 
in a nucleus than in a proton. If the momentum fraction of gluons is 
conserved, there must be a corresponding enhancement in the large $x$ 
region. However, there can also be momentum transfers from quarks and 
antiquarks to the gluons.  Here we consider nuclei with
$A\sim 200$, for which we expect an overall increase in the 
gluonic momentum fraction, $\epsilon_A$, to be only about 4\%
\cite{FS88,CQR89,KE93} as constrained by the experimental data
on quark (anti-quark) shadowing in deep inelastic scatterings.
These two sources of momentum flow will result in anti-shadowing
(enhancement of parton densities) at large $x>x_0^A$. To account
for the anti-shadowing we assume $g(x,Q_0^2)=a_A g_{\rm CTEQ}(x,Q_0^2)$
for $x>x_0^A$ with $g_{\rm CTEQ}(x,Q_0^2)$ the non-shadowed gluon
distribution and $a_A>1$ to be determined by the momentum sum rule,
\begin{eqnarray}
\int_0^{x_0^A}dx\, xg(x,Q_0^2)\bigg|_{C}
&+&a_A\int_{x_0^A}^1dx\, xg_{\rm CTEQ}(x,Q_0^2)         \nonumber\\
&=&f_0(1+\epsilon_A).
\label{momP} 
\end{eqnarray}
The right-hand-side of the equation is the gluonic momentum fraction
of the non-shadowed parton distributions plus the momentum transfer
$\epsilon_A$ from quarks and anti-quarks during the recombination.
One can solve the above equation iteratively for $a_A$ with the 
boundary condition $C:\, g_A(x_0^A,Q_0^2) =  a_A g_{\rm CTEQ}(x_0^A,Q_0^2)$.
Typically, $a_A\sim 1.1$ for $A\sim 200$.

\begin{figure}
\vspace{-0.8in}
\centerline{\psfig{figure=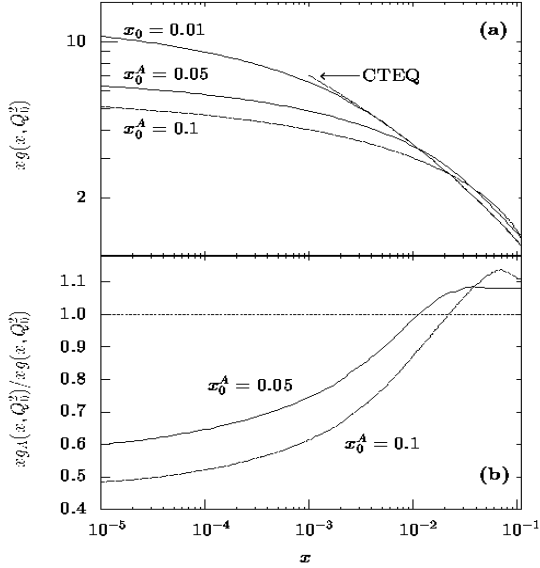,width=4in,height=3.8in}}
\caption{ (a) The gluon distributions $xg(x,Q_0^2)$ at $Q_0=2$ GeV vs.~$x$. 
         The result for proton is labeled by $x_0$, and the results for 
         $A\sim 200$ by $x_0^A$, respectively. The CTEQ gluon 
         distribution \protect\cite{CTEQ93} is labeled by ``CTEQ''. 
         (b) The ratio $xg_A(x,Q_0^2)/xg(x,Q_0^2)$ of the shadowed gluon 
         distributions  vs. $\protect x$, demonstrating a strong 
         perturbative nuclear shadowing in heavy nuclei.}
\label{fig10}
\end{figure}

In Fig.~\ref{fig10}(a), nucleon and effective nuclear gluon 
distributions for a nucleus of $A=200$ are compared with the 
input CTEQ gluon distribution at $Q_0=2$ GeV. Notice 
the $\sim$20 \% uncertainty in the nuclear case resulting from
varying $x_0^A$ from 0.05 to 0.1. To demonstrate the formation 
of strong perturbative nuclear shadowing, corresponding to 
the {\it relative} depletion of gluon distributions in a nucleus, 
we plot the ratio $G_A(x,Q_0^2)/G(x,Q_0^2)$ in Fig.~\ref{fig10}(b).
Notice also that as $x$ decreases, the gluon distribution in a proton
increases much faster, or approaches saturation at a much
smaller $x$ than that in a nucleus.  Therefore, as shown 
in Fig.~\ref{fig10}(b) the ratio saturates only when
the gluons in a {\it proton} do so. Thus, saturation of 
the perturbative nuclear shadowing reflects actually the 
behavior of the gluons in a proton. We see that, 
due to the enhanced gluon recombination in a heavy nucleus, 
a $\sim$50\% nuclear shadowing in small-$x$ region is generated
perturbatively through the modified QCD evolution, accompanied 
by a $\sim$10 \% antishadowing from momentum conservation. 

We have not included the non-perturbative, {\it e.g.}, 
Glauber-derived \cite{BRLU} contribution to the shadowing.
Inclusion of such a contribution is equivalent to changing the 
initial values of gluon distribution at scale $Q_0$ 
and will slightly reduce the 
perturbative contribution to the shadowing through QCD evolution.
We also have not considered the shadowing of quark distributions
in this study of perturbative shadowing.
However, experimental data \cite{EMC90} on quark shadowing
are roughly consistent with the perturbative gluon shadowing 
we have just estimated. The shadowing, at least for quarks,
has been shown to depend weakly on the scale $Q$ \cite{KE93},
also consistent with experimental data.

\begin{figure}
\centerline{\rotateright{\psfig{figure=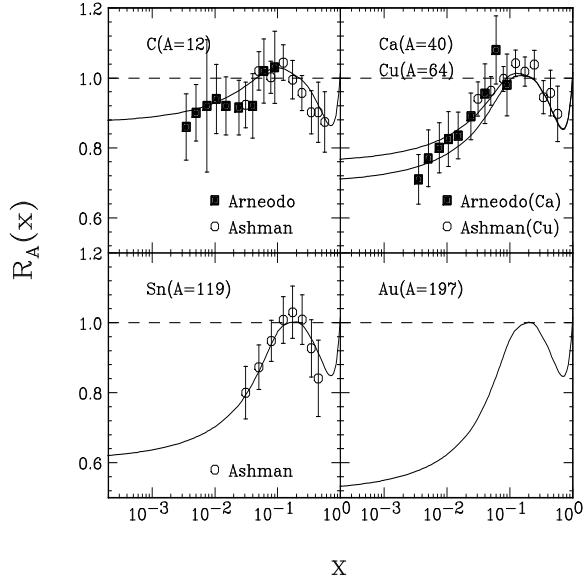,width=3in,height=3in}}}
\caption{The ratio of quark structure functions 
        $R_A(x)\equiv F_2^A(x)/AF_2^N(x)$ as a function of $x$
        in small and medium $x$ regions for different nuclear mass numbers $A$.
        The data are from Ref.~\protect\cite{EMC90} and curves are the
        parametrization in Eqs.~(\protect\ref{eq:shadow}) 
        and (\protect\ref{eq:shadow1}).}
\label{fig11}
\end{figure}

For a practical implementation of the nuclear shadowing in HIJING, 
we consider that quarks and gluons are
shadowed by the same amount inside a nucleus and use the
following parametrized form:
\begin{eqnarray}
        R_A(x)&\equiv&\frac{f_{a/A}(x)}{Af_{a/N}(x)} \nonumber\\
          &=&1+1.19\ln^{1/6}\!A\,[x^3-1.5(x_0+x_L)x^2+3x_0x_Lx]\nonumber\\
          & &-\left[\alpha_A-\frac{1.08(A^{1/3}-1)}{\ln(A+1)}\sqrt{x}
              \right]e^{-x^2/x_0^2},\label{eq:shadow}\\
        \alpha_A&=&0.1(A^{1/3}-1),\label{eq:shadow1}
\end{eqnarray}
where $x_0=0.1$ and $x_L=0.7$. The term proportional to $\alpha_A$ in 
Eq.~(\ref{eq:shadow}) determines the shadowing for $x<x_0$ with the 
most important nuclear dependence, while the rest gives the overall 
nuclear effect on the structure function in $x>x_0$ with a very 
weak $A$ dependence. As shown in Fig.~\ref{fig11} , 
this parametrization reproduces the measured  overall 
nuclear effect on the quark structure function in
the small and medium $x$ regions. However, I should emphasize
that this parametrization does not satisfy the momentum sum rule
and does not include the weak scale dependence which was found 
by Eskola in a  detailed study in Ref.~\cite{KE93}.

Eq.~(\ref{eq:shadow}) represents only the average nuclear dependence of the 
structure function.  However, in $pA$ or $AA$ collisions, we have to 
calculate the effective jet cross sections at
the nucleon-nucleon level for each impact parameter.
Physically, it is natural to expect that the nuclear effects
on the structure functions could depend on the local
nuclear thickness at each impact parameter \cite{KE91}.
Eq.~(\ref{eq:shadow1}) is consistent with the assumption that the shadowing 
parameter $\alpha_A(r)$ is proportional to the longitudinal thickness of 
the nucleus at impact parameter $r$. We therefore parameterize the 
impact parameter dependence of $\alpha_A$ in Eq.~(\ref{eq:shadow}) as
\begin{equation}
        \alpha_A(r)=0.1(A^{1/3}-1)\frac{4}{3}\sqrt{1-r^2/R_A^2},
                        \label{eq:rshadow}
\end{equation}
where $r$ is the transverse distance of the interacting nucleon from its 
nucleus center and $R_A$ is the radius of the nucleus. For a sharp-sphere
nucleus with thickness 
function $T_A(r)=\frac{3A}{2\pi R_A^2}\sqrt{1-r^2/R_A^2}$,
the averaged $\alpha_A(r)$ is 
$\int_0^{R_A^2}\pi dr^2 T_A(r)\alpha_A(r)/A=\alpha_A$, with
$\alpha_A$ from Eq.~(\ref{eq:shadow1}). Because the rest
of Eq.~(\ref{eq:shadow}) has a weaker $A$ dependence, 
we  only consider the impact parameter dependence of $\alpha_A$. 

        To simplify the calculation during the Monte Carlo simulations, we
can decompose $R_A(x,r)$ into two parts,
\begin{equation}
        R_A(x,r)\equiv R_A^0(x)-\alpha_A(r)R_A^s(x),
\end{equation}
where $\alpha_A(r)R_A^s(x)$ is the term proportional to $\alpha_A(r)$ 
in Eq.~(\ref{eq:shadow}) with  $\alpha_A(r)$ given in Eq.~(\ref{eq:rshadow}) 
and $R_A^0(x)$ is the rest of $R_A(x,r)$. Both $R_A^0(x)$ and $R_A^s(x)$ 
are now independent of $r$. The effective jet production cross section
of a binary nucleon-nucleon  interaction in $A+B$ nuclear collisions 
can therefore be decomposed as \cite{KE91}
\begin{equation}
\sigma_{\rm jet}^{eff}(r_A,r_B)=\sigma_{\rm jet}^0
            -\alpha_A(r_A)\sigma_{\rm jet}^A
            -\alpha_B(r_B)\sigma_{\rm jet}^B
        +\alpha_A(r_A)\alpha_B(r_B)\sigma_{\rm jet}^{AB},\label{eq:sjetab}
\end{equation}
where $\sigma_{\rm jet}^0$, $\sigma_{\rm jet}^A$, $\sigma_{\rm jet}^B$ and
$\sigma_{\rm jet}^{AB}$ can be calculated through Eq.~(\ref{eq:sjet1}) by 
multiplying \\
$f_a(x_1,p_T^2)f_b(x_2,p_T^2)$ in the integrand with 
$R_A^0(x_1)R_B^0(x_2)$, $R_A^s(x_1)R_B^0(x_2)$, \\
$R_A^0(x_1)R_B^s(x_2)$ and $R_A^s(x_1)R_B^s(x_2)$, respectively. 

        In central $Au+Au$ collisions at $\sqrt{s}=200$ AGeV,
the average parton fractional momentum is $x=2p_T/\sqrt{s}\simeq 0.02$
for mini-jets with $p_T\geq p_0=2$ GeV. The impact-parameter 
dependent parton shadowing reduces the averaged 
inclusive mini-jet cross section by 50\% from its value
in $pp$. This estimate will be somewhat modified if
one includes the scale dependence of the shadowing effect.
However, at RHIC energies we can see from Fig.~\ref{fig11} that
mini-jet production with $x\simeq 0.02$ is still not in the 
deep-shadowed region of $x$.
For sufficiently high energies, most of the mini-jets come from
$x\lsim 0.01$ region so that the effective mini-jet
cross section may be reduced by a factor 3
in central $Au+Au$ collisions. Note that this large reduction of mini-jet
multiplicity by gluon shadowing may increase the limit for
independent multi-jet production up to $\sqrt{s}=50$ ATeV for
$Au+Au$ collisions.

\subsection{Disappearance of Cronin Effect at High Energies}

In hadron-nucleus and nucleus-nucleus collisions, multiple
scatterings happen both at hadronic and partonic levels.
Due to the interference between multiple scattering
amplitudes, a hadron can only ``see'' the surface
of a nucleus, leading to a total cross section $\sim R^2_A$.
Parton fusions which cause nuclear shadowing of the
parton distribution functions can also be viewed as
multiple parton scattering in the laboratory frame.
The parton depletion arises from the destructive
interference among different scattering amplitudes \cite{BRLU}.
These multiple scatterings and their interference will also
affect the momentum spectra of the produced partons.
In general, as observed first by Cronin {\it et al.} \cite{CRON1},
particle production at large $p_T$ will be enhanced.
However, recent experiments \cite{CRON2} show that
the enhancement decreases both with $p_T$ and  energy $\sqrt{s}$.

I would like to emphasize that the initial multiple scatterings
discussed here are very different from the late final parton
rescatterings which are responsible for parton thermalization.
The initial partons in general have very small transverse
momenta but large center-of-mass (c.m.) energies. As
demonstrated in Ref.~\cite{GWLPM1}, multiple scatterings among
these initial partons cannot be treated as semiclassical
cascades which might be valid for final parton rescatterings. 
The eikonal limit is more relevant for the initial partons 
scatterings which can be treated by a Glauber multiple scattering 
analysis.

Following the same procedure that leads to the hadron-hadron
scattering cross section due to exchange of multiple Pomerons,
Eq.~(\ref{eq:incr}), one can also obtain the inelastic cross 
section for $pA$ collisions,
\begin{equation}
\sigma_{pA}(s)=\int d^2b [1 - e^{-\sigma_{NN}(s)T_A(b,s)}], \label{eq:pacr}
\end{equation}
where $\sigma_{NN}$ is the total nucleon-nucleon cross section and
\begin{equation}
T_A(b)=\int_{-\infty}^{\infty}dz \rho(b,z),
\end{equation}
is the thickness function of the nucleus at impact parameter $b$.
The nuclear density, $\rho(b,z)$, is normalized to $\int d^2b T_A(b)=A$.

Let us assume that the total nucleon-nucleon cross section
can be described by parton scatterings with given cross
sections. Define the differential cross section for a parton-nucleon 
scattering, $i+N\rightarrow j+ X$, as $h^{ij}(p_i,p_j)$,
with the total cross section
\begin{equation}
  \sigma_i(p_i)=\frac{1}{2}\sum_j\frac{d^3p_j}{E_j}h^{ij}(p_i,p_j).
\end{equation}

Let us also define
\begin{equation}
T^{\pm}_A(b,z)=\pm\int_z^{\pm\infty}dz^{\prime}\rho(b,z^{\prime}).
\end{equation}
Note that
\begin{equation}
T_A(b)=T_A^-(b,z)+T_A^+(b,z). \label{eq:thick}
\end{equation}

Now according to a Glauber multiple scattering interpretation
of the $pA$ cross section Eq.~(\ref{eq:pacr}),
the probabilities for parton $i$ not to interact with the nucleons
up to $z$ and for $j$ not to interact after $z$, are respectively,
\begin{equation}
e^{-\sigma_i(p_i)T_A^-(b,z)},\;\;\;\;
e^{-\sigma_j(p_j)T_A^+(b,z)}.
\end{equation}
Then, the probability for one parton-nucleon scattering, 
$i+A\rightarrow j +X$, is
\begin{eqnarray}
  \frac{dH^{ij}_{(1)}}{d^2b}&=&\int_{-\infty}^{\infty}dz 
  e^{-\sigma_iT_A^-(b,z)}\rho(b,z)h^{ij} e^{-\sigma_jT_A^+(b,z)} \nonumber\\
&=&\frac{h^{ij}}{\sigma_j-\sigma_i}
\left[e^{-\sigma_iT_A(b)}-e^{-\sigma_jT_A(b)}\right]\nonumber\\
&=&h^{ij} T_A(b)\left[1-\frac{\sigma_i+\sigma_j}{2}T_A(b)\right]
+\cdots \;\;,
\end{eqnarray}
where the definition of the thickness functions $T_A(b)$ and
Eq.~(\ref{eq:thick}) have been used, and an expansion in terms 
of $\sigma_{i,j}T_A(b)$ has been made in the last step.

We can use the above result to calculate the probability 
for a parton $k$ to have only one scattering after $z$, by
replacing $T_A(b)$ with $T_A^+(b,z)$. Then the contribution
to $i+A\rightarrow j+X$ from two parton-nucleon scatterings is
\begin{eqnarray}
  \frac{dH^{ij}_{(2)}}{d^2b}&=&\sum_k \int_{-\infty}^{\infty}dz
  e^{-\sigma_iT_A^-(b,z)}\rho(b,z) 
  \frac{h^{ik}h^{kj}}{\sigma_j-\sigma_k}\left[
  e^{-\sigma_kT_A^+(b,z)}-e^{-\sigma_jT_A^+(b,z)}\right]\frac{d^3p_k}{E_k}
  \nonumber \\
&=&\sum_k\frac{d^3p_k}{E_k} \frac{h^{ik}h^{kj}}{\sigma_j-\sigma_k}\left[
\frac{e^{-\sigma_iT_A^(b)}-e^{-\sigma_kT_A(b)}}{\sigma_k-\sigma_i}-
\frac{e^{-\sigma_iT_A(b)}-e^{-\sigma_jT_A(b)}}{\sigma_j-\sigma_i}\right]
\nonumber \\
&=&\frac{1}{2}T_A^2(b)\sum_k\frac{d^3p_k}{E_k}h^{ik}h^{kj}+\cdots \;\;\;.
\end{eqnarray}

Therefore, up to second order in $\sigma T_A(b)$, we have
the cross section for $i+A\rightarrow j+X$,
\begin{equation}
  \frac{dH^{ij}}{d^2b}\approx h^{ij}T_A(b)+\frac{1}{2}T_A^2(b)
  \left[\sum_k\frac{d^3p_k}{E_k}h^{ik}h^{kj}
  -(\sigma_i+\sigma_j)h^{ij}\right].
\end{equation}
In the second term in the above equation, we can already
see that the interference between double and single scattering
comes into play.  If we assume the parton distribution inside
the projectile nucleon to be $f_{i/N}(p_i)$, then the differential
cross section for $N+N\rightarrow j+X$ is given by
\begin{equation}
\sigma^j_{NN}\equiv \sum_i\int\frac{d^3p_i}{E_i}f_{i/N}(p_i)h^{ij}.
\end{equation}
Assuming a hard sphere for the nuclear density $\rho(r)$ with
a radius $R_A=R_0 A^{1/3}$, we then obtain the ratio between
the differential cross sections of $N+A\rightarrow j+X$ 
and $N+N\rightarrow j+X$,
\begin{equation}
  R_A\equiv\frac{\sigma^j_{NA}}{A\sigma^j_{NN}}=
  1+\frac{9A^{1/3}}{16\pi R_0^2}\frac{1}{\sigma^j_{NN}}\sum_i\int
  \frac{d^3p_i}{E_i}f_{i/N}(p_i)\left[\sum_k\frac{d^3p_k}{E_k}h^{ik}h^{kj}
  -(\sigma_i+\sigma_j)h^{ij}\right]. \label{eq:cronin1}
\end{equation}
Compared to the additive model of hard scatterings, parton
production in $NA$ collisions is enhanced due to multiple
scattering. The enhancement is proportional to $A^{1/3}$, which
is the average number of nucleons inside a nucleus
along the beam direction. This formula was used successfully \cite{LP83} 
to explain the enhancement of large $p_T$ particle
production in $pA$ collisions.

To demonstrate how this enhancement
depends on the transverse momentum and colliding energy,
let us assume a simple power-law form for the parton-nucleon
differential cross section,
\begin{equation}
h^{ij}\equiv d\sigma/dydp_T^2=C/p_T^n, \;\;(|y|<\Delta Y/2, p_T>p_0).
\end{equation}
Assuming that all partons are identical, the total cross section is then,
\begin{equation}
  \sigma=\frac{1}{2}\int dyd^2p_T\frac{C}{p_T^n}
  =\frac{\pi\Delta Y C}{(n-2)p_0^{n-2}}.
\end{equation}
For large $p_T\gg p_0$, one can evaluate the integral,
\begin{eqnarray}
  \int\frac{d^3p_k}{E_k}h^{ik}h^{kj}&=&\int dyd^2q_T\frac{C}{q_T^n}
  \frac{C}{({\bf p}_T-{\bf q}_T)^n} \nonumber \\
  &=&\frac{C}{p_T^n}\sigma\left[2+\frac{n(n-2)}{2}(p_0/p_T)^2
  +{\cal O}((p_0/p_T)^4)\right].
\end{eqnarray}
Substituting the above cross sections into Eq.~(\ref{eq:cronin1}),
we have,
\begin{equation}
  R_A=1 + \frac{9A^{1/3}}{16\pi R_0^2}\sigma\frac{n(n-2)}{2}(p_0/p_T)^2
    +{\cal O}((p_0/p_T)^4)
\end{equation}

There are a few interesting features in the above estimate. The
nuclear enhancement of jet cross sections decreases with $p_T$,
a general feature of high-twist processes. It also depends on
the power $n$ of the differential parton cross section.
This feature is a direct consequence of the interference effect,
{\em i.e.}, the cancellation between terms in the single and
double scattering amplitudes.
From both pQCD calculations and experimental data, we know
that the power $n$ decreases with the colliding energy.
Thus, the nuclear enhancement of large $p_T$ parton production
will decrease, and eventually it will disappear at high energies.
This trend has already been observed in experiments \cite{CRON2} 
in the energy range $\sqrt{s}=20 - 40$ GeV. A simple way to 
understand this energy dependence is the following.
At low energies, the differential cross section for a single
hard scattering decreases very fast with $p_T$ (large power).
It is then  easier for the incident parton to acquire
a large $p_T$ through two successive scatterings, each
with small transverse momentum transfer, than through
a single large $p_T$ scattering. This is why multiple
scatterings in $pA$ collisions cause the enhancement
of large $p_T$ parton production. As energy increases,
the differential cross section for a single scattering
decreases less rapidly with $p_T$ as compared to
low energies. It is  no longer more economical to
produce a large $p_T$ parton through double scatterings
than through a single scattering. At extremely high energies,
the Cronin effect will disappear. Therefore, we will not
consider multiple initial scatterings in the following
since we only consider parton production in heavy-ion 
collisions at future collider energies (RHIC and LHC).

\subsection{Monte Carlo Simulations}

Since we have argued that multiple initial parton scatterings
will become unimportant at high energies, we can assume
a binary approximation for hard scatterings. In that case, 
multiple hard processes involve only independent pairs of partons.
Only very rarely does a given parton suffer two high $p_T$ 
scatterings in one event. As the energy increases the number of 
partons that can participate in moderate $p_T>p_0$ processes 
increases rapidly and the nuclear shadowing phenomenon will 
also become important at small $x$. However, for the bulk
of parton and transverse energy production, the basic 
independent binary nature of the multiple mini-jet production 
rate is expected to remain a good  approximation as 
long as Eq.~(\ref{indepaa}) holds. 

For soft interactions, a nucleus-nucleus collision is also
decomposed into binary collisions involving in general excited or
wounded nucleons. Wounded nucleons are assumed to be $q-qq$ string-like
configurations that decay on a slow time scale compared to the overlapping 
time of the nuclei. In the FRITIOF \cite{FRIT} scheme wounded nucleon 
interactions follow the same excitation law as the original 
hadrons. In the DPM \cite{DPM} scheme subsequent collisions 
essentially differ from the first since they are assumed to involve
sea partons instead of valence ones. In HIJING \cite{HIJING} we 
adopted  a hybrid scheme, iterating string-string 
collisions as in FRITIOF but utilizing DPM-like distributions 
as in Eqs.~(\ref{eq:xdistr1}) and (\ref{eq:xdistr2}). Another
difference in the way soft interactions are treated in HIJING is that
string-string interactions are also allowed to de-excite as well as 
to excite
the strings further within the kinematic limits. In contrast, in the FRITIOF
model multiple interactions are assumed to lead only to excitations of 
strings with greater mass. 
Many variations of the algorithm for multiple soft interaction
are of course possible as emphasized before. The one implemented in HIJING is
simply a minimal model which reproduces essential features of moderate
energy $pA$ and $AA$ data. 

The number of binary collisions at a given nuclear impact parameter is
determined by Glauber geometry. We employ three-parameter Woods-Saxon 
nuclear densities determined by electron scattering data \cite{WDSX} 
to compute that geometry as done in ATTILA \cite{ATTILA}. 

For each binary collision, we use the eikonal formalism as given in
the previous section to determine the collision probability, elastic or
inelastic, and the number of jets it produces.  After a hard scattering, 
the energy of the scattered partons is subtracted from the nucleon and only 
the remaining energy is used to process the soft string excitation. 
The excited string system minus the scattered hard partons
suffers further collisions according to the geometric probabilities.

        For each hard scattering, one then has to take
into account the corrections due to initial and final
state radiations.  In an axial gauge and in the leading logarithmic
approximation, the interference terms of the radiation
drop out. The amplitude for successive radiations has then a
simple ladder structure and the probability for multiple
emissions becomes the product of each emission \cite{FIELD}. 
The virtualities of the radiating partons are ordered along 
the tree, decreasing until a final value $\mu_0^2$ is reached
below which pQCD is no longer valid. This provides the framework for a 
Monte Carlo simulation of parton showering and its space-time 
interpretation \cite{ODOR,WEBB}.

        At a given vertex of the branching tree, the probability
for the off-shell parton $a$ of virtuality $q^2<q^2_{\rm max}$
to branch into partons $b$ and $c$ with fractions $z$ and $1-z$
of the light-cone momentum is given by \cite{ODOR,WEBB}

\begin{equation}
        d{\cal P}_{a\rightarrow bc}(q^2,z)= \frac{dq^2}{q^2}
        dz\,P_{a\rightarrow bc}(z)\frac{\alpha_s[z(1-z)q^2]}{2\pi}
        \frac{{\cal S}_a(q^2_{\rm max})}{{\cal S}_a(q^2)},
                 \label{eq:shr1}
\end{equation}
where $P_{a\rightarrow bc}(z)$ is the Altarelli-Parisi 
splitting function \cite{GLAP} for the process $a\rightarrow bc$. 
By requiring the relative transverse momentum $q_T$ of $b$ and $c$
to be real,
\begin{equation}
        q_T^2=z(1-z)\left(q^2-\frac{q_b^2}{z}-\frac{q_c^2}{1-z}
        \right)\geq 0, \label{eq:shr2}
\end{equation}
and a minimum virtuality $q_b^2,\,q_c^2\geq\mu_0^2$, 
the kinematically allowed region of phase space is
then,
\begin{eqnarray}
        4\mu_0^2< & q^2 & <q^2_{\rm max}; \nonumber \\
        \epsilon(q)< & z & <1-\epsilon(q),\ \ 
        \epsilon(q)=\frac{1}{2}(1-\sqrt{1-4\mu_0^2/q^2}).
\end{eqnarray}
Note that the `$+$function' and $\delta$-function due 
to virtual corrections in the splitting functions in 
Eqs.~(\ref{eq:split1})-(\ref{eq:split4}) are not in effect in 
the allowed phase space. Their role has been replaced by the 
Sudakov form factor ${\cal S}_a(q^2)$ which is defined as \cite{ODOR,WEBB}
\begin{equation}
        {\cal S}_a(q^2)=\exp\left\{
        -\int_{4\mu_0^2}^{q^2}\frac{dk^2}{k^2}
         \int_{\epsilon(k)}^{1-\epsilon(k)}dz\sum_{b,c}
        P_{a\rightarrow bc}(z)\frac{\alpha_s[z(1-z)k^2]}{2\pi}
        \right\}, \label{eq:sdk}
\end{equation}
so that $\Pi_a(q^2_{\rm max},q^2)={\cal S}_a(q^2_{\rm max})/{\cal S}_a(q^2)$ 
is the probability for parton $a$ not to have any branching
between $q^2_{\rm max}$ and $q^2$. Therefore, the Sudakov
form factor in the Monte Carlo simulation is essential
to include virtual corrections and ensure unitarity.
Since the probability of parton emission between $q^2$ and $q^2-dq^2$ is
\begin{equation}
  \frac{dq^2}{q^2} \int_{\epsilon(k)}^{1-\epsilon(k)}dz 
  \sum_{b,c}P_{a\rightarrow bc}(z)\frac{\alpha_s[z(1-z)q^2]}{2\pi}\; ,
\end{equation}
the probability of no parton emission, by unitarity, will be
\begin{equation}
1-\frac{dq^2}{q^2} \int_{\epsilon(k)}^{1-\epsilon(k)}dz 
  \sum_{b,c}P_{a\rightarrow bc}(z)\frac{\alpha_s[z(1-z)q^2]}{2\pi}\; ,
\end{equation}
between $q^2$ and $q^2-dq^2$.
The probability of no parton emission between $q^2_{\rm max}$
and $q^2-dq^2$ then will be the product of the two probabilities,
\begin{eqnarray}
  \Pi_a(q^2_{\rm max},q^2-dq^2)&=&\Pi_a(q^2_{\rm max},q^2)
  -d\Pi_a(q^2_{\rm max},q^2) \nonumber\\
  &=&\Pi_a(q^2_{\rm max},q^2)\left\{1-
        \frac{dq^2}{q^2} \int_{\epsilon(k)}^{1-\epsilon(k)}dz 
        \sum_{b,c}P_{a\rightarrow bc}(z)
        \frac{\alpha_s[z(1-z)q^2]}{2\pi}\right\}\; .
\end{eqnarray}
This is just another form of Eq.~(\ref{eq:shr1}) integrated over
$z$. One can see that the Sudakov form factor \cite{ODOR} is the
solution of the above equation in which unitarity plays an important
role.

        In principle, one can perform the simulation of initial state 
radiation processes in a similar way. The partons inside a nucleon
can initiate a space-like branching increasing their 
virtuality from some initial value $Q_0^2$. A hard scattering
can be considered as a probe which can only resolve
partons with virtuality up to the scale of the hard scattering.
Otherwise without the scattering, the off-shell partons are only
virtual fluctuations inside the hadron and they will reassemble
back to the initial partons. In PYTHIA, which uses backward evolution,
a hard scattering is selected first with the known QCD-evolved
structure function at that scale, and then the initial branching
processes are reconstructed down to the initial scale $Q_0^2$. The
evolution equations are essentially the same as in final state
radiation except that one has to convolute with the parton
structure functions \cite{PYTHIA}.
HIJING explicitly uses subroutines from PYTHIA to simulate each
hard parton scattering and the associated initial and final state
radiations. The initial virtuality for the initial state 
evolution is  set to be $Q_0=$2 GeV/$c$, and the
minimum virtuality for the final state radiation is $\mu_0=$ 0.5 GeV/$c$.
The maximum virtuality for the associated radiations in a
hard scattering with transverse momentum transfer $p_T$
is chosen to be $q_{\rm max}=2p_T$. Angular ordering can also be 
enforced in PYTHIA to take into account the soft gluon 
interference \cite{WEBB} in the final state radiation.

After all binary collisions are processed, the scattered gluons from
each nucleon are arranged according to their rapidities and connected 
to the valence quarks and diquarks of that nucleon in the collision. 
The rare hard scatterings of $q$-$\bar{q}$ pairs with opposite flavors 
are treated as a special case and processed as independent strings. 

As studied in Refs.~\cite{MGMP90,GPTW,PWG93}, large $p_T$ partons
must propagate transversely through the whole excited matter.
They will suffer both elastic and radiative energy loss.
The energy loss $dE/dz$ is very sensitive to the
Debye screening mass in the medium. Thus, the study of
the energy loss of the produced hard jets or jet quenching
can provide us with a unique probe of the dense matter.
To test the sensitivity of final observables to jet quenching, 
we used a simple gluon splitting scheme (an effective 
induced radiation) in HIJING, given $dE/dz$ and the 
mean free path of the interaction $\lambda_f$. The  interactions 
are mostly soft between both soft and hard partons
in the medium. These interactions and induced radiation, in
some way, mimic pre-equilibrium cascading and semi-thermalization
of the produced partons.

Induced radiative energy loss is modeled in HIJING
by determining first the points of final state interaction
of hard partons in the transverse direction
and performing a collinear gluon splitting at every point.
We assume that interactions only occur with the locally comoving 
matter in the transverse direction. Interactions with
the nuclear fragments are negligible, because the two
nuclear discs pass each other on a very short 
time scale $2R_A/\gamma \ll 1$ fm.
The interaction points are determined by a probability with
a constant mean free path $\lambda_f$, 
\begin{equation}
        dP=\frac{d\ell}{\lambda_f}e^{-\ell/\lambda_f},
\end{equation}
where $\ell$ is the distance the jet has traveled in the
transverse direction after its last
interaction. 

Since the pre-hadronization state in HIJING is represented
by connected groups (strings) of valence partons and gluons (kinks),
interactions can be easily simulated by transferring a part 
of the parton energy, $ \Delta E(\ell)=\ell dE/dz$, from one 
string configuration to another. Collinear gluon splitting 
results in a net jet quenching at the stage of hadronization,
because  the original hard parton energy is shared among several 
independent strings. This simple mechanism of course conserves 
energy and momentum and is numerically simple. A more dynamical 
parton cascade approach involving the space-time development of 
parton showering and collisions between the produced partons
has been made in PCM \cite{PCM}.

\begin{figure}
\centerline{\rotateright{\psfig{figure=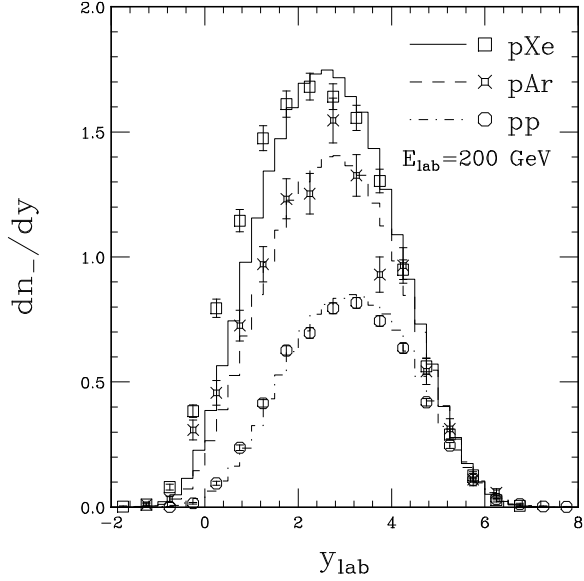,width=3in,height=3in}}}
\caption{Rapidity distributions for negative particles in $pp$
        (circle, dot-dashed histogram), $pAr$ (crossed-square, 
        dashed histogram)and $pXe$ (square, solid histogram) 
        collisions at $E_{\rm lab}=200$ GeV. The points are data 
        from Ref.~\protect\cite{DMARZO} and histograms are 
        from HIJING calculation.}
\label{fig12}
\end{figure}

To make sure that the model incorporates the right physics down to low
energies we have compared the numerical results
in $pA$ and $AA$ collisions with data at SPS energies.
Shown in Fig.~\ref{fig12} are the calculated
rapidity distributions of negative charged particles in $pp$ (dot-dashed 
histogram), $pAr$ (dashed histogram) and $pXe$ (solid histogram) collisions 
at $E_{\rm lab}=200$ GeV. The data are from Ref.~\cite{DMARZO}.
Because jet production is negligible at this 
energy, particle production occurs  mainly through soft excitations of 
projectile and target nucleons. The HIJING low $p_T$ algorithm
reproduces an increase of particle production in the central region 
with the number of participating target nucleons. The peak of the 
rapidity distribution is shifted back towards the target region
and its height is proportional to the target atomic number. 
In the target region, HIJING under-predicts particle production 
due to the neglect of final state cascading.

\begin{figure}
\centerline{\rotateright{\psfig{figure=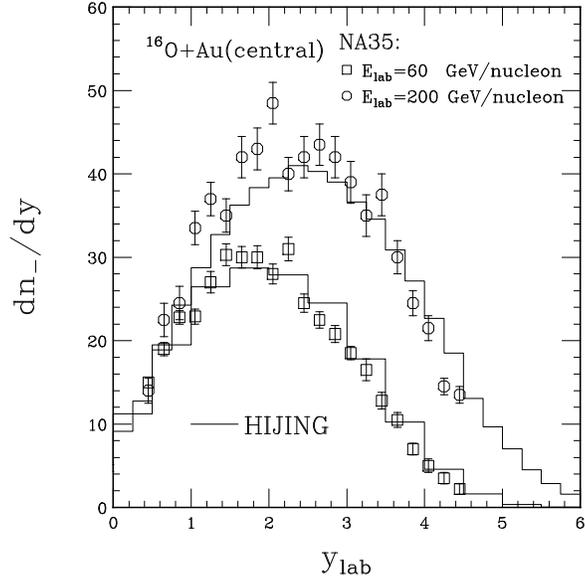,width=3in,height=3in}}}
\caption{Rapidity distributions for negative particles in central
        $O+Au$ collisions at $E_{\rm lab}=60$ and 200 AGeV. The data are 
        from Ref.~\protect\cite{NA35} and histograms are from 
        HIJING calculation.}
\label{fig13}
\end{figure}

\begin{figure}
\centerline{\psfig{figure=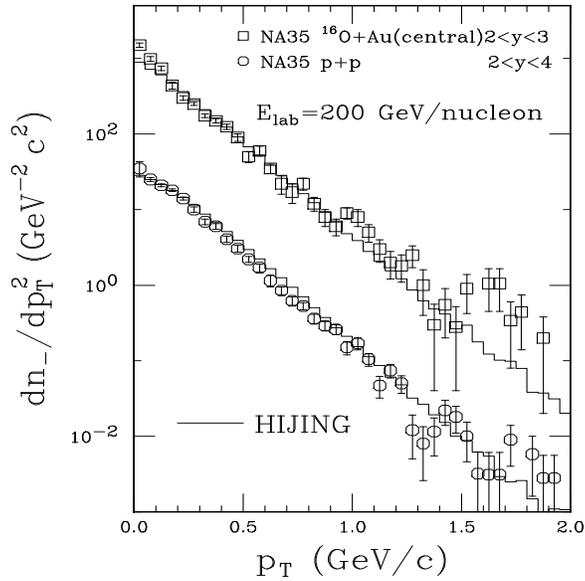,width=3in,height=3in}}
\caption{$p_T$ distributions for negative particles in $pp$
        and central $O+Au$ collisions at $E_{\rm lab}=200$ AGeV. The data
        are from Ref.~\protect\cite{NA35} and histograms are from HIJING
        calculation.}
\label{fig14}
\end{figure}

Shown in Figs.~\ref{fig13} and \ref{fig14} are the 
calculated rapidity distributions of
negative particles in central $O+Au$ collisions at $E_{\rm lab}=60$ and 
200 AGeV, and the transverse momentum
distributions of negative particles in $p+p$ and central $O+Au$ collisions
at $E_{\rm lab}=200$ AGeV. The overall features
of the data\cite{NA35} are well accounted for except
for the enhancement at low $p_T<0.2$ GeV/$c$
in $O+Au$. That enhancement is currently believed to originate also from 
final state interactions \cite{GR91,WEBE91}.
The data for the $O+Au$ collisions 
are taken with a central trigger. In HIJING simulations,
central events are selected which can give the corresponding
averaged multiplicity. 

\begin{figure}
\centerline{\rotateright{\psfig{figure=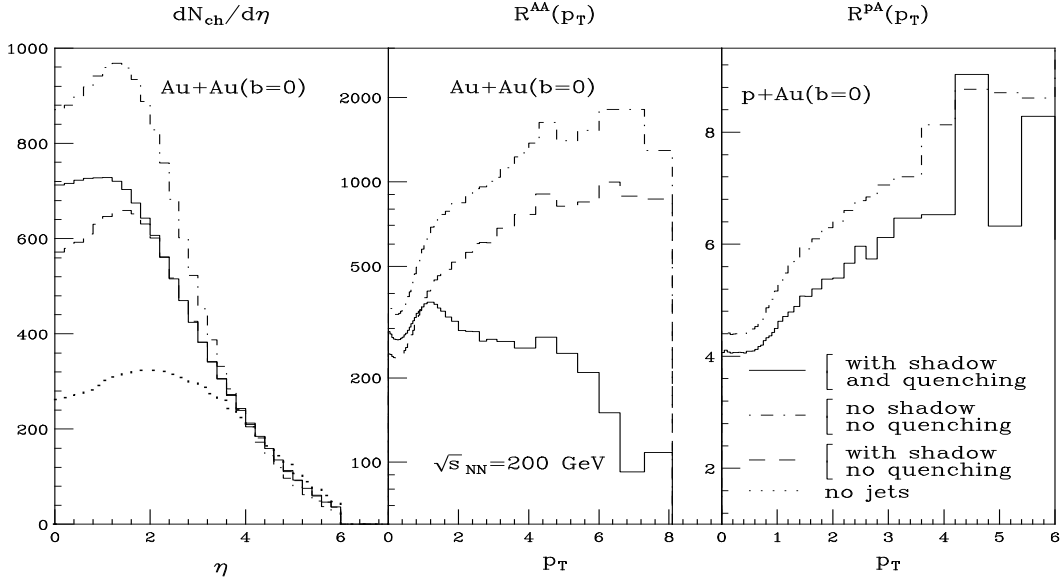,width=3in,height=5.5in}}}
\caption{HIJING results of the dependence of the inclusive charged 
         hadron spectra in central $Au+Au$ and $p+Au$ collisions 
         on minijet production (dash-dotted), gluon shadowing (dashed),
         and jet quenching (solid) assuming that gluon shadowing is 
         identical to that of quarks and $dE/dz=$2 GeV/fm with 
         $\lambda_f=$1 fm. $R^{AB}(p_T)$ is the ratio of the inclusive 
         $p_T$ spectrum of charged hadrons in $A+B$ collisions to that 
         of $p+p$.}
\label{fig15}
\end{figure}

The calculated results for central $Au+Au$ collisions 
at $\sqrt{s}=$200 AGeV are shown in Fig.~\ref{fig15}.
The left panel shows the pseudorapidity distributions of charged
particles. Note that without minijets (dotted), the $2A$
soft beam jets in HIJING lead to $dN_{AA}/d\eta\approx A dN_{pp}/d\eta$.
Each  beam jet contributes about 0.75 to the central
rapidity density almost independent of energy \cite{HIJINGPP}.
Without gluon shadowing(dash-dotted), minijets are found
to approximately triple the rapidity density
due to beam jets. However, if gluon shadowing
is of the same magnitude as that for quarks and antiquarks, 
then the minijet contribution to the rapidity density is reduced 
by approximately a half (dashed).  The solid histogram
shows that the effect of jet quenching on the rapidity density
is small for $dE/dz=2$ GeV/fm and $\lambda_f=$1 fm.

Plotted in the middle panel of Fig.~\ref{fig15}, is the ratio,
\begin{equation}
        R^{AB}(p_T)=\frac{d^2N_{AB}/dp_T^2/d\eta}
                {d^2N_{pp}/dp_T^2/d\eta}, \label{eq:rab}
\end{equation}
of the inclusive $p_T$ spectrum of charged particles in central
$Au+Au$ collisions to that of $p+p$. For $p_T>2$ GeV/$c$, both
shadowing and quenching are seen to  reduce dramatically the inclusive
hadron production. In the absence of shadowing and jet quenching
(dash-dotted) the ratio approaches the number of
binary $pp$ collisions in the reaction. Shadowing alone (dashed) 
suppresses moderate $p_T$ hadrons by a factor of about 2. 
Inclusion of jet quenching with $dE/dz=$ 2 GeV/fm reduces
that yield by another factor of about 3--5. It is remarkable
and encouraging that the {\em single} inclusive hadron spectrum appears
so sensitive to nuclear effects on jet production. Clearly
 higher order correlations, {\em e.g.}, high $p_T$ back-to-back
hadron pairs, will be even more sensitive to these effects.

From the above, however, we see that $A+A$ data alone would 
not be sufficient to disentangle the effects of shadowing
and jet quenching. To separate the two, $p+A$ collisions
{\em must} be studied at the same energy! In those reactions, 
the density of comoving hadrons is so low that final state 
interactions leading to jet quenching should be
negligible. Therefore, any observed suppression of 
moderate $p_T$ hadrons could be attributed to  gluon shadowing alone. 
The right panel in Fig.~\ref{fig15} shows the 
ratio as defined in Eq.~(\ref{eq:rab}) for
central $p+Au$ collisions. Without shadowing (dash-dotted), 
the high $p_T$ limit  again reaches the number of binary 
collisions, which is about 8 in this case. The low $p_T$
limit is, on the other hand, controlled by the number
of {\em pairs} of beam jets, which is approximately 4
in central collisions. We see from this figure
that the ratio $R^{pA}(p_T)$ at moderate $p_T$ is indeed sensitive
to gluon shadowing. Therefore, a systematic measurement of the
inclusive hadron spectra in $p+p$ and $p+A$ at the
same energy is essential in order to determine the magnitude
of gluon shadowing. Since the Cronin effect on jet production in
$p+A$ collisions at $\sqrt{s}\gsim 100$ AGeV is expected to be 
small, the spectra of moderate $p_T$
hadrons will provide information on the gluon
structure function  complementary to that deduced by the conventional
direct photon technique that is limited to higher $x$ gluons
due to the high  $\pi^0\rightarrow \gamma \gamma$ background.
Once the magnitude of gluon shadowing is determined,
its contribution to the observed suppression 
of moderate $p_T$ hadrons in  $A+A$ data
can be calculated, thereby making it possible to isolate the 
the contribution due to jet quenching.

\subsection{Space-Time Structure of Parton Production}

\subsubsection{Space-time history}

So far I have discussed the general features of a pQCD-inspired model
for parton and particle production in hadronic and nuclear
collisions. However, in order to estimate the initial number and
energy density of the produced partons, one must know their 
formation time. The total formation time must include the 
interaction time of the semihard processes and the formation time 
for initial and final state radiations.  As demonstrated in 
Refs.~\cite{GLAUBER,GWLPM1}, formation time is a consequence of interference
in multiple scattering processes, which can also be obtained from the
uncertainty principle. Destructive interference suppresses
those secondary scatterings which happen within the formation
time of the previous interactions. Inclusion of the formation
time effects is necessary to treat the secondary scatterings properly.
Otherwise, one might overestimate the interaction rates and
parton production. One source of overestimate may arise 
if one neglects the interaction time of the scatterings. 
During the interaction time, which depends on whether the 
scattering is hard, semihard or soft, the participating and 
therefore the produced partons cannot scatter again immediately 
with the beam partons. Another overestimate has to do 
with the initially radiated partons before the hard or 
semihard scatterings. In calculating jet cross sections and 
the number of hard scatterings, parton 
structure functions of a nucleus $f_{a/A}(x,Q^2)$ evaluated
at the scale $Q^2=p_T^2$ are used. These parton distribution
functions then have already included the effect of QCD evolution,
producing more partons at small $x$ and larger $Q^2$. Therefore,
in Monte Carlo simulations in which initial radiations are treated
by backward evolution, the radiated partons should not participate
in any interactions before the end of the corresponding hard scattering.
This is also in accord with the factorization theorem \cite{COLL}
of pQCD.

Formation time is usually related to a radiative process.
According to the uncertainty principle, an off-shell parton
can be considered as a virtual fluctuation and it can only
live for a finite time, $\Delta t$, determined by its 
virtuality $q^2$,
\begin{equation}
        \Delta t\approx q_0/q^2, \label{eq:dt}
\end{equation}
where $q_0$ is the energy of the parton. This $\Delta t$
is exactly the formation time of the
radiation which will follow. Thus, after $\Delta t$, the 
off-shell parton will then branch or ``decay'' into other
partons which can further initiate branchings until a minimum
virtuality $\mu_0$ is reached. At $q^2\leq\mu_0^2$, pQCD is not 
considered to be valid anymore and the process of nonperturbative 
hadronization takes over. Following this tree of 
branching (which also includes
initial space-like radiation) and assuming a classical straight line trajectory
for partons, one can then follow the space and time evolution
of the initial parton production \cite{KEXW94a}. 

        The life time of a virtual parton in Eq.~(\ref{eq:dt}) is
only a rough estimate according to the uncertainty principle. One
could also estimate it via a pQCD analysis of the radiative amplitudes 
induced by multiple scatterings \cite{GWLPM1,GWLPM2}. Thus, $q_0/q^2$ 
should be a good estimate of $\Delta t$ in magnitude. However,
this formation time or the life time of a virtual parton cannot
be identified with the decay rate of a virtual parton as has 
been done in PCM \cite{PCM} because the later has an additional dependence
on $\alpha_s$. The decay rate is only related to the radiation
intensity, not the formation time of the radiated partons.

Eq.~(\ref{eq:dt}) can also be used to estimate the interaction
time for each hard scattering with $q$ being the sum of the initial
or final four-momenta of the colliding partons. If the fractional
momenta of the partons are $x_1$ and $x_2$, then the interaction
time can be estimated as
\begin{equation}
                \Delta t_i\approx \frac{x_1+x_2}{2\,x_1\,x_2\sqrt{s}}.
\end{equation}
In this case, the asymmetric scatterings ($x_1\gg x_2$ or $x_1\ll x_2$)
have longer interaction time than the symmetric ones ($x_1\sim x_2$)
for fixed parton-parton center-of-mass energy $x_1\,x_2\sqrt{s}$.
We will also assume that the interaction time is the same for all
channels.

\begin{figure}
\centerline{\psfig{figure=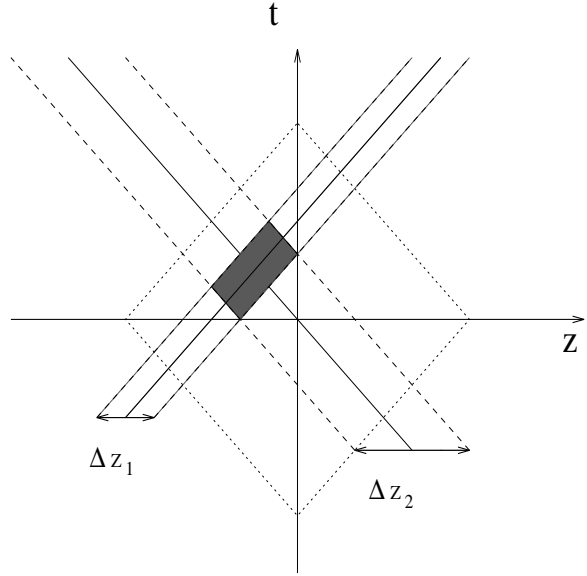,width=3in,height=3in}}
\caption{The overlap region in space and time of two incoming
         partons each with spatial spread of $\Delta z_1$ and $\Delta z_2$,
         respectively. The interaction point is chosen randomly inside
         the shaded region. Solid lines show the trajectories of the
         parent nucleons and dotted lines represent the overlap
         region of the two nuclei.}
\label{fig16}
\end{figure}

        In the rest frame of each nucleus, three-parameter Woods-Saxon 
nuclear densities are used to construct the nucleon distribution
inside the nucleus. The system is then boosted to the 
center-of-mass frame of the two colliding nuclei. 
Due to the fact that gluons, sea quarks and antiquarks are 
only quantum fluctuations before they really 
suffer scatterings, their longitudinal distribution
around the center of the nucleon is still governed by 
the uncertainty principle in any boosted frame \cite{AM88}.
One usually refers to this distribution as the ``contracted 
distribution'', in which a parton with $x_i$ fractional momentum 
has a finite spatial spread,
\begin{equation}
        \Delta z_i\approx 2/x_i\sqrt{s}. \label{eq:dz}
\end{equation}
Transversely, partons are distributed around their parent 
nucleons according to the Fourier transform of a dipole 
form factor \cite{WANG91}. If we define $t=0$ as 
the moment when the two {\em nuclei} have complete
overlap, the overlap of two colliding {\em nucleons}
can happen anywhere inside the diamond-shaped region (with dotted lines)
in {}Fig.{}~\ref{fig16}. The diagonal length of the diamond
is $2R_A \gamma$ with $\gamma$ the Lorentz boost factor.
The interaction point of two partons in a $t$-$z$ plane 
can be anywhere within the shaded area in {}Fig.{}~\ref{fig16}.
The solid lines are the trajectories of the two parent
nucleons which spread around the nuclei according to a
longitudinally contracted Woods-Saxon distribution.

\begin{figure}
\centerline{\psfig{figure=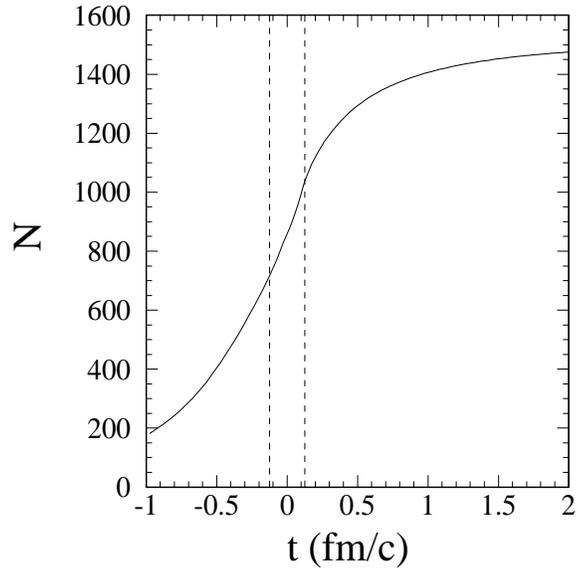,width=3in,height=3in}}
\caption{The total number of produced partons $N$ as a function of 
         time $t$, with $t=0$ defined as when the two nuclei have 
         complete overlap.}
\label{fig17}
\end{figure}

To study the space-time history of parton production,
let us look at the results of HIJING simulations for
$Au+Au$ at the highest RHIC energy, $\sqrt{s}=200$ AGeV.
Shown in {}Fig.~\ref{fig17} is the total number of produced 
partons, on-shell as well as off-shell, as a function of time. 
We see that long  before the two nuclei overlap and 
hard scatterings take place, partons have already 
been produced via initial state radiation. 
Note that, if the coherence is not broken by the hard
scattering, partons which would have been emitted from 
the initial state radiation will not emerge as produced
partons. In Fig.~\ref{fig17} the initiators of the 
space-like branching are also included as produced partons. 
Therefore, a parton is defined to be a produced one before the hard
scattering if it has initial state radiation.
If a parton does not have initial state radiations, 
it will only become a produced parton after the hard scattering.
{}From Fig.{}~\ref{fig17},
we can see that about $2/3$ of the total number of partons
are produced between $t=-0.5$ and 0.5 fm/$c$ while about $200$
semihard scatterings happen between $t=-0.1$ and 0.1 fm/$c$ as
indicated by dashed lines. We find also that about $2/3$ of the
total number of partons are produced in initial and final
state radiations. The fraction of partons from branching
should increase with the colliding energy and with smaller
choices for $\mu_0$.

\begin{figure}
\centerline{\psfig{figure=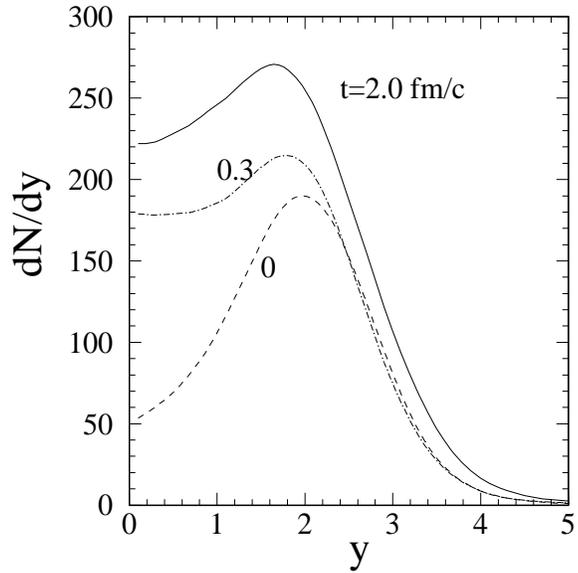,width=3in,height=3in}}
\caption{The rapidity distribution $dN/dy$ of produced partons at different
         times $t$.}
\label{fig18}
\end{figure}

        To see how hard or semihard scatterings and initial 
and final state bremsstrahlung contribute to parton 
production, the rapidity distribution of produced partons at 
different time is shown in Fig.{}~\ref{fig18}. Before $t=0$,
most of the partons come from initial state radiation.
Since the radiations are almost collinear, these partons move
along the beam direction and therefore have large rapidities.
The semihard scatterings then produce partons uniformly over
a rapidity plateau and fill up the middle rapidity region.
Final state radiations, which happen after the semihard
scatterings, will also produce partons uniformly in the
central rapidity region. It is quite clear from Fig.{}~\ref{fig18}
that the dip of $dN/dy$ in middle rapidity is actually
caused by the parton production from initial state
bremsstrahlung at large rapidity. These partons
with large longitudinal momenta will move away
from the interaction region after the semihard scatterings. 
They do not rescatter with the beam partons in the
leading twist approximation.

\begin{figure}
\centerline{\psfig{figure=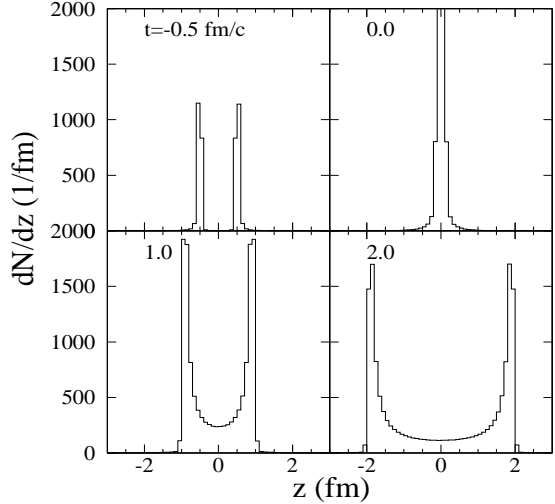,width=3in,height=3in}}
\caption{Parton distribution along the $z$-axis at different times.}
\label{fig19}
\end{figure}

        {}Fig.{}~\ref{fig19} contains snapshots of $dN/dz$ 
at different times to illustrate how the parton production
evolves in space and time. At $t=-0.7$ fm/$c$, as the two nuclei 
approach toward each other before they actually overlap, 
initial state radiations have already begun. These partons
have large rapidities and are Lorentz contracted with an
average spread in $z$,
\begin{equation}
        \Delta z\approx 1/p_0+2R_A\frac{2m_N}{\sqrt{s}}
        \approx 0.25 \ {\rm fm}, \ \ \sqrt{s}=200 {\rm GeV},
\end{equation}
where $p_0=2$ GeV/$c$ is the $p_T$ cutoff for semihard
scattering, $R_A$ is the nuclear radius of $Au$ and
$m_N$ is the nucleon mass.  After the hard scatterings
and during the interaction time, partons are produced
uniformly in the central rapidity region. Afterwards,
partons follow straight lines by free-streaming and are
distributed evenly in $z$ between the two receding pancakes
of beam partons (partons from initial state radiations).
The hyperbolic-shaped equal-time parton distribution in $z$ 
corresponds approximately to a boost invariant parton density
in the central region in this free-streaming picture. 

\begin{figure}
\centerline{\rotateleft{\psfig{figure=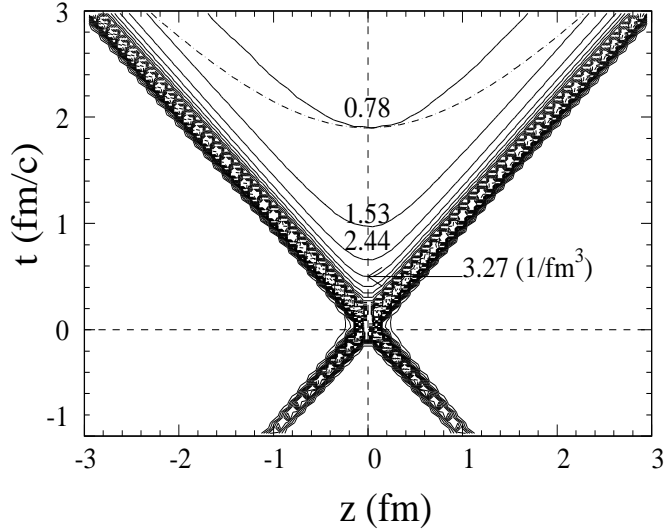,width=3in,height=4in}}}
\caption{Contour plot in $z-t$ plane of the parton density $n$ of
        Eq.~(\protect\ref{eq:rho1}), as indicated by the numbers. The wavy 
        structure along the light-cone is only an artifact of the 
        plotting program.}
\label{fig20}
\end{figure}

        Another illustrative way to study the evolution 
of local parton density is to make a contour plot of parton 
density $n$ in $z$ and $t$ as shown in Fig.{}~\ref{fig20}.
Here, $n$ is defined as
\begin{equation}
        n=\frac{1}{\pi R_A^2}\frac{dN}{dz}, \label{eq:rho1}
\end{equation}
and a sharp sphere distribution is assumed for the nuclear density.
One can clearly see that partons inside the two approaching nuclei
have a spatial spread of $\Delta z=0.25$ fm in $z$. This spread 
continues for partons from the initial 
state radiation as they escape from the interaction 
region along the beam direction with large rapidities. The
interaction region where semihard scatterings happen lasts
for about 0.5 fm/$c$, from $t=-0.25$ to 0.25 fm/$c$.  Because
of the contracted distribution, some of the partons lie 
outside the light-cone which is defined with respect to
the overlapping point of the two nuclei.

        If one assumes boost invariance \cite{BJOR},
the parton density can be estimated as
\begin{equation}
        n=\frac{1}{\pi R_A^2\tau}\frac{dN}{dy}, \label{eq:den1}
\end{equation}
where $\tau$ is the proper time and $n$ 
should be a function of $\tau$ only.  By comparing the 
contour of constant density with the hyperbola of 
constant $\tau$ (dot-dashed line) in Fig.~\ref{fig20},
we see that this is approximately true. We also see that 
the density, as indicated by the numbers, decreases 
like $1/\tau$ due to free-streaming.

\subsubsection{Local isotropy and chemical composition}
\label{sec:chem}

        Since, as we have demonstrated, there are numerous 
partons produced within a rather short time, the initial parton
density is very high at $\tau \ge 0.25$ fm/$c$, immediately after 
the interaction region (see Fig.~\ref{fig20}). Within such a dense
system, secondary parton scatterings and production are 
inevitable at a time scale much larger than the average formation
time. The equilibration time for the system can be 
estimated by solving a set of rate equations \cite{BDMTW} as I
will show later. In this approach, 
one must make sure that there is 
approximately local isotropy in momentum space. This
could be achieved through secondary parton scatterings
as has been investigated in Refs. \cite{PCM,SHUR}. One
can also use free-streaming to estimate the upper bound
of the thermalization time \cite{BDMTW,HWA},
by studying the momentum distribution of partons in a
cell of the size of the mean free path $\lambda_f$,
assuming that parton scatterings are frequent enough
to achieve thermalization within this upper bound.

\begin{figure}
\centerline{\psfig{figure=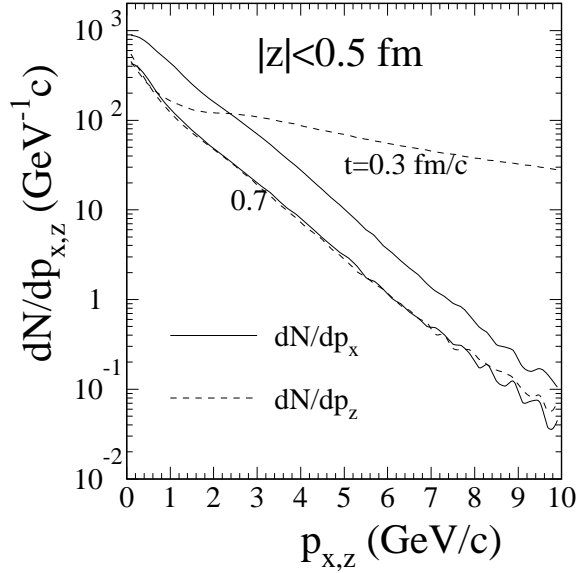,width=3in,height=3in}}
\caption{$p_x$ and $p_z$ distributions at different times for 
         partons in the central slice $|z|<0.5$ fm.}
\label{fig21}
\end{figure}

        Let's concentrate on the central slice at $z=0$ with
$|z|<0.5$ fm. At the very early stage during the interaction
region, produced  partons with different rapidities are confined 
to a highly compressed slab with $\Delta z \approx 0.5$ fm.
As the system expands, partons with large rapidities will escape
from the central slice while partons with small rapidities
remain. Shown in Fig.{}~\ref{fig21} is the evolution of momentum 
distributions  in $p_x$ (solid lines) and $p_z$ (dashed lines) 
at different times (indicated by the number on each line). 
The slope of the $p_z$ distribution decreases because partons with large
longitudinal momenta gradually escape from the central region.
At $t=0.7$ fm/$c$, the slopes of $p_x$ and $p_z$ distributions become
the same. We should note that the isotropy in momentum
distribution via free-streaming at this moment is only
a transient phenomenon which will disappear quickly afterwards.
What really determines the thermalization time is the
relaxation time due to parton scatterings \cite{HHXW,WONG}.
If the relaxation time is short, thermalization can be
achieved even before the transient momentum isotropy.
On the other hand, for a longer relaxation time, the momentum distribution
will become anisotropic again and will eventually approach
thermal equilibration afterwards. To calculate the relaxation time,
one has to consider both elastic \cite{HHXW} and 
inelastic \cite{WONG} parton scatterings and include
resummation of hot thermal loops. Unfortunately, a complete
calculation including higher order processes in a 
nonequilibrium system is not available.
In the following we will simply assume that the time for
local momentum isotropy can be replaced by the time 
for the central slab of about 1 fm to achieve isotropy in 
momentum space via free-streaming,
\begin{equation}
        \tau_{\rm iso}\approx 0.7 \ \ {\rm fm}/c.
\end{equation}
The readers should be reminded that such assumption will
give rise to corrections from the neglected viscosities 
to the results concerning parton equilibration.

Unlike quarks and gluons in an ideal gas with chemical equilibrium, 
partons produced in the initial stage of heavy-ion collisions,
which are determined by the parton structure functions, the hard
scattering cross sections and the radiation processes in pQCD,
are far from chemical equilibrium.
Due to the difference in the numbers of degrees of freedom in
color space, cross sections involving gluons are always larger
than those of quarks.
Similarly, both initial and final state radiations produce
more gluons than quarks and anti-quarks. Therefore in pQCD,
the ratio of produced quarks to gluons is much smaller
than the ratio in an ideal gas, which is $9/4$ for three quark
flavors.

\begin{figure}
\centerline{\psfig{figure=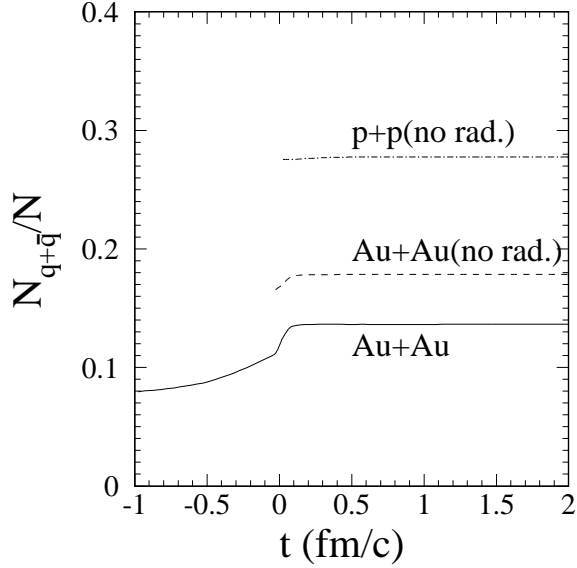,width=3in,height=3in}}
\caption{The fractional number of produced quarks and anti-quarks
         as a function of time for $Au+Au$ (solid)collisions, $Au+Au$ 
         without radiations (dashed), and $p+p$ without 
         radiation (dot-dashed).}
\label{fig22}
\end{figure}

        Shown in Fig.{}~\ref{fig22} are the fractions of produced 
quarks and anti-quarks as functions of time $t$ in $pp$ and $AA$ 
collisions at $\sqrt{s}=200$ AGeV. For $Au+Au$ collisions, 
hard scatterings with $p_T>2$ GeV/$c$ produce about 13\% quarks 
and anti-quarks. If initial and final state radiations are
not included the ratio jumps to 18\%, because radiations
produce more gluons than quarks and antiquarks. In $pp$ collisions, 
about 28\% of the partons produced via hard scatterings without 
radiations are quarks and anti-quarks. 

\begin{figure}
\centerline{\psfig{figure=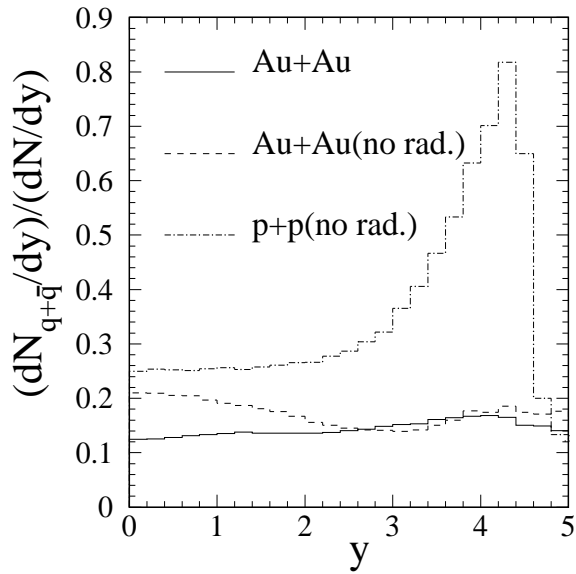,width=3in,height=3in}}
\caption{The final fractional number of produced quarks and anti-quarks
         as functions of rapidity for $Au+Au$ (solid)collisions, $Au+Au$ 
         without radiations (dashed), and $p+p$ without 
         radiation (dot-dashed).}
\label{fig23}
\end{figure}

        The difference between $Au+Au$ and $pp$ collisions is 
due to the different $A$ dependence of the valence and sea quark 
production. Since the partons are produced via binary collisions,
the number of produced gluons and sea quarks and antiquarks scales like 
$A^{4/3}$. On the other hand, baryon  number conservation requires 
valence quark production to scale like $A$. Given the fraction of 
total produced quarks $q_{pp}=0.28$ and valence quarks $v_{pp}=0.14$
in $pp$ collisions, one can find for $Au+Au$ collisions,
\begin{equation}
        q_{AA}=\frac{q_{pp}-(1-A^{-1/3})v_{pp}}{1-(1-A^{-1/3})v_{pp}},
\end{equation}
which gives $q_{AA}=0.18$ for $A=197$, as we obtained from 
Fig.{}~\ref{fig22}  of the numerical calculation. To demonstrate the
effective valence quark production, we plot in Fig.{}~\ref{fig23} 
the rapidity dependence of the fractional quark number. 
In $pp$ collisions, the valence quark production peaks at
large rapidities. In nucleus-nucleus collisions, the valence 
quark production has a different scaling in $A$ than gluons and
sea quarks. This is why the fraction of quark and antiquark
production is suppressed more at large rapidity than in the
central rapidity region. The relative quark and antiquark 
production in the central rapidity region is further 
suppressed by final state radiation. 

The relative quark production found in PCM \cite{KGJK93}
is about five times larger than what I show here. This
is due to an overestimate of intrinsic quark production
through flavor excitations in PCM. Such flavor excitations
should be suppressed by interference among pQCD amplitudes 
at the same order \cite{CSS86}. For the same reason, other
heavy quark production is also overestimated in PCM \cite{KG93}.

        The small number of produced quarks relative to gluons 
within pQCD has important consequences for chemical equilibration of
the partonic system. Because of the small initial relative
quark density and small quark production cross section
as compared to gluon production, it takes a very long time
for the system to achieve chemical equilibrium, if it ever does.
If this time is longer than the phase transition time,
a fully equilibrated QGP may never be formed \cite{BDMTW}.

\subsubsection{Multiple initial parton scatterings ?}

I have already demonstrated earlier in Sec. 3.2 that
multiple parton scatterings become less important at 
high energies and eventually will disappear. This phenomenon
can also be understood in a space-time picture. But before 
doing that, I should emphasize again that the multiple parton 
scatterings in this context always refer to those during
the overlapping time of two colliding nuclei. They are
different from the final parton rescatterings which are
responsible for parton thermalization.

        We have explicitly taken into account the interaction
time for the semi-hard scatterings, which is roughly
$\tau_i\sim 1/p_T$. Inside a highly Lorentz contracted nucleus, 
the spatial spread for partons which could participate in a hard 
scattering with transverse momentum transfer $p_T$ is also
about $\Delta z\sim 1/p_T$. This would leave the produced
partons no time to have another hard scattering of $p_T$ 
with the incoming beam partons. For finitely contracted nuclei
at relatively low energies, this 
kind of double high $p_T$ scattering is still possible. However,
these  higher twist processes should be 
suppressed by a factor of $1/p_T^2$. It is also 
possible for a parton to go through a hard and a soft 
scattering  subsequently since the soft partons always have 
a spatial spread of 1 fm.  This kind of hard-soft multiple
interactions constitutes the leading contribution to higher
twist corrections to hard processes in nuclear
collisions \cite{QIU}. However, double semihard 
scatterings at high energies  with $p_T\sim p_0$ will be 
suppressed due to finite interaction time.

        The large $p_T$ enhancement of both Drell-Yan dilepton
production and single hadrons in $pA$ collisions at $\sqrt{s}\leq 50$
GeV is considered as a result of multiple parton interactions
\cite{BROD}. At these energies, the contracted length of a heavy
nucleus is still relatively larger than the interaction and formation
time. The factorized form of the parton model is then modified
due to the finite beam energy. However, a collection of 
experimental data \cite{CRON1,CRON2} even in this intermediate
energy range has already shown the effects of the finite
interaction time. As the energy increases, the interaction
time becomes more important as compared to the size of an
increasingly contracted nucleus. The partons then have less
time for secondary scatterings. This then leads to the
observed decrease of the large $p_T$ enhancement in $pA$
interactions as the beam energy increases. At ultrarelativistic
energies, one should therefore expect the suppression of multiple
semihard scatterings. As I have argued earlier, partons from 
the associated initial state radiation will not scatter 
again with the incoming beam partons due to factorization.

\section{Parton Equilibration}

The initial semihard interactions between two beams 
of partons happen in a short
period of time. After the leading partons leave the interaction
region, a partonic system will be formed in the central region.
In this section, I will discuss how this partonic system
equilibrates and describe the physics involved, based on a microscopical 
description. The thermalization and equilibration in
this description are driven by multiple scatterings and parton
production (or annihilation) by induced radiation (or absorption).  
Though an exact quantum field treatment of multiple parton 
interaction is not possible with present technology, a 
semiclassical approximation has been undertaken \cite{PCM}.
The most important and difficult effort in a semiclassical 
approximation is to simulate the quantum effects, like 
interference, coherence, and especially, nonperturbative 
phenomena in QCD. Under extreme conditions at high temperature
and densities, these quantum effects become very important
and may dictate the evolution of the partonic system.
In this section, I will also discuss the interference effects 
associated with induced radiations, since radiative processes 
produce extra partons, leading to fast parton equilibration.

\subsection{Initial Conditions: a Hot and Undersaturated Gluonic Gas}

As we have discussed earlier, the kinematic separation of 
partons with different rapidities in a central slab of about 1 fm
give rise to a transient local momentum isotropy at the time 
of the order of $\tau_{\rm iso}=0.7$ fm/$c$. 
We will further assume this is the actual kinetic equilibration 
(or thermalization) time for the partonic system. 
Though this assumption is not well founded, this is the best
we can do before a complete calculation of the relaxation
time due to parton rescatterings and radiations is available.
After this time the momentum distribution of partons
is assumed to be  locally isotropic  and approximately exponential.
What we are interested next is how this system approaches
chemical equilibrium.

Let us first estimate the initial conditions for the equilibration
from the previous chapters. 
Since we are here primarily interested in the chemical equilibration 
of the parton gas, we shall assume that the parton distributions 
can be approximated by thermal phase space distributions with 
non-equilibrium fugacities $\lambda_i$:
\begin{equation}
f(k;T,\lambda_i)  = \lambda_i\left( e^{\beta u\cdot k} \pm
\lambda_i\right)^{-1}, \label{eq:eq1}
\end{equation}
where $\beta$ is the inverse temperature and $u^{\mu}$ is the 
four-velocity of the local comoving reference frame.  When the 
parton fugacities $\lambda_i$ are much less than unity as 
may happen during the early evolution of the parton system,
we can neglect the quantum statistics in Eq.~(\ref{eq:eq1}) and 
write the momentum distributions in the factorized form,
\begin{equation}
\label{19}
f(k;T,\lambda _i)=\lambda _i\left (e^{\beta u\cdot k}\pm 1\right)^{-1}.
\end{equation}
Using this form for the distributions, one has the parton and energy
densities,
\begin{equation}
n = (\lambda_g a_1 +\lambda_q b_1)T^3, \quad \varepsilon 
= (\lambda_g a_2 +\lambda_q b_2) T^4. \label{eq:eq2}
\end{equation}
where $a_1=16\zeta (3)/\pi^2\approx 1.95$, $a_2=8\pi^2/15\approx 5.26$,
for a Bose distribution, $b_1=9\zeta (3)N_f/\pi^2\approx 2.20$ and
$b_2=7\pi^2N_f/20 \approx 6.9$ for a Dirac distribution for a baryon
symmetric system, $\lambda_q=\lambda_{\bar q}$. Since we 
have demonstrated that boost invariance is a good approximation for 
the initially produced partons, we can then estimate the initial
parton fugacities, $\lambda_{g,q}^0$ and temperature $T_0$ 
from [cf. Eq.~(\ref{eq:den1})]
\begin{equation}
n_0 = \frac{1}{\pi R^2_{A} \tau_{\rm iso}} \frac{dN}{dy}\; ,
\quad \varepsilon_0 = n_0 \frac{4}{\pi}\langle k_T\rangle, \label{eq:eq3}
\end{equation}
where $\langle k_T\rangle$ is the average transverse momentum.
The quark fugacity is taken as $\lambda_q^0 = 0.16 \lambda_g^0$, 
corresponding to a ratio 0.14 of the initial quark(antiquark) 
number to the total number of partons.  Table \ref{table1} 
shows these relevant quantities at the moment $\tau_{\rm iso}$,
for Au + Au collisions at RHIC and LHC energies. One can observe
that the initial parton gas is rather hot, reflecting the large
average transverse momentum. However, the parton gas is very far 
from saturation in phase space as compared to the ideal gas 
at the same temperature. The gas is also far away from chemical 
equilibrium with respect to parton species (gluons and quarks with
different flavors) since it is dominated by gluons. 
I should emphasize that the initial 
conditions listed here result from HIJING 
calculation of parton production through semihard scatterings.
Soft partons, {\em e.g.}, due to parton production from the color
field \cite{KEMG}, are not included.

\begin{table}
\begin{center}
\begin{tabular}{lrr} \hline\hline
&\makebox[1in]{$\;\;\;$RHIC} &\makebox[1in]{$\;\;$LHC} \hfill \\ \hline
$\tau_{\rm iso}$ (fm/$c$) &0.7\phantom{000}   &0.5\phantom{000} \hfill \\
$\varepsilon_0$ (GeV/fm$^3$)  &3.2\phantom{000}   &40\phantom{.0000} \hfill \\
$n_0$ (fm$^{-3}$)             &2.15\phantom{00}   &18\phantom{.0000}  \\
$\langle k_T\rangle$ (GeV)    &1.17\phantom{00}  &1.76\phantom{00} \hfill\\
$T_0$ (GeV)                   &0.57\phantom{00}  &0.83\phantom{00} \hfill\\
$\lambda_g^0$                 & 0.05\phantom{00} &0.124\phantom{0} \hfill\\ 
$\lambda_q^0$                 &0.008\phantom{0} &0.02\phantom{00}\hfill \\ 
\hline\hline
\end{tabular}

\caption{Values of the relevant parameters characterizing the parton
plasma at the moment $\tau_{\rm iso}$, when local isotropy 
of the momentum distribution is first reached.}
\label{table1}
\end{center}
\end{table}

\subsection{Master Rate Equations}

In general, chemical reactions among partons can be quite complicated
because of the possibility of initial and final-state gluon radiations.
As we have discussed earlier, interference effects due to multiple
scatterings inside a dense medium lead to a strong suppression
of soft gluon radiation. One lesson we learned from the LPM effect
is that the radiation between two successive scatterings is the 
sum, {\em on the amplitude level}, of both the initial state 
radiation from the first scattering and the final state radiation 
from the second one. Since the off-shell parton is space-like
in the first amplitude and time-like in the second, the picture
of a time-like parton propagating inside a medium in the parton 
cascades simulations \cite{PCM} breaks down. Instead,
we shall here only consider processes where a single additional 
gluon is radiated, such as $gg\to ggg$ between two scatterings,
in which we can include the LPM effect by a radiation suppression
factor. The analysis of QCD LPM effect in Refs.~\cite{GWLPM1,GWLPM2}
has been done for a fast parton traveling inside a parton
plasma. We will use the results here for radiation off thermal 
partons whose average energy is about $T$, since we expect
the same physics to happen.

In order to permit the approach to chemical equilibrium, the reverse 
process, {\em i.e.}, gluon absorption, has to be included as well. This is 
easily achieved by making  use of detailed balance.  Closer inspection 
shows that gluon radiation is dominated by the process $gg\to ggg$, 
because radiative processes involving quarks have substantially smaller 
cross sections in pQCD, and quarks are considerably less 
abundant than gluons in the initial phase of the chemical evolution of 
the parton gas.  Here we are interested in understanding the basic 
mechanisms underlying the formation of a chemically equilibrated 
quark-gluon plasma, and the essential time-scales.  We hence restrict 
our considerations to the dominant reaction mechanisms for the 
equilibration of each parton flavor.  These are just four 
processes \cite{MSM86}:
\begin{equation}
gg \leftrightarrow ggg, \quad gg\leftrightarrow
q\overline{q}.\label{eq:eq4}
\end{equation}
Other scattering processes ensure the maintenance of thermal
equilibrium $(gg\leftrightarrow gg, \; gq \leftrightarrow gq$, {\it etc.})
or yield corrections to the dominant reaction rates 
$(gq\leftrightarrow qgg$, {\it etc.}).

Restricting ourselves to the reactions Eq.~(\ref{eq:eq4}), 
the evolution of the parton densities is governed by the 
master equations:
\begin{eqnarray}
\partial_{\mu}(n_gu^{\mu}) &= &
 \frac{1}{2}\sigma_3 n_g^2 \left( 1-\frac{n_g}{\tilde n_g}\right)
 -\frac{1}{2}\sigma_2 n_g^2 \left( 1 - \frac{n_q^2 \tilde n_g^2}
 {\tilde n_q^2 n_g^2}\right), \label{eq:eq5}\\
 \partial_{\mu} (n_qu^{\mu}) &= &\partial_{\mu} (n_{\bar{q}} u^{\mu})
 \frac{1}{2}\sigma_2 n_g^2 \left( 1 - \frac{n_q^2 \tilde n_g^2}
 {\tilde n_q^2 n_g^2}\right), \label{eq:eq6}
\end{eqnarray}
where ${\tilde n_i}\equiv n_i/\lambda_i$ denote the densities 
with unit fugacities, $\lambda_i=1$, $\sigma_3$ and $\sigma_2$ 
are thermally averaged, velocity weighted cross sections,
\begin{equation}
\sigma_3 = \langle\sigma(gg\to ggg)v\rangle, \quad \sigma_2 =
\langle \sigma (gg\to q\bar q)v\rangle. \label{eq:eq7}
\end{equation}
We have also assumed detailed balance and a baryon symmetric
matter, $n_q=n_{\bar q}$. If we assume that parton scatterings
are sufficiently rapid to maintain local thermal equilibrium,
and therefore we can neglect effects of viscosity 
due to elastic \cite{KEMG,VISC} and inelastic \cite{WONG} scatterings, 
we then have the hydrodynamic equation

\begin{equation}
\partial_{\mu} (\varepsilon u^{\mu}) + P\;\partial_{\mu} u^{\mu} = 0,
\label{eq:eq8}
\end{equation}
which determines the evolution of the energy density.
For a more complete study, one can include the viscosity
corrections to the energy-momentum tensor and modify the above
hydrodynamic equation. This is beyond the scope of this review.

For a time scale $\tau\ll R_A$, we can neglect the transverse
expansion and consider a purely longitudinal expansion of 
the parton plasma, which leads to  Bjorken's scaling 
solution \cite{BJOR} of the hydrodynamic equation:
\begin{equation}
{d\varepsilon\over d\tau} + {\varepsilon+P\over\tau} = 0. \label{eq:eq9}
\end{equation}

We further assume an ultrarelativistic equation of state, 
$\varepsilon=3 P$, with $n_i$ and $\varepsilon$ given by Eq.~(\ref{eq:eq2}).
We can then solve the hydrodynamic equation,
\begin{equation}
  [\lambda_g + \frac{b_2}{a_2}\lambda_q]^{3/4} T^3\tau = \hbox{const.} \;\; , 
  \label{eq:eq10}
\end{equation}
and rewrite the rate equation in terms of the time evolution of the
parameters $T(\tau)$, $\lambda_g(\tau)$ and $\lambda_q(\tau)$,
\begin{eqnarray}
\frac{\dot\lambda_g}{\lambda_g} + 3\frac{\dot T}{T} + \frac{1}{\tau} &=
&R_3 (1-\lambda_g)-2R_2 \left(1- \frac{\lambda_q^2}{\lambda_g^2} \right)
        \label{eq:eq11} \\
\frac{\dot\lambda_q}{\lambda_q} + 3\frac{\dot T}{T} + \frac{1}{\tau} &=
&R_2 {a_1\over b_1} \left( \frac{\lambda_g}{\lambda_q} -
\frac{\lambda_q}{\lambda_g}\right), \label{eq:eq12}
\end{eqnarray}
where the density weighted reaction rates $R_3$ and $R_2$ are defined as
\begin{equation}
R_3 = \textstyle{{1\over 2}} \sigma_3 n_g, \quad
R_2 = \textstyle{{1\over 2}} \sigma_2 n_g.  \label{eq:eq13}
\end{equation}
Notice that for a fully equilibrated system ($\lambda_g=\lambda_q=1$),
Eq. (\ref{eq:eq10}) corresponds to the Bjorken solution,
$T(\tau)/T_0=(\tau_0/\tau)^{1/3}$.

\subsection{Parton Equilibration Rates}

Often referred to as Landau-Pomeranchuk-Migdal (LPM) effect,
the interference in radiation induced by multiple scatterings
was first investigated in QED for induced photon radiation off a 
charged particle propagating through a dense medium \cite{LPM,SHUL}. 
It suppresses those soft photons whose formation time is much larger 
than the mean free path of the fast particle inside the medium. 
In QCD, however, the interference pattern is different 
from the QED case, since multiple scatterings through gluon 
exchanges will force a quark to change its color along its path. 
In recent studies \cite{GWLPM1,GWLPM2,DOKS}, pQCD analyses indeed 
indicated that the LPM effect suppresses soft gluon radiation by multiple 
scatterings. Due to color exchanges, the interference has 
some non-Abelian behavior which actually depends on the 
color representation of the jet parton. The same mechanism
also causes the color conductivity in a quark-gluon 
plasma to be small \cite{ASMG,HEIS}.

To account for the LPM effect in the calculation of
the reaction rate $R_3$ for $gg\rightarrow ggg$, we will use
the triple differential cross section from 
Refs.~\cite{GWLPM1,GWLPM2} and apply the LPM suppression to
the radiation of gluons whose effective formation time $\tau_{\rm QCD}$
are much longer than the mean free path $\lambda_f$ of multiple 
scatterings. At the same time, the LPM effect also regularizes
the infrared divergency associated with QCD radiation amplitude. 
However, $\sigma_3$ still contains infrared singularities in the gluon
propagators associated with QCD scatterings. 
For an equilibrium system one can in principle apply
the resummation technique developed by Braaten and Pisarski \cite{BP90}
to regularize the electric part of the propagators, though the
magnetic sector still has to be regulated by an unknown magnetic
screening mass which can only be calculated nonperturbatively \cite{TBBM93}
up to now. Since we are dealing with a nonequilibrium system,
Braaten and Pisarski's resummation is not well defined. As an
approximation, we will use \cite{BMW92} the Debye screening mass,
\begin{equation}
\mu_D^2 = {6g^2\over \pi^2} \int_0^{\infty} kf(k) dk
=4\pi\alpha_s T^2\lambda_g, \label{eq:eq14}
\end{equation}
to regularize all singularities in both the scattering cross
sections and the radiation amplitude.

\begin{figure}
\centerline{\psfig{figure=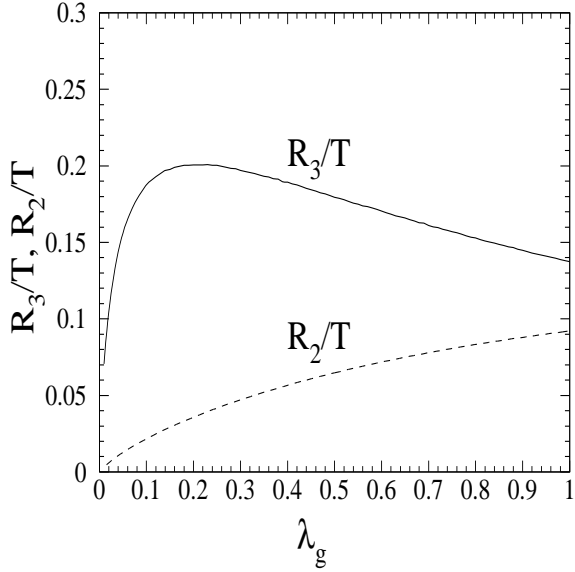,width=3in,height=3in}}
\caption{The scaled gluon production rate $R_3/T$ (solid line) for 
         $gg\rightarrow ggg$ and the quark production rate 
         $R_2/T$ (dashed line) for $gg\rightarrow q\bar{q}$
         are shown as function of the gluon fugacity $\lambda_g$ for 
         $\alpha_s$ = 0.3.}
\label{fig26}
\end{figure}

To further simplify the calculation we approximate the
LPM suppression factor \cite{GWLPM2} by a $\theta$-function,
$\theta(\lambda_f-\tau_{\rm QCD})$. The modified
differential cross section for $gg\rightarrow ggg$ is then
\begin{equation}
  \frac{d\sigma_3}{dq_{\perp}^2 dy d^2k_{\perp}}
  =\frac{d\sigma_{\rm el}^{gg}}{dq_{\perp}^2}\frac{dn_g}{dy d^2k_{\perp}}
  \theta(\lambda_f-\tau_{\rm QCD})\theta(\sqrt{s}-k_{\perp}\cosh y),
\end{equation}
where 
\begin{equation}
\tau_{\rm QCD}=\frac{C_A}{2C_2}\frac{2\cosh y}{k_{\perp}} \label{eq:ftime}
\end{equation}
is the effective formation time \cite{GWLPM1,GWLPM2}. The second
step-function accounts for energy conservation, and 
$s=18T^2$ is the average squared center-of-mass energy of the two 
gluons in the thermal gas. The regularized gluon density distribution
induced by a single scattering is,
\begin{equation}
  \frac{dn_g}{dy d^2k_{\perp}} =\frac{C_A\alpha_s}{\pi^2}
  \frac{q_{\perp}^2}{k_{\perp}^2[({\bf k}_{\perp}
    -{\bf q}_{\perp})^2 +\mu_D^2]}. \label{eq:dng}
\end{equation}
Similarly, the regularized small angle $gg$ scattering cross section is
\begin{equation}
  \frac{d\sigma_{\rm el}^{gg}}{dq_{\perp}^2}
  =\frac{9}{4}\frac{2\pi\alpha_s^2}{(q_{\perp}^2+\mu_D^2)^2}.
\end{equation}
The mean free path for elastic scatterings is then
\begin{equation}
  \lambda_f^{-1}\equiv n_g\int_0^{s/4}dq_{\perp}^2
  \frac{d\sigma_{\rm el}^{gg}}{dq_{\perp}^2}
  =\frac{9}{8}a_1\alpha_s T\frac{1}{1+8\pi\alpha_s\lambda_g/9},
\end{equation}
which depends very weakly on the gluon fugacity $\lambda_g$,
where again $a_1=16\zeta (3)/\pi^2\approx 1.95$. Using
\begin{equation}
  \int_0^{2\pi}d\phi \frac{1}{({\bf k}_{\perp}-{\bf q}_{\perp})^2+\mu_D^2}
    =\frac{2\pi}{\sqrt{(k_{\perp}^2+q_{\perp}^2+\mu_D^2)^2
        -4q_{\perp}^2k_{\perp}^2}},
\end{equation}
we can complete part of the integrations and have
\begin{equation}
  R_3/T=\frac{32}{3a_1}\alpha_s\lambda_g(1+8\pi\alpha_s\lambda_g/9)^2
  {\cal I}(\lambda_g).
\end{equation}
Here, ${\cal I}(\lambda_g)$ is a function of $\lambda_g$:
\begin{eqnarray}
{\cal I}(\lambda_g)=\int_1^{\sqrt{s}\lambda_f}dx
\int_0^{s/4\mu_D^2}&dz& \frac{z}{(1+z)^2}
  \left\{ {\cosh^{-1}(\sqrt{x}) \over
  x\sqrt{[x+(1+z)x_D]^2-4x\;z\;x_D}}\right. \nonumber \\
  &+& \left. \frac{1}{s\lambda_f^2}{\cosh^{-1}(\sqrt{x}) \over
  \sqrt{[1+x(1+z)y_D]^2-4x\;z\;y_D}}\right\},
\end{eqnarray}
where $x_D=\mu_D^2\lambda_f^2$ and $y_D=\mu_D^2/s$.
We can evaluate the integration numerically and find out
the dependence of $R_3/T$ on the gluon fugacity $\lambda_g$.
In Fig.~\ref{fig26}, $R_3/T$ is plotted versus $\lambda_g$
for a coupling constant $\alpha_s=0.3$. The gluon production
rate increases with $\lambda_g$ and then saturates when
the system is in equilibrium.

The calculation of the quark equilibration rate $R_2$ for
$gg\rightarrow q\bar{q}$ is more straightforward. But one
also has to use effective quark and gluon propagators to
regularize the singularities for massless quark 
production \cite{TADS93}. We can also estimate the
quark equilibration rate by using the effective thermal
quark mass \cite{KM82},
\begin{equation}
M_q^2  = \left(\lambda_g + {1\over 2}\lambda_q\right)
{4\pi\over 9} \alpha_s T^2,  \label{eq:mq}
\end{equation}
as a cutoff in the divergent integral over momentum
transfer. After averaging over the thermal gluon distribution,
we have \cite{BDMTW}
\begin{equation}
R_2 = {1\over 2}\sigma_2 n_g \approx 0.24 N_f \;\alpha_s^2 \lambda_g T
\ln (5.5/\lambda_g). \label{56}
\end{equation}
The dashed line in Fig.~\ref{fig26} shows the normalized 
rate $R_2/T$ for $N_f=2.5$, which takes into account the reduced 
phase space of strange quarks at moderate temperatures, as 
a function of the gluon fugacity.

\subsection{Evolution of the Parton Plasma}

\begin{figure}
\centerline{\psfig{figure=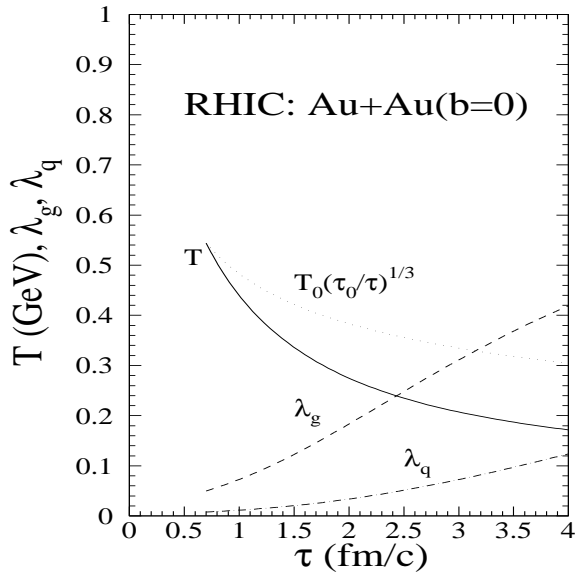,width=3in,height=3in}}
\caption{Time evolution of the temperature $T$ and the
         fugacities $\lambda_g$ and $\lambda_q$ of gluons and quarks in the
         parton plasma created in Au + Au collisions at the RHIC energy
         of $\protect\sqrt{s}=200$ AGeV. The initial values for 
         $T,\; \lambda_g$ and $\lambda_q$ are determined from 
         HIJING simulations and are listed in Table~\protect\ref{table1}.}
\label{fig27}
\end{figure}

\begin{figure}
\centerline{\psfig{figure=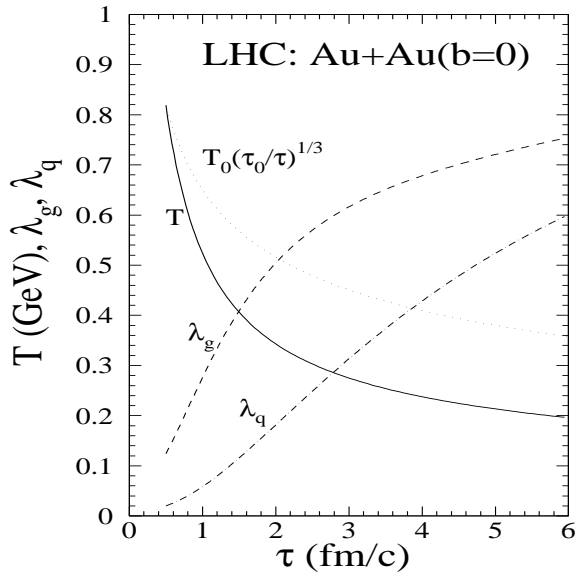,width=3in,height=3in}}
\caption{The same as in Fig.~\protect\ref{fig27}, except for LHC energy,
         $\protect\sqrt{s}=5.5 $ ATeV.}
\label{fig28}
\end{figure}

With the parton equilibration rates which in turn depend
on the parton fugacity, we can solve the master equations 
self-consistently and obtain the time evolution of the temperature
and the fugacities. Shown in Figs.~\ref{fig27} and \ref{fig28}
are the time dependence of $T$, $\lambda_g$, and $\lambda_q$
for the initial conditions listed in Table~\ref{table1} at RHIC and
LHC energies. We find that the parton gas cools considerably 
faster than predicted by Bjorken's scaling solution 
$(T^3\tau$ = const.) shown by the dotted lines, because 
the production of additional partons 
during the approach to chemical equilibrium consumes an appreciable 
amount of energy. The increased cooling, in turn, impedes the 
chemical equilibration process; this is more apparent at the RHIC 
than at the LHC energies. Therefore, the parton system can hardly
reach its equilibrium state before the effective temperature
drops below $T_c \approx 200$ MeV in the short period of time of
1-2 fm/$c$ at the RHIC energy. At the LHC energy, however,
the parton gas, especially the gluon component, becomes very close to 
equilibrium (in terms of phase space occupation)
and the plasma may exist in a deconfined phase for as long as 4-5 fm/$c$.
Another important observation is that quarks never
reach chemical equilibrium at both energies.
This is partly due to the small initial quark fugacity
and  partly due to the small quark equilibration rate.

I should emphasize that the initial conditions used here result
from the HIJING model calculation in which only initial direct parton
scatterings are taken into account.  Due to the fact that HIJING is
a QCD motivated phenomenological model, there are some uncertainties 
related to the initial parton production.

(1) As I have discussed in the second section, the total jet cross 
section and initial parton production depend on the $p_T$ cutoff $p_0$.
We only included those partons which are produced through a hard
or semihard scattering with transverse momentum transfer $p_T$ larger
than $p_0$. Soft partons through soft interactions with $p_T<p_0$
are locked up in the phenomenological strings which will fragment
into hadrons later. How one should include these soft partons
in the parton equilibration processes and what value one should
choose for $p_0$ cannot be answered within the model. The initial
parton density is especially sensitive to the value of $p_0$.

(2) The dominant initial gluon production is also sensitive to
the gluon distribution at small $x$. But so far there is no 
precise measurement of the gluon distribution in the small $x$ region 
where most of the minijets originate at RHIC and especially at the LHC
energies. The constraints
from recent HERA data \cite{HERA} suggest a more singular behavior of 
parton distributions than the original Duke-Owen parametrization which 
we have used here.  A study in Ref.~\cite{EKV94} shows that these
more singular gluon distributions can give rise to a much larger amount
of produced partons at the LHC energy. 
Furthermore, we have no handle experimentally so 
far on the nuclear  shadowing of the gluon structure function. In 
HIJING we have assumed the same shadowing for gluons and quarks, 
which has reduced the parton production by about half inside a heavy 
nucleus.

(3) Secondary elastic parton scatterings before $\tau_{\rm iso}$
have not been considered here, which could speed up the thermalization
process. For example, the parton cascade model \cite{PCM} claims
a shorter thermalization time than what we have used here. Changes 
in the initial thermalization time, therefore, will also lead to
uncertainties in the initial energy and parton densities. 

\begin{figure}
\centerline{\psfig{figure=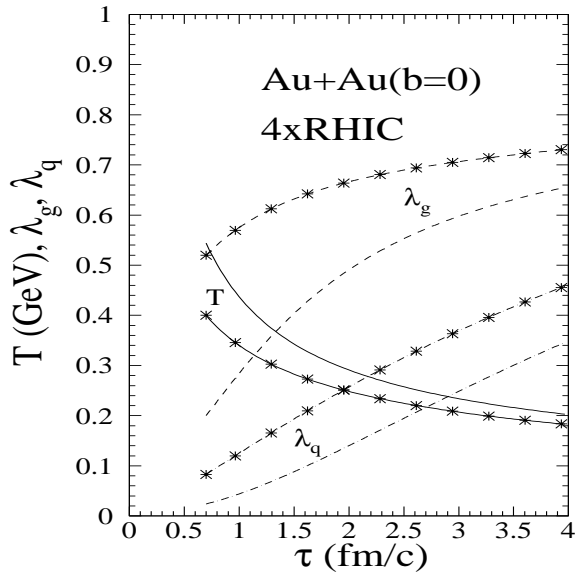,width=3in,height=3in}}
\caption{The same as in Fig.~\protect\ref{fig27}, except that the initial
         parton densities are 4 times higher than given in 
         Table~\protect\ref{table1} with the same (ordinary lines),
         or reduced initial temperature, $T_0=0.4$ GeV (lines with stars)}
\label{fig29}
\end{figure}

We can estimate the effect of the uncertainties in the 
initial conditions on the parton gas evolution by multiplying
the initial energy and parton number densities at RHIC energy
by a factor of 4. This will result in the initial fugacities,
$\lambda_g^0=0.2$ and $\lambda_q^0=0.024$. With these high
initial densities, the parton plasma can evolve into a
nearly equilibrated gluon gas, as shown in Fig.~\ref{fig29}.
The deconfined phase will also last longer, for about 4 fm/$c$.
However, the system is still dominated by gluons, and thus has fewer quarks
and antiquarks than expected in a fully chemical equilibrated
system. If the uncertainties in the initial conditions
are caused by the soft parton production from the color
mean fields, the initial effective temperature will decrease
because soft partons are generally less energetic.
Therefore, we can alternatively increase the initial 
parton density by a factor of 4 and decrease $T_0$ to
0.4 GeV at the same time. This leads to higher initial fugacities,
$\lambda_g^0=0.52$ and $\lambda_q^0=0.083$. As shown in
Fig.~\ref{fig29} by the curves with stars, this system evolves
faster toward equilibrium, however, with shorter life-time
in the deconfined phase due to the reduced initial
temperature.
We thus can conclude that perturbative parton
production and scatterings are very likely to produce
a quark-gluon plasma (or gluon plasma to be more accurate)
in ultrarelativistic heavy-ion collisions at LHC energy.
However, the fate of the quark-gluon plasma at RHIC energy
has to be determined by a more careful examination of the
uncertainties in the initial conditions.

\subsection{Effects of $gg\rightarrow (n-2)g$ processes}

So far we have only considered $gg\rightarrow ggg$ process in
parton equilibration. In fact, higher order processes like
$gg \rightarrow (n-2)g$ for $n>5$ should also contribute,
where $n$ is the total number of gluons involved in the processes.
Even though these processes are higher order in $\alpha_s$,
the infrared behavior of QCD radiation gives rise to
a logarithmic factor $\ln Q^2$ after integration which
can offset the logarithmically decreasing coupling constant
and make the contributions relatively important for all orders 
in $\alpha_s(Q)$. This is especially true for interactions with
large momentum transfer $Q$.

The complete amplitude involving $n$-gluon processes in general
has contributions from different helicity transitions. 
Parke and Taylor \cite{PT86} have derived the exact form
of the maximum helicity violation amplitude for the
$n$-gluon processes,
\begin{equation}
  |{\cal M}_n^{\rm PT}|^2 = \frac{(N_c\alpha_s/2\pi)^{n-2}}{N_c^2-1}
  \sum_{i>j}s_{ij}^4\sum_{\rm P}\frac{1}{s_{12}s_{23}\cdots s_{n1}}, 
\label{eq:ng1}
\end{equation}
where $s_{ij}=(p_i+p_j)^2$ and the summation P is over the $(n-2)!/2$
non-cyclic permutation of $(1\cdots n)$.  There is no known exact 
expression for the other chiral amplitudes. Kunszt and Stirling \cite{KS88}
proposed to add  a factor to the Parke-Taylor amplitude,
\begin{equation}
  |{\cal M}_n^{\rm KS}|^2= C_{\rm KS}(n)|{\cal M}_n^{\rm PT}|^2, \;\;
  {\rm with} \;\; C_{\rm KS}(n)=\frac{2^n-2(n+1)}{n(n-1)}, \label{eq:ng2}
\end{equation}
to get the total $n$-gluon amplitude. This procedure can be checked
to recover the analytic results for $n=4$ and 5 \cite{BKCG81}.

Using the above tree-level $n$-gluon amplitude, Xiong and Shuryak \cite{XS93}
have calculated the $n$-gluon reaction rate,
\begin{eqnarray}
R_{gg\rightarrow (n-2)g}& =&\frac{1}{2! \, (n-2)!} \int 
\frac{d^3k_1}{(2\pi)^3\,2\omega_1}f(k_1)
\frac{d^3k_2}{(2\pi)^3\,2\omega_2}f(k_2) \nonumber \\
& & |{\cal M}_n^{\rm KS}|^2 (2\pi)^4\delta^4(k_1+k_2-\sum_{i=3}^n k_i)
\prod_{i=3}^n [1+f(k_i)]\frac{d^3k_i}{(2\pi)^3\,2\omega_i} .
\end{eqnarray}
In order to regulate the infrared and collinear divergencies
they impose a cut-off,
\begin{equation}
  s_{ij} \geq s_0=\eta^2 T^2,
\end{equation}
in the integration over the phase space. If such a cut-off is
provided by the Debye screening mass, then $\eta^2 =\alpha_s/4\pi$.
Neglecting the temperature dependence of the coupling constant,
the rates scale with temperature as
\begin{equation}
  R_{gg\rightarrow (n-2)g}= \alpha_n(\eta) T^4,
\end{equation}
where the dimensionless coefficients $\alpha_n(\eta)$'s
depend only on the screening parameter $\eta$.
Xiong and Shuryak found that for $\eta=1.1$, $\alpha_n(\eta)$
decreases with $n$. They also found that the total
contribution to the reaction rate from $n\geq 6$ processes is 
\begin{equation}
  \sum_{n=6}^8 \alpha_n(\eta=1.1)=2.56,
\end{equation}
which is comparable in magnitude to $\alpha_5(\eta=1.1)=2.32$ 
for the  $gg\rightarrow ggg$ processes. Therefore, they found that 
$n$-gluon processes contribute significantly to gluon equilibration.

Two remarks can be made here regarding $n$-gluon processes.
As I have discussed earlier, the Landau-Pomeranchuk-Midgal
effect is very important to regulate the infrared behavior
of parton  bremsstrahlung and to suppress soft parton
radiation induced by multiple scatterings. It thus
reduces the reaction rate of $gg\rightarrow ggg$.
One would expect that this effect should also be important
in multiple gluon production processes. Unfortunately,
the analysis of the LPM effect has not been done so far for multiple
particle radiation. In order to do so, one has to understand
the space-time structure of the $n$-gluon processes.

Secondly, the Parke-Taylor amplitude is only the tree-level
amplitude for $n$-gluon processes. A complete calculation
should also include all loop corrections to the same order
in $\alpha_s$. Take the soft radiation limit for an
illustration. In this case the Parke-Taylor amplitude can
be factorized. Assume the $n$th soft gluon is radiated with 
a small angle (corresponding to a transverse momentum $p_T$) 
along an energetic one and a fraction (z) of its momentum.  
The cross section for the $n$-gluon process can then be 
expressed in terms of that of $(n-1)$-gluon,
\begin{equation}
  \sigma_n\approx \int \sigma_{n-1} \frac{\alpha_s}{2\pi} \frac{2N_c}{z}
 dz  \frac{dp_T^2}{p_T^2}.
\end{equation} 
With the gluon splitting function, $P_{g \rightarrow gg}(z) \approx 2N_c/z$
for $z\ll 1$,  this is much like the formula for successive 
initial and final state  radiation we have encountered earlier 
in the Monte Carlo simulation  [cf. Eq.~(\ref{eq:shr1})] except that
the Sudakov form factors are missing, which account for the virtual 
corrections. One can thus include virtual corrections in this
soft limit via multiplying the cross section by Sudakov form factors 
or introducing the `$+$ function' and $\delta$-function in the
splitting function. The virtual corrections for the
complete amplitude are very complicated and so far no one has
attempted to calculate them. However, once included, they will
make the total cross section unitary and likely reduce
the gluon multiplication  rate calculated from tree level
amplitudes.

\section{Hard Probes of the Equilibrating Parton Plasma}

So far I have discussed parton production and quark-gluon
plasma formation in the framework of a pQCD-based
model. In this framework, hard or semihard scatterings
among partons play an essential role in liberating partons
from the individual confining nucleons and driving the initially 
produced parton system toward equilibrium. The same kind of hard
processes can also be used as direct probes of the early
parton dynamics and the evolution of the quark-gluon plasma.
Among these hard probes, electromagnetic signals, like
direct photons and dileptons, are considered
more direct since they can escape the dense matter without further
interaction. They can thus reveal the dynamics of initial
parton production and evolution. Similarly, open charm
production, jet quenching due to energy loss and $J/\psi$
suppression can all provide us information about 
parton scattering and thermalization inside a parton
plasma. In particular, preequilibrium $J/\psi$ suppression
can reveal evidence of the deconfinement of the parton gas
since the $J/\psi$ dissociation cross section with
a deconfined parton gas is very different from that
with a hadronic gas. For a recent review on signals
of quark-gluon plasma, see Ref.~\cite{BM94} by M\"uller. 
I will here only discuss hard probes in the framework
of an equilibrating parton plasma.

\subsection{Open charm production}

Unlike strange quarks, charm quarks cannot be easily produced
during the mixed and hadronic phases of the dense matter
since the charm mass is much larger than the corresponding 
temperature scale. The only period when charm quarks can
be easily produced is during the early stage of the parton 
evolution when the effective temperature is still high. At this
stage, the parton gas is still not fully equilibrated yet so that
the temperature is only an effective parameter describing
the average momentum scale. By measuring this pre-equilibrium
charm production, one can thus probe the initial parton
density in phase space and shed light on the equilibration
time \cite{BMXW92}.

There are three different contributions to charm production in the
history of the evolution of the parton system: (1) initial production 
similar to minijets; (2) pre-thermal production from secondary 
parton scatterings during the thermalization, $\tau<\tau_{\rm iso}$;
(3) and thermal production during the parton equilibration,
$\tau>\tau_{\rm iso}$.  I will discuss them
in the reversed order, starting with the charm production during
the final stage of parton equilibration.

\subsubsection{Thermal production during equilibration}

Similar to light quark equilibration, charm quarks are 
produced through gluon fusion $gg\rightarrow c\bar{c}$ and 
quark antiquark annihilation $q\bar{q}\rightarrow c\bar{c}$ 
during the evolution of the parton plasma. However, since 
the number of charm quarks is very small as compared 
to gluons and light quarks, we can neglect the back reactions, 
$c\bar{c}\rightarrow gg,\ q\bar{q}$ and their effect on the parton 
evolution. Given the phase-space density of the equilibrating
partons, $f_i(k)$, the differential production rate is then \cite{LMW94},
\begin{eqnarray}
  E\frac{d^3A}{d^3p}&=&\frac{1}{16(2\pi)^8}\int \frac{d^3k_1}{\omega_1}
  \frac{d^3k_2}{\omega_2}\frac{d^3p_2}{E_2}\delta^{(4)}(k_1+k_2-p-p_2)
  \nonumber \\
  & & \left[\frac{1}{2}g_G^2f_g(k_1)f_g(k_2)
  |\overline{\cal M}_{gg\rightarrow c\bar{c}}|^2
  +g_q^2 N_f f_q(k_1)f_{\bar{q}}(k_2)
  |\overline{\cal M}_{q\bar{q}\rightarrow c\bar{c}}|^2\right]
  , \label{eq:therm1}
\end{eqnarray}
where $g_G=2(N_c^2-1)$, $g_q=2N_c$, are the degeneracy factors for gluons and
quarks (antiquarks) respectively, 
$|\overline{\cal M}_{gg\rightarrow c\bar{c}}|^2$,
$|\overline{\cal M}_{q\bar{q}\rightarrow c\bar{c}}|^2$ are
the {\em averaged} matrix elements for $gg\rightarrow c\bar{c}$
and $q\bar{q}\rightarrow c\bar{c}$ processes, respectively, 
\begin{eqnarray}
  \frac{|\overline{\cal M}_{gg\rightarrow c\bar{c}}|^2}{\pi^2\alpha_s^2} &=&
  \frac{12}{\hat{s}^2}(M^2-\hat{t})(M^2-\hat{u})+\frac{8}{3}
  \left(\frac{M^2-\hat{u}}{M^2-\hat{t}}
  +\frac{M^2-\hat{t}}{M^2-\hat{u}}\right) \nonumber \\
  &-&\frac{16M^2}{3} \left[ \frac{M^2+\hat{t}}{(M^2-\hat{t})^2}
+\frac{M^2+\hat{u}}{(M^2-\hat{u})^2} \right]
  -\frac{6}{\hat{s}}(2M^2-\hat{t}-\hat{u}) \nonumber \\
  &+&\frac{6}{\hat{s}}\frac{M^2(\hat{t}-\hat{u})^2}
  {(M^2-\hat{t})(M^2-\hat{u})}
  -\frac{2}{3}\frac{M^2(\hat{s}-4M^2)}{(M^2-\hat{t})(M^2-\hat{u})},\\
\frac{|\overline{\cal M}_{q\bar{q}\rightarrow c\bar{c}}|^2}{\pi^2\alpha_s^2}
&=&\frac{64}{9\hat{s}^2}
  \left[(M^2-\hat{t})^2+(M^2-\hat{u})^2+2M^2\hat{s}\right],
\end{eqnarray}
Due to the small charm density, we can neglect the Pauli blocking 
of the final charm quarks. For large charm quark mass, $M$, we
can approximate the phase-space density $f_i(k)$ by a Boltzmann
distribution. We further assume that the distributions are
boost invariant, {\em i.e.},
\begin{equation}
  f_i(k)\approx e^{-k_T\cosh(y-\eta)},
\end{equation}
where $\eta=0.5\ln(t+z)/(t-z)$ is the spatial rapidity of a 
space-time cells at $(t,z)$. Neglecting the transverse
expansion, the above assumption implies that the
space-time cell at $(t,z)$ have a flow velocity, $u=(\cosh\eta,\sinh\eta)$.
We can now complete the integral over $\eta$ in
$\int d^4x=\pi R_A^2\int d\eta d\tau$ and obtain,
\begin{eqnarray}
  \frac{dN_{\rm th}}{dyd^2p_T}&=&\frac{\pi R_A^2}{16(2\pi)^8}
\int_{\tau_{\rm iso}}^{\tau_c}\tau d\tau \int p_{T2}dp_{T2} d\phi_2 dy_2
d\phi_{k1} dy_{k1} \frac{2k_{T1}^2}{\hat{s}}2K_0(Q_T/T) \nonumber \\
& &\left[\frac{1}{2}g_G^2\lambda_g^2
|\overline{\cal M}_{gg\rightarrow c\bar{c}}|^2 +g_q^2 N_f \lambda_q^2|
\overline{\cal M}_{q\bar{q}\rightarrow c\bar{c}}|^2\right],
\label{eq:therm2}
\end{eqnarray}
where $K_0$ is the modified Bassel function and $\tau_c$ is the 
time when the temperature, $T$, drops below 200 MeV. 
The kinematic variables are chosen such that,
\begin{eqnarray}
p_2&=&(M_{T2}\cosh y_2,p_{T2}\cos\phi_2,p_{T2}\sin\phi_2, M_{T2}\sinh y_2),
\;\; M_{T2}=\sqrt{M^2+p_{T2}^2}; \nonumber \\
k_i&=&k_{Ti}(\cosh y_{ki},\cos\phi_{ki},\sin\phi_{ki},\sinh y_{ki}),
 \ \ i=1,2 \ .
\end{eqnarray}
The center-of-mass momentum,$Q=(Q_T\cosh y_Q,{\bf q_T},Q_T\sinh y_Q)$, 
is defined as $Q=p+p2=k1+k2$, and
\begin{eqnarray}
  Q^2&=&\hat{s}=2[M^2+M_{T}M_{T2}\cosh(y-y_2)-p_Tp_{T2}\cos\phi_2],\nonumber\\
  q_T^2&=&p_T^2+p_{T2}^2+2p_Tp_{T2}\cos\phi_2,\nonumber \\
  Q_T^2&=&Q^2+q_T^2=M_{T}^2+M_{T2}^2+2M_{T}M_{T2}\cosh(y-y_2).
\end{eqnarray}
Using these variables and the energy-momentum conservation, we have,
\begin{eqnarray}
k_{T1}&=&\frac{Q^2/2}{M_T\cosh(y-y_{k1})+M_{T2}\cosh(y_2-y_{k1})
  -q_T\cos\phi_{1q}}, \nonumber \\
\cos\phi_{1q}&=&[p_T\cos\phi_{k1}+p_{T2}\cos(\phi_2-\phi_{k1})]/q_T.
\end{eqnarray}
In the integral over $\tau$, we shall use the time evolution of
the temperature, $T(\tau)$, and fugacities, $\lambda_i(\tau)$,
as given in the previous sections.

\subsubsection{Pre-thermal production}

Before the parton distributions reach local isotropy in momentum
space so that the rate equations can be applied to describe
the equilibration of the parton system, scatterings among
free-streaming partons can also lead to charm production. Since
the system during this period consists dominantly of gluons,
we shall only consider gluon fusions. To model the phase-space
distribution, we take into account the distribution of the
initial production points which spread over a region
of width 
\begin{equation}
\Delta_k\approx \frac{2}{k_T\cosh y}
\end{equation}
in $z$ coordinate. Assuming
free-streaming until $\tau_{\rm iso}$ and neglecting the expansion in
the transverse direction, the correlated phase-space distribution
function is given by
\begin{equation}
  f(k,x)=\frac{1}{g_G\pi R_A^2}g(k_T,y)
  D({\bf x}-\hat{\bf z}t\tanh y)\theta(\tau_{\rm iso}\cosh y-t),
  \label{eq:phase}
\end{equation}
where
\begin{equation}
D({\bf x})=\frac{e^{-z^2/2\Delta_k^2}}{\sqrt{2\pi}\Delta_k}\theta(R_A-r)
\end{equation}
is the initial spatial distribution at $t=0$, and 
\begin{equation}
  g(k_T,y)=\frac{(2\pi)^3}{k}\frac{dN_g}{dyd^2k_T}
  =\frac{(2\pi)^2}{k}h(k_T)\frac{1}{2Y}\theta(Y^2-y^2)
  \label{eq:phase2}
\end{equation}
is the parametrization of the parton spectrum given by HIJING 
simulations. The phase-space distribution is normalized such that
$\lim_{t\rightarrow\infty}\int d^3 x f(k,x)/(2\pi)^3 =d^3N_g/d^3k$. 
The function $h(k_T)$ 
and the rapidity width $Y$ are given in Table~\ref{table2} for 
central $Au+Au$ collisions at RHIC and LHC energies.

\begin{table}
\begin{center}
\begin{tabular}{cll} \hline\hline
\makebox[1in]{$\sqrt{s}$ (TeV)} & $Y$ &\makebox[1in]{$h(k_T)$ (GeV$^-2$)} 
            \hfill \\ \hline
& &  \\
0.2 & 2.5 & $1754.4e^{-k_T/0.9}/(k_T+0.3)$ \hfill \\
& & \\
5.5 & 5.0 & $2.66\times 10^7/(k_T+2.9)^{6.4}$ \hfill \\
& & \\
\hline\hline
\end{tabular}

\caption{Parametrizations of the momentum spectra of the initially
produced partons in HIJING calculation. The unit of the transverse momentum
$k_T$ is GeV}
\label{table2}
\end{center}
\end{table}

Substituting the phase-space distribution, into Eq.~(\ref{eq:therm1}),
and integrating over space and time, we obtain the charm production
distribution in the pre-thermal period,
\begin{eqnarray}
  \frac{dN_{\rm pre}}{dyd^2p_T}&=&\frac{1}{16(2\pi)^8\pi R_A^2}
  \int p_{T2}dp_{T2} d\phi_2 dy_2 d\phi_{k1} dy_{k1}\frac{2k_{T1}^2}{\hat{s}}
  g(k_{T1},y_{k1})g(k_{T2},y_{k2})\nonumber \\
  & &\frac{1}{2}|\overline{\cal M}_{gg\rightarrow c\bar{c}}|^2
  \frac{1}{\sqrt{2\pi}\Delta_{\rm tot}}\int_0^{t_f}dt
  e^{-t^2(\tanh y_{k1} -\tanh y_{k2} )/2\Delta^2_{\rm tot}},
\label{eq:preth}
\end{eqnarray}
\begin{equation}
  t_f=\tau_{\rm iso}\min(\cosh y_{k1},\cosh y_{k2}),
  \;\; \Delta_{\rm tot}=\sqrt{\Delta_{k1}^2+\Delta_{k2}^2},
\end{equation}
where the kinematic variables are similarly defined as in
Eq.~(\ref{eq:therm2}), and in addition,
\begin{eqnarray}
  k_{T2}^2&=&Q_T^2+k_{T1}^2-2k_{T1}[M_T\cosh(y-y_{k1})
  +M_{T2}\cosh(y_2-y_{k1})], \nonumber \\
  \cosh y_{k2}&=&[M_T\cosh y +M_{T2}\cosh y_2-k_{T1}\cosh y_{k1}]/k_{T2}.
\end{eqnarray}

Note that the correlation between momentum and space-time in the
phase-space distribution is very important as discussed by
Lin and Gyulassy in Ref.~\cite{LG94}, where formation time effects
are also included. As we will see, this correlation will reduce
the pre-thermal charm production as compared to the result
in Ref.~\cite{BMXW92}.

\subsubsection{Initial fusion}

\begin{figure}
\centerline{\psfig{figure=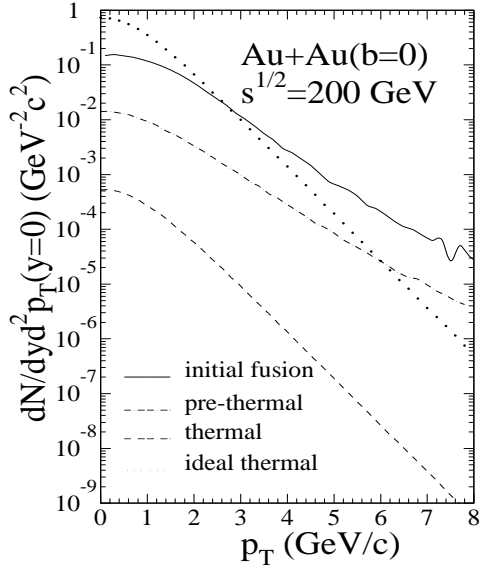,width=2.5in,height=3in}}
\caption{The $p_T$ distributions of initial (solid), prethermal (dot-dashed),
         and thermal (dashed) charm production for central $Au+Au$ collisions
         at RHIC energy, $\protect\sqrt{s}=200$ AGeV with initial 
         conditions given in Tables \protect\ref{table1} and 
         \protect\ref{table2}. The dotted line is the
         thermal production assuming an initial fully equilibrated QGP
         at the same temperature.}
\label{fig30}
\end{figure}

\begin{figure}
\centerline{\psfig{figure=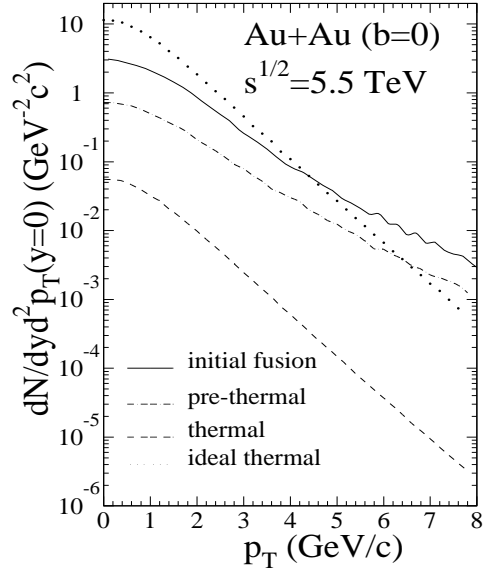,width=2.5in,height=3in}}
\caption{The same as in Fig.~\protect\ref{fig30}, except at LHC energy,
         $\protect\sqrt{s}=5.5$ ATeV.}
\label{fig31}
\end{figure}

During the initial interaction period, charm quarks are produced 
together with minijets through gluon fusion and quark anti-quark 
annihilation. Like gluon and light quark production, charm 
production through the initial fusion is very sensitive to the 
parton distributions inside nuclei. In addition, the cross
section is also very sensitive to the value of the charm quark
mass, $M$. If higher order corrections are taken into account, 
the production cross section depends also on the choices of 
the renormalization and factorization scales. Detailed studies 
of the next-leading-order calculation\cite{VOGT} shows, however, 
that higher order corrections to the total charm 
production cross section can be accounted for by a constant $K$ 
factor of about 2. This is what we will use next. For consistency
we use $M=1.5$ GeV for all calculations. Shown in Figs.~\ref{fig30}
and \ref{fig31} as solid lines are the initial charm
production given by HIJING calculations at RHIC and LHC energies,
with MRSD$-'$ \cite{MRS} parton  distributions. The corresponding total 
integrated cross sections are, $\sigma_{c\bar{c}}=0.16$ (5.75) mb 
at RHIC (LHC) energy, where nuclear shadowing of the gluon distribution
function is also taken into account. In the HIJING calculations, higher
order corrections are included via parton cascade in both initial 
and final state radiations. The resultant distributions in
$c\bar{c}$-pair momentum are very close to the explicit
higher order calculations \cite{VOGT}.

Plotted in Figs.~\ref{fig30} and \ref{fig31} as
dot-dashed and dashed lines are the pre-thermal and thermal
production. In the calculation, a factor of 2 is also
multiplied to the lowest order matrix elements of charm
production. Both contributions are
much smaller than the initial charm production at both
energies. The pre-thermal contributions shown are
also much smaller than what was found in Ref.~\cite{BMXW92}.
This is because momentum and space-time correlation
was not taken into account in Ref.~\cite{BMXW92} which
suppresses the pre-thermal charm production. Similar
results are also found in a study by Lin and Gyulassy \cite{LG94}.
However, a fully equilibrated parton plasma
($\lambda_g=\lambda_q=1$) at the same initial temperature
would give an enhancement of charm production about 4 times
higher than the initial production, shown as dotted lines
in Figs.~\ref{fig30} and \ref{fig31}. In this
case, the enhancement not only comes from higher parton
densities, but also from the much longer life time of
the parton plasma (cf. Figs.~\ref{fig27} and \ref{fig28}).

\begin{figure}
\centerline{\psfig{figure=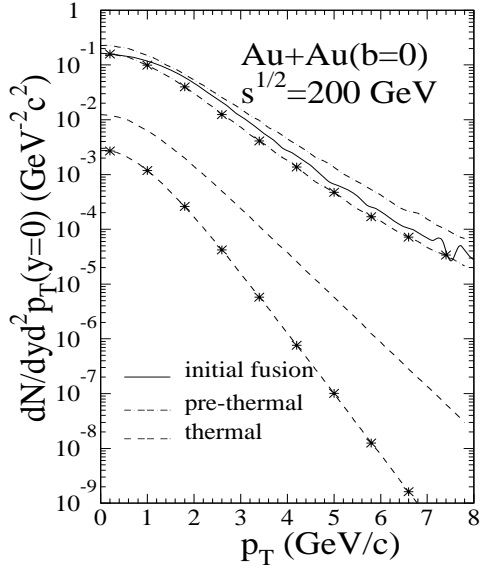,width=2.5in,height=3in}}
\caption{The $p_{\perp}$ distribution of the initial (solid), pre-thermal
  (dot-dashed) and thermal (dashed) charm production for
  initial parton densities 4 times higher than the HIJING estimate
  given in Table \protect\ref{table1}, with the same (ordinary lines) or
  reduced initial temperature, $T_0=0.4$ GeV (lines with stars).}
\label{fig31b}
\end{figure}

As we have already discussed, the initial conditions
in Tables~\ref{table1} and \ref{table2} given by HIJING
calculations have many uncertainties. If one increases the
initial parton number density at RHIC energy by a factor 
of 4 with the same initial temperature, charm production
from both pre-thermal and thermal sources will increase
about a factor of 16 as shown in Fig.~\ref{fig31b}, 
leading to a total secondary
contribution comparable to the initial charm production.
Alternatively, as we have discussed in the Section on parton evolution,
we can consider the initial parton densities to be 4 times higher than
given in Table~\ref{table1} at RHIC energy but with lower 
initial temperature, $T_0=0.4$ GeV. 
Accordingly, the initial phase-space distribution
is also modified to: $h(k_{\perp})=9649.2 e^{-k_{\perp}/0.65}/(k_{\perp}+0.3)$
from the one in Table~\ref{table2}, which gives 4 times the
initial parton density but a smaller average transverse
momentum, $\langle k_{\perp}\rangle = 0.85$ GeV. The
reduced average transverse momentum corresponds to 
a lower initial effective temperature. This system with higher initial
fugacities evolves faster toward equilibrium but the
life-time of the deconfined phase is shorter due to
the reduced temperature as we have discussed. The
corresponding open charm production is shown in Fig.~\ref{fig31b}
by the lines with stars. We observe that open charm production
from both pre-thermal and thermal contribution is reduced due
to the reduction in initial temperature and life-time of the
parton plasma, even though the initial fugacities are much
higher and the evolution toward equilibrium is faster.
Thus, open charm production is much more sensitive to the
change in the initial temperature than to the parton fugacities.

We note from Eqs.~(\ref{eq:therm2}) and (\ref{eq:preth})
that the pre-thermal and thermal charm production depends
on the thermalization time $\tau_{\rm iso}$ and the life
time of the parton plasma. Therefore, by measuring the
charm enhancement, we can probe both the initial
parton phase-space distribution and the thermalization
and equilibration time. However,
in the most optimistic case, the charm production through
thermal and pre-thermal scatterings, cannot be 40 to 50
times higher than the initial direct charm production, as
claimed in PCM calculation \cite{KG93}.
A detailed examination exhibited that the copious
charm production in Ref.~\cite{KG93} is due to an overestimate
of the intrinsic charm production. Though the intrinsic
charm production is important in the forward direction at
large $x_f$ \cite{VBH}, it is strongly suppressed in the
mid-rapidity region due to interference among pQCD
amplitudes to the same order \cite{CSS86}.

\subsection{Dilepton and Photon Production}

The above calculation of open charm distribution does not include
the effects of secondary scatterings suffered by the produced charm
quarks. Though the secondary scatterings do not change the
total number of charm quarks, they do modify the final momentum 
distribution. Therefore, there is no straightforward connection
between the shape of charm distribution and the evolution history
of the parton plasma. Dileptons and photons, on the other hand,
are not subject to final state scatterings due to their week
interaction with the medium. Their momentum distribution will
remain intact throughout the lifetime of the system, and thus
can reveal the evolution history. However, unlike open charm
production, low-mass dileptons and small $p_T$ photons can be
easily produced during the hadronic phase together with
contributions from many resonance decays. Recently, it
has been argued that the  enhancement of low-mass dileptons
at CERN SPS energy \cite{CERES} can be interpreted by the the mass-shift
of the $\rho$ mesons \cite{KO}. Readers can find detailed discussions 
about hadronic contributions to dilepton production in
Ref.~\cite{RUUS_DL}. At very high dilepton mass or large photon $p_T$,
the production is again dominated by initial hard scatterings. 
Therefore, to study dilepton and photon production from the parton 
phase, one has to find a window in phase space in which
the thermal partonic contribution is dominant. We will find
that this is not necessarily possible for all initial conditions.
I will not discuss in detail on how to find this window and
other proposed techniques \cite{yuki} to single out dileptons
and photons from thermal parton scatterings. Instead, I will
only concentrate on how to calculate dilepton and photon
production from an equilibrating parton plasma.

Similar to open charm production, dilepton and photon
production from parton scatterings can also be divided
into initial, pre-thermal and thermal production. 
The thermal dilepton production rate can be written as
\begin{equation}
\frac{dN_{\rm therm}}{d^4x}
=\frac{1}{(2\pi)^6}\int \frac{d^3k_1}{2\omega_1}
\frac{d^3k_2}{2\omega_2}2M^2\sum_q\sigma_q(M)f_q(k_1)f_{\bar{q}}(k_2),
\label{eq:dlth1}
\end{equation}
where $\sigma_q(M)=12e_q^2 4\pi\alpha^2/3M^2$ is the cross section
of $q\bar{q}\rightarrow \ell^+\ell^-(M)$ summed over spin and color.
Insert $\int d^4P\delta^4(P-k_1-k_2)$ and use $d^4P=(1/2)dM^2d^2P_Tdy$,
we can obtain the differential rate as,
\begin{equation}
\frac{dN_{\rm therm}}{d^4xdM^2dy}=\frac{1}{(2\pi)^6}\sum_q M^2\sigma_q(M)
\int d^2P_Tdy_1d\phi_1\frac{k_{T1}^2}{2M^2}f_q(k_1)f_{\bar{q}}(P-k_1),
\label{eq:dlth2}
\end{equation}
where $k_{T1}=M^2/2[M_T\cosh(y-y_1)-P_T\cos\phi_1]$ from the requirement
that $(P-k_1)^2=0$, and $M_T=\sqrt{M^2+P_T^2}$. For large values
of $M\gg T$, one can approximate a boost invariant distribution such
that
\begin{equation}
  f_q(k_1)f_{\bar{q}}(P-k_1)\approx \lambda_q^2 e^{-M_T\cosh(y-\eta)/T}.
\end{equation}
One can then complete the integration in momentum and space-time,
$d^4x=\tau d\tau d\eta d^2x_T$. The total number of thermal
dileptons produced during the evolution is, similar to
the thermal charm production in Eq.~(\ref{eq:therm2}),
\begin{equation}
\frac{dN_{\rm therm}}{dM^2dy}=\frac{\pi R_A^2}{2(2\pi)^4}
\int_{\tau_{\rm iso}}^{\tau_c} \tau d\tau M^3 T K_1(M/T)
\sum_q \lambda_q^2\sigma_q(M). \label{eq:dlth3}
\end{equation}
This number can be calculated numerically given the time-dependence
of the temperature and quark fugacity.

{}Following the calculation of pre-thermal charm production, we
can also compute pre-thermal dilepton yield by substituting
$f_q(k)$ with a correlated phase-space distribution function
similar to the one in Eq.~(\ref{eq:phase}) for a free-streaming
system. From Eq.~(\ref{eq:dlth2}), we can write down the
dilepton yield during the pre-thermal period,
\begin{eqnarray}
\frac{dN_{\rm pre}}{dM^2dy}&=&\frac{M^2}{(2\pi)^5\pi R_A^2 g_q^2}
\sum_q\sigma_q(M) \int P_TdP_T dy_1 d\phi_1 \frac{k_{T1}^2}{2M^2} 
q(k_{T1},y_1)q(k_{T2},y_2) \nonumber \\
 &\mbox{}& \frac{1}{\sqrt{2\pi}\Delta_{\rm tot}}\int_0^{t_f}dt
e^{-t^2(\tanh y_1-\tanh y_2)^2/2\Delta^2_{\rm tot}},
\end{eqnarray}
where the time integral is similarly defines as in Eq.~(\ref{eq:preth}),
$g_q=12$ is the degeneracy of a quark (or anti-quark) with 2 flavors, and
\begin{eqnarray}
  k_{T2}^2&=&P_T^2-2P_T k_{T1}\cos \phi_1 + k_{T1}^2 \nonumber \\
  \sinh y_2&=&(M_T\sinh y - k_{T1}\sinh y_1)/k_{T2} .
\end{eqnarray}
The quark distribution $q(k_T,y)$ is obtained by multiplying
the gluon distribution $g(k_T,y)$ in Eq.~(\ref{eq:phase2}) by
0.07 which is determined by the the quark-to-gluon ratio
in  HIJING simulation (see Fig.~\ref{fig23}).

To the lowest order, the initial dilepton production for a 
central $AA$ collision is,
\begin{equation}
\frac{dN_{\rm initial}}{dydM^2}= \frac{T_{AA}(0)}{s}
\frac{4\pi \alpha^2}{9M^2}\sum_{q}e_q^2 [f_q(x_1,M^2)f_{\bar{q}}(x_2,M^2)
 + f_q(x_2,M^2)f_{\bar{q}}(x_1,M^2)],
\end{equation}
where $M$ is the invariant mass of the dilepton, $e_q$ the fractional
charge of each quark $q$, and $x_{1,2}=e^{\pm y}M/\sqrt{s}$ are
the fractional momenta carried by the quark and antiquark.
Here we neglect the nuclear effects due to initial multiple
interactions. One can also calculate the next-to-leading order (NLO)
corrections which generally give a $K$-factor about 1.1--1.4,
depending on $M$.

\begin{figure}
\centerline{\psfig{figure=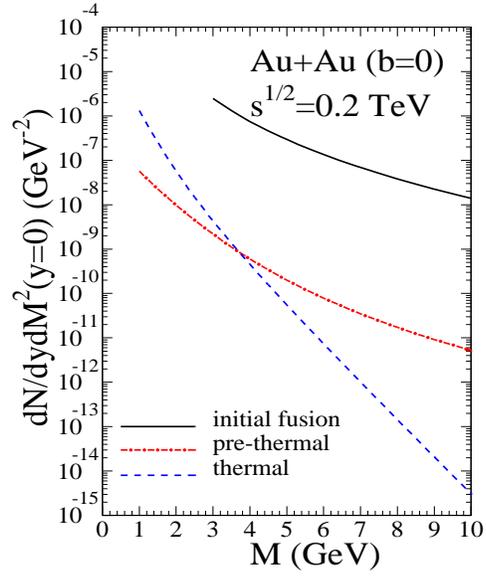,width=2.5in,height=3in}}
\caption{The distributions of initial (solid), pre-thermal (dot-dashed),
         and thermal (dashed) dilepton production as functions of the
         invariant mass $M$ for central $Au+Au$ collisions
         at RHIC energy, $\protect\sqrt{s}=200$ AGeV with initial 
         conditions given in Tables \protect\ref{table1} and 
         \protect\ref{table2}.}
\label{fig:dl1}
\end{figure}

\begin{figure}
\centerline{\psfig{figure=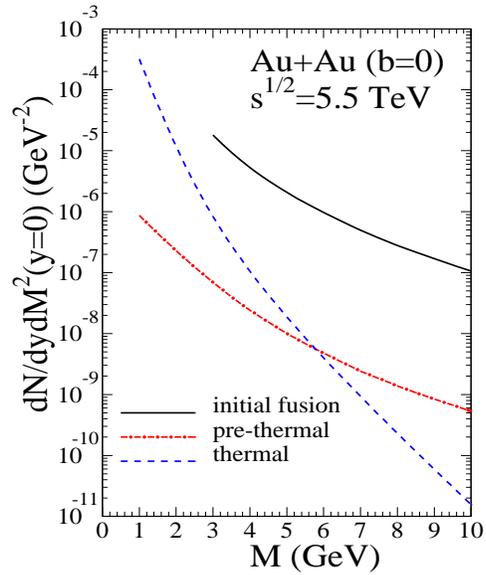,width=2.5in,height=3in}}
\caption{The same as in Fig.~\protect\ref{fig:dl1}, except at LHC energy,
         $\protect\sqrt{s}=5.5$ ATeV.}
\label{fig:dl2}
\end{figure}

The numerical results of the above dilepton yields during
three different stages are plotted in Figs.~\ref{fig:dl1}
and \ref{fig:dl2} for central $Au+Au$ collisions at RHIC
and LHC energies. The results of initial dilepton production
are taken from a NLO calculation in Ref.~\cite{HPDL} with
MRS D$-'$ parton distributions. Similar to open charm
production, we can see that dilepton production during the
equilibration is much smaller than from initial
parton scatterings due to small quark number
density and the correlation between space and momentum
for the produced partons. One also notices that thermal
production at low masses becomes larger than the pre-thermal
one because of the contribution at later time when the
system is approaching chemical equilibrium and the quark
fugacity is much larger than its initial value. We can
also increase the initial parton density by a factor of 4 as before.
The corresponding thermal and pre-thermal dilepton yields will 
also increase by roughly 16 times, which is, nevertheless,
still smaller than the initial production. Only in the extreme
limit of complete thermal and chemical equilibrium initial
condition can the thermal dilepton yield overcome the
shadow of the initial Drell-Yan background in the low and intermediate
mass range.

This result is also consistent with earlier
estimate by Eskola and Lindfors \cite{ESLIN}. My estimate
here is smaller than those given by Strickland \cite{STRICKLAND}
and Shuryak and Xiong \cite{SXDL93} because of the smaller quark
fugacity I used here. The effective quark fugacity in  
Ref.~\cite{SXDL93} is much larger because it is estimated
simply from  the parton distribution function in a nucleon
without taking into account smaller quark production cross
section as compared to the gluon production. Using a parton
cascade model, Geiger and Kapusta \cite{GKDL93} predicted
a total dilepton yield which is 5 times larger than
the underlying Drell-Yan even at $M=8$ GeV region. A detailed
analysis is needed to completely understand their results.
Similarly as I discussed in multiple scattering and open
charm production, it is very likely that the effect of finite
formation time is also at play in this case.

Photon production during parton equilibration can
be calculated in the same way as done in Ref.\cite{STRICKLAND}.
One problem which is unique in photon production is that
the quark screening mass has to be included to regulate the
infrared divergency \cite{KLS91}. Such a regularization,
however, becomes technically difficult in the calculation
of pre-thermal photon production. However, the final
result would be qualitatively similar to dilepton production.

One last note on photon and dilepton production is that 
background from hadronic decays has to be taken into account
in experimental measurements. In particular, dilepton background
from charm decays is very important even at relatively
large invariant masses \cite{VJMR}. One can even use this
background, properly identified experimentally, to measure
open charm production in $pA$ collisions and the effect
of nuclear shadowing \cite{LIGY95}.

\subsection{$J/\psi$ Suppression}

I have demonstrated so far that open charm, dilepton and
photon production during the equilibration are very sensitive
to the initial parton density and the chemical composition
of the initially produced parton system. However, as a careful
reader might notice, I have not addressed the question how sensitive
these probes are to the color deconfinement of the parton plasma. In
principle, charm and dilepton production in a hadronic
gas are suppressed by form-factors as compared to a deconfined
parton plasma at the same temperature. This is why 
dilepton production in the hadronic phase contributes mostly
to the low mass region while charm production from
the hadronic phase can be practically neglected. As we have
learned, dileptons and direct photons are only useful when the initial
parton density is high enough to produce a signal
much larger than the background from the initial parton scatterings.
Let us consider another process, $J/\psi$ suppression,
which might be more sensitive to the color deconfinement
but less dependent on the initial parton density.

Using $J/\psi$ suppression as a probe of the color deconfinement \cite{ms}
requires that the interactions of $J/\psi$ with hadrons and
deconfined partons are different \cite{ks94}. Because
of its small size, a heavy quarkonium can probe the short-distance
properties of light hadrons. It is thus possible to make a parton-based
calculation of the $J/\psi$-hadron cross section via an operator product
expansion method similar to that used in deeply-inelastic
lepton-hadron scatterings \cite{ks94,peskin,kaid}. The resulting
$J/\psi$-hadron cross section can be related to the distribution 
function of gluons inside a hadron.
The energy dependence of the cross section near the
threshold for the break-up of a $J/\psi$ is determined by the
large $x$ behavior of the gluon distribution function, giving rise
to a very small break-up cross section at low energies. Only at very
high energies, this cross section will reach its
asymptotic value of a few mb.
Therefore, the  dissociation can only occur if the gluon
from the light hadron wave function is hard enough in the $J/\psi$'s 
rest frame to overcome the binding energy threshold.
A hadron gas with temperature below 0.5 GeV
cannot provide such energetic gluons to break up the $J/\psi$.
Therefore, a slow $J/\psi$ is very unlikely to be absorbed inside
a hadron gas of reasonable temperature \cite{ks94}.

\begin{figure}
\centerline{\psfig{figure=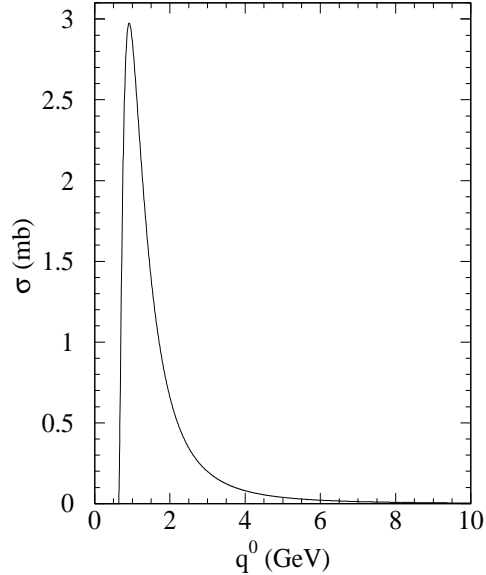,width=2.5in,height=3in}}
\caption{Gluon-$J/ \psi$ dissociation cross section as a function
  of the gluon energy $q^0$ in the rest frame of the $J/\psi$.}
\label{fig:jp1}
\end{figure}

On the other hand, a deconfined partonic system contains much
harder gluons which can easily break up a $J/\psi$
\cite{ks94,ks95}. The operator product expansion allows one 
to express the hadron-$J/\psi$ inelastic cross section in terms 
of the convolution of the gluon-$J/\psi$ dissociation cross section
with the gluon distribution inside the hadron \cite{ks94}.
The gluon-$J/\psi$ dissociation cross section is given by \cite{ks95}
\begin{equation}
  \sigma (q^0)=\frac{2\pi}{3}\left(\frac{32}{3}\right)^2
  \frac{1}{m_Q(\epsilon_0 m_Q)^{1/2}}
\frac{(q^0/\epsilon_0-1)^{3/2}}{(q^0/\epsilon_0)^5}\;, \label{eq:jp1}
\end{equation}
with $q_0$ the gluon energy in the rest frame of the $J/\psi$,
$m_Q$ the heavy quark mass, $\epsilon_0=2M_D-M_{J/\psi}$ 
the binding energy, where $M_{J/\psi}$ and $M_D$ are
the $J/\psi$ and $D$ meson masses, respectively.
As shown in Fig.~\ref{fig:jp1}, the inelastic cross section 
shows a strong peak just above the break-up threshold of the 
gluon energy, with a maximum value of about 3 mb. Low-momentum 
gluons have neither the resolution to distinguish the heavy constituent 
quarks nor the energy to excite them to the continuum.

In the pre-equilibrium stage, i.e., before the partons have reached
equilibrium, the average parton transverse momentum is sufficiently
large to break up a $J/\psi$, provided the partons
are deconfined. The dissociation of $J/\psi$ will
continue during the whole equilibration process until the
effective temperature drops below a certain value or the
beginning of hadronization, whichever takes place first.
Therefore measurements of $J/\psi$
suppression can probe the deconfinement of the early partonic
system and shed light on the subsequent equilibration process,
provided that possible nuclear effects on the production of $Q{\bar Q}$
pairs and on pre-resonance charmonium states
are understood and taken into account.

Let us consider $J/\psi$ suppression in the central rapidity
region ($y_{J/\psi}\simeq 0$). In this case, the $J/ \psi$ will
move in the transverse direction with a four-velocity
\begin{equation}
u=(M_T, \vec{P_T}, 0)/M_{J/\psi}, \label{eq:jp2}
\end{equation}
where $M_T=\sqrt{P_T^2+M^2_{J/ \psi}}$ is defined as the $J/\psi$'s
transverse mass. A gluon with a four-momentum $k=(k^0,\vec{k})$
in the rest frame of the parton gas has an energy $q^0=k\cdot u$
in the rest frame of the $J/\psi$. Using Eq.~(\ref{eq:jp1}), one
can find out the thermal averaged gluon-$J/\psi$ dissociation 
cross section \cite{XKSW},
\begin{eqnarray}
  \langle v_{\rm rel}\sigma(k\cdot u)\rangle_k&=&
(\frac{8}{3})^3\frac{\pi}{\zeta(3)}\frac{M_{J/\psi}^2}{P_TM_T T^3}
(\frac{\epsilon_0}{m_Q})^{3/2} \nonumber  \\
&\mbox{}& \sum_{n=1}^{\infty}T_n
\int_1^{\infty} dx\frac{(x-1)^{3/2}}{x^4}(e^{-a_n^-x}-e^{-a_n^+x})\; ,
\label{eq:jp4}
\end{eqnarray}
with $T_n=T/n$ and
\begin{equation}
  a_n^{\pm}=\frac{\epsilon_0}{T_n}\frac{M_T\pm P_T}{M_{J/\psi}}
  \; . \label{eq:jp5}
\end{equation}

Using the thermal cross section just obtained, we can now
calculate the survival probability of $J/ \psi$ in an
equilibrating parton plasma, Again, we will neglect the
transverse expansion and consider only longitudinal expansion.
A $J/\psi$ produced at point $\vec{r}$
with velocity $\vec{v}$ in the transverse
direction will travel a distance
\begin{equation}
  d=-r\cos\phi + \sqrt{R_A^2-r^2(1-\cos^2\phi)} \label{eq:jp5b}
\end{equation}
in the time interval $t_{\psi}=M_T d/P_T$ before it escapes 
from a gluon gas of transverse extension $R_A$; here
$\cos\phi= \hat{\vec{v}}\cdot\hat{\vec{r}}$. Suppose the
system evolves in a deconfined state until the temperature
drops below a certain value, which we assume to be 200 MeV.
The total amount of time the $J/\psi$ remains inside a 
deconfined parton gas is the smaller one of the two times
$t_{\psi}$ and $t_f$, the life-time of the parton gas. Assume 
that the initial production rate of the $J/\psi$ is proportional 
to the number of binary nucleon-nucleon interactions at 
impact-parameter $r$, $N_A(r)=A^2(1-r^2/R_A^2)/2\pi R_A^2$. 
The survival probability of the $J/\psi$ averaged over its 
initial position and direction in an equilibrating parton gas is
\begin{equation}
  S(P_T)=\frac{\int d^2r (R_A^2-r^2) \exp{[-\int^{t_{\rm min}}_0 d\tau
      n_g(\tau)\langle v_{\rm rel}\sigma(k\cdot u)\rangle_k]}}
  {\int d^2r (R_A^2-r^2)} \; , \label{eq:jp6}
\end{equation}
where
\begin{equation}
  t_{\rm min}=\min (t_{\psi},t_f) \; , \label{eq:jp7}
\end{equation}
and $n_g(\tau)$ is the gluon number density at a given time  $\tau$.
With the time dependence of the gluon fugacity and temperature
we can then evaluate numerically the survival  probability of
a $J/\psi$ in an equilibrating parton plasma.

\begin{figure}
\centerline{\psfig{figure=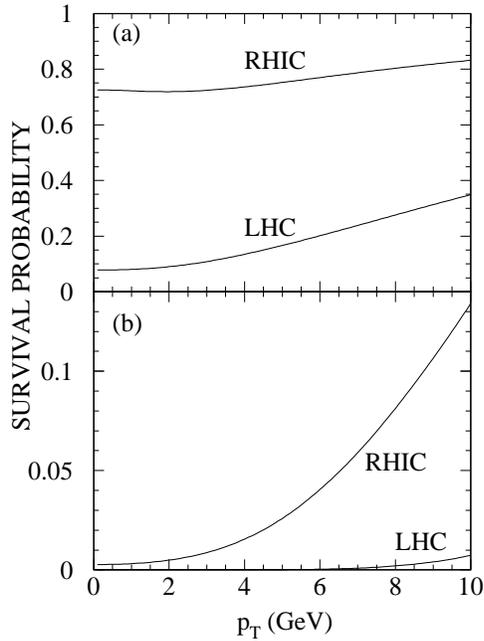,width=2.5in,height=3in}}
\vspace{22pt}
\caption{ (a) The survival probability of $J/\psi$ in an 
  equilibrating parton plasma at RHIC and LHC energies with initial 
  conditions given by HIJING simulation in \protect\ref{table1} (b) 
  and for an initially equilibrated plasma at the same temperatures.}
\label{fig:jp2}
\end{figure}

Shown in Fig.~\ref{fig:jp2}a are the $J/\psi$ survival
probabilities in the deconfined and equilibrating parton
plasma at RHIC and LHC energies. We  find that there is stronger
$J/\psi$ suppression at LHC than at RHIC energy, due both to the
higher initial parton densities and longer life-time of the
parton plasma. The increase of the survival probabilities with
$J/\psi$'s transverse momentum is a consequence of the
decrease of the thermal cross section with increasing $P_T$
at high temperatures, and the shorter time
spent by a higher-$P_T$ $J/\psi$ inside the parton plasma, an 
effect first considered in Ref.\cite{kp}.

\begin{figure}
\centerline{\psfig{figure=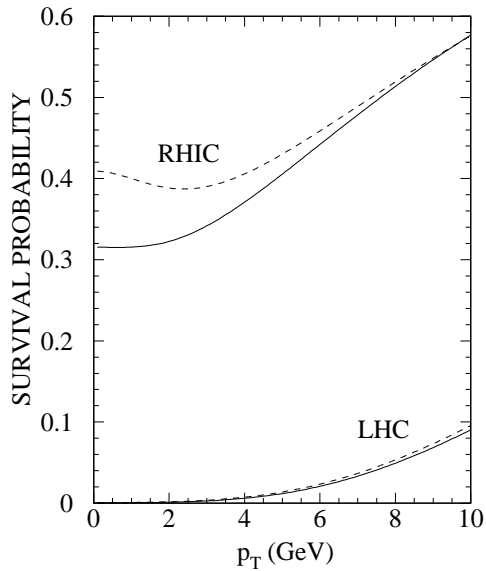,width=2.5in,height=3in}}
\caption{The survival probability of $J/\psi$ in an equilibrating parton
  plasma with initial parton densities 4 times higher than HIJING estimate
  given in \protect\ref{table1} but with the same (solid line), or
  reduced initial temperature, $T_0=0.4$ GeV (dashed line).}
\label{fig:jp3}
\end{figure}

As in open charm and dilepton production, we multiply initial 
parton number densities by a factor of 4, thus increasing 
initial parton fugacities to test the sensitivity of $J/\psi$
suppression to the uncertainties in the initial conditions.
The corresponding survival probabilities, shown
in Fig.~\ref{fig:jp3} as solid lines, are much lower. 
If the uncertainties in initial conditions are caused by the soft parton
production from the color mean fields, the initial effective
temperature will decrease. Therefore, we can alternatively
increase the initial parton density by a factor of 4 and
at the same time decrease $T_0$ to 0.4 and 0.72 GeV at RHIC and
LHC energies, respectively. This leads also to higher initial fugacities.
Comparing the solid and dashed lines shows that the $J/\psi$
suppression is less sensitive to the variation of the initial
temperature and fugacities as long as the parton densities
are fixed. In the most extreme case of a fully equilibrated
parton plasma (unit initial fugacities),  $J/\psi$ is much more 
suppressed, as shown in Fig.~\ref{fig:jp2}b, because of both 
the higher parton density and the longer life time of the system.

At high temperature, the thermal-averaged gluon-$J/\psi$
dissociation cross section decreases monotonically with
$P_T$ since the averaged collision energy is already
above the break-up threshold. One can translate this
into a  $J/\psi$ survival probability monotonically increasing
with $P_T$. However, at low temperature
blow 300 MeV, the $P_T$ dependence of the thermally averaged
gluon-$J/\psi$ dissociation cross section also displays
the threshold peak as in Fig.~\ref{fig:jp1}. The peak
moves to larger values of $P_T$ at a lower temperature \cite{XKSW}.
Therefore, due to the large values of the cross section
around the threshold peak, the resultant survival probability
integrated over the evolution time at low temperatures
is quite flat at small $P_T$. The final shape of the
$P_T$ dependence of the survival probability at small
$P_T$ depends on the entire evolution history of the
parton system, especially the relative lengths of time
the system spends in the high and low temperature stages.
The smaller $P_T$ dependence at small $P_T$ seen 
in Figs.~\ref{fig:jp2} and \ref{fig:jp3} is therefore due to
the $J/\psi$ dissociation during the late stage of the evolution
when the temperature is below 300 MeV \cite{XKSW}.
For a parton system with a low initial temperature (below 300 MeV),
the $P_T$ dependence of the survival probability should be
even flatter. One can therefore use the $P_T$ dependence to
shed light on the initial temperature and the evolution history
of the system.

We have seen that the large average momentum in the hot gluon
gas enables gluons to break up the $J/\psi$, while
hadron matter at reasonable temperature does not provide sufficiently
hard gluons.  A substantial $J/\psi$ suppression occurs in such 
a non-equilibrium partonic medium, though smaller than that 
in a fully equilibrated parton plasma.
I should emphasize that, in addition to the $J/\psi$ 
dissociation during the equilibration
of the parton plasma, there are other possible sources of suppression
for the actually observed $J/\psi$'s, such as nuclear
modifications of the $Q{\bar Q}$ production process e.g., through
modified gluon distributions in a nucleus \cite{MQ86,EQW94}, multiple
scattering accompanied by energy loss \cite{gm92}, or a suppression of
the nascent $J/\psi$ before it forms an actual physical resonance
\cite{ksc,rwuh}. Such effects would
cause $J/\psi$ suppression in addition to what we have obtained
from the equilibrating parton plasma and modify the transverse
momentum dependence of $J/\psi$ suppression \cite{ggj}.
Moreover, there should also be $J/\psi$ pre-thermal suppression.
On the other hand, gluon fusion could also result in
$J/\psi$ production during the evolution of the parton system,
similar to the enhancement of open charm. A
consistent study of $J/\psi$ suppression should include
the pre-equilibrium production in a form of a master rate
equation.

\subsection{Jet Quenching and Monojet Production}

Complementary to $J/\psi$ suppression, a study of high $p_T$ 
jets and their propagation inside the medium can also
probe the structure of the dense matter and possibly the phase
transition, since these high $p_T$ jets
are produced on a very short time scale as compared to the
soft processes and their production rates and
spectrum can be reliably calculated via pQCD.
For example, an enhanced acoplanarity
of two back-to-back jets in nuclear collisions can be used to study
multiple scatterings of a parton inside a dense medium \cite{ACOP}.
What jets could also probe in high energy heavy ion collisions is
the stopping power, $dE/dz$, of the dense matter for high energy
quarks and gluons \cite{WG92,MGMP90,GPTW}. That stopping
power in turn is controlled by the color screening mass $\mu_D$
in that medium. A possible rapid change of $\mu_D$ near the phase
transition point could lead to a variation of jet quenching
phenomena which may serve as signatures of QGP formation \cite{GPTW}.
The energy loss, $dE/dz$, of partons through interaction is also 
closely related to the thermalization and equilibration of
partonic system as we have discussed.

\subsubsection{Energy loss of a fast parton in QGP}

When an energetic parton propagates through a quark-gluon plasma,
it suffers both elastic energy loss, $dE_{\rm el}/dz$ due to simple
scatterings, and radiative energy loss $dE_{\rm rad}/dz$ due to 
induced radiations from multiple
scatterings. Studies \cite{BJ82,THOMA} of the elastic energy loss at 
finite temperatures give a logarithmic energy dependence 
of $dE_{\rm el}/dz$. Its value is also quite small, for example,
0.2 $-$ 0.5 GeV/fm for a quark jet with $E=30$ GeV. The dominant source
of energy loss is expected to be caused by induced radiations.
The estimates of $dE_{\rm rad}/dz$ in the past vary 
widely \cite{RYSKI,gm92,SOREN,BRODS} from energy independent 
to $dE/dz\propto E$. M.~Gyulassy and I have analyzed 
soft gluon bremsstrahlung ~\cite{GWLPM1,GWLPM2} 
and found that interference due to multiple scatterings is 
very important to calculate the radiative energy loss. Similarly to
the parton equilibration rate, the LPM effect, {\em i.e.}, 
the suppression of soft gluons whose formation times are much 
larger than the mean free path of parton scatterings, also reduces
the radiative energy loss inside a dense QCD medium. Dokshitzer
{\it et al.} \cite{DOKS} have made a more complete analysis
of the QCD radiation amplitude, including also the rescatterings
of the radiated gluon. Their results also shown suppression of
soft gluon radiation but with a square-root energy-dependence of the
energy loss. Their energy loss, $dE/dz$, is also proportional
to the distance $z$ the propagating parton has traveled. To
demonstrate the LPM effect on energy loss and jet quenching,
we will use a simple estimate in this paper.

To estimate the radiative energy loss, we can use the regularized 
gluon spectrum induced by a single scattering in Eq.~(\ref{eq:dng}).
We then again take into account the LPM effect in QCD due to color 
interference by approximating the suppression factor with a step
function, $\theta(\lambda_f-\tau_{\rm QCD})$. This step function
restricts the phase space of the radiation to a region in which 
the effective formation time $\tau_{\rm QCD}(k)$ must be smaller 
than the mean free path $\lambda_f$. Radiations not fulfilling 
this requirement will be suppressed by the destructive interference. 
The radiative energy loss suffered by the jet  parton per 
collision is then,
\begin{equation}
\Delta E_{\rm rad}=\int d^2k_{\perp}dy\frac{dn_g}{d^2k_{\perp}dy}
k_{\perp}\cosh y\theta(\lambda_f-\tau_{\rm QCD}(k))
\theta(E-k_{\perp}\cosh y), \label{eq:de1}
\end{equation}
where $\tau_{\rm QCD}(k)$ is given by Eq.~(\ref{eq:ftime}), $E$ is
the jet parton energy, and the regularized gluon density distribution, 
$dn_g/dyd^2k_{\perp}$, induced by a single scattering is given 
by Eq.~(\ref{eq:dng}). Since the transverse momentum transfer 
$q_{\perp}$ is the result of elastic scatterings, we have to 
average any function  $f({\bf q}_{\perp})$ of ${\bf q}_{\perp}$ 
by the elastic cross section,
\begin{equation}
\langle f({\bf q}_{\perp})\rangle=\frac{1}{\sigma_i}
\int_{\mu_D^2}^{s/4}dq^2_{\perp}
\frac{d\sigma_i}{dq^2_{\perp}}f({\bf q}_{\perp}),
\end{equation}
where $s\approx 6ET$ is the average $c.m.$ energy squared for the 
scattering of a jet parton with energy $E$ off the thermal partons
at temperature $T$. Since we now only consider a fully equilibrated
quark-gluon plasma, the Debye screening mass will be given
by Eq.~(\ref{eq:eq14}) with $\lambda_g=1$, {\em i.e.}, 
$\mu_D^2=4\pi\alpha_s T^2$.
For the dominant small angle scattering, the elastic cross sections are,
\begin{equation}
\frac{d\sigma_i}{dq^2_{\perp}}\cong C_{\rm el}^i 
\frac{2\pi \alpha^2_s}{q^4_{\perp}}, \label{eq:els}
\end{equation}
where $C_{\rm el}^i=9/4,\;1,\;4/9$ respectively for $gg$, 
$gq$ and $qq$ scatterings. This average can be approximated by 
replacing $q^2_{\perp}$ in the numerator of Eq.~(\ref{eq:dng}) 
with its average value,
\begin{equation}
  \langle q^2_{\perp}\rangle=\mu_D^2\ln\frac{3ET}{2\mu_D^2}. \label{eq:avq}
\end{equation}
In the denominator, we simply replace ${\bf q_{\perp}^2}$
by $\mu_D^2$ after the angular integration.
The remaining  integration in Eq.~(\ref{eq:de1}) over the restricted 
phase space approximately leads to the simple analytic formula,
 \begin{equation}
\Delta E_{\rm rad}\approx \frac{C_A\alpha_s}{\pi}
\langle q^2_{\perp}\rangle\left(\frac{\lambda_f}{2r_2}{\cal I}_1
+\frac{E}{2\mu_D^2}{\cal I}_2\right), \label{eq:de2}
\end{equation}
\begin{eqnarray}
{\cal I}_1&=&\ln\left[\frac{r_2E}{\mu_D^2\lambda_f}+
\sqrt{1+\left(\frac{r_2E}{\mu_D^2\lambda_f}\right)^2}\,\right]-
\ln\left[\left(\frac{r_2\sqrt{2}}{\mu_D\lambda_f}\right)^2+
\sqrt{1+\left(\frac{r_2\sqrt{2}}{\mu_D\lambda_f}\right)^4}\,\right], 
     \label{eq:I1} \\
{\cal I}_2&=&\ln\left[\frac{\mu_D^2\lambda_f}{r_2E}+
\sqrt{1+\left(\frac{\mu_D^2\lambda_f}{r_2E}\right)^2}\,\right]-\ln\left[
\frac{2\mu_D^2}{E^2}+\sqrt{1+\left(\frac{2\mu_D^2}{E^2}\right)^2}\,\right]
.\label{eq:I2}
\end{eqnarray}
In small $k_{\perp}$ regime, the phase space is mainly restricted by 
a small effective formation time, $\tau_{\rm QCD}<\lambda_f$, which 
gives the first term proportional to $\lambda_f$. For large $k_{\perp}$, 
the radiation becomes additive in a restricted phase space 
constrained by energy conservation. That region  contributes to the 
second term which appears to be proportional to the incident energy $E$.
However, in the high energy limit the function ${\cal I}_2\propto 1/E$ 
and hence the radiated energy loss grows only as $\log^2 E$.

The above derivation assumed that the mean free path 
is much larger than the interaction range specified 
by $1/\mu_D$. As we shall discuss below, this is satisfied 
in a quark gluon plasma at least in the weak coupling limit. 
Therefore, we can neglect the second term in ${\cal I}_1$. 
{}For a high energy jet parton, $E\gg\mu_D$, we can also 
neglect the second term in ${\cal I}_2$. The resulting
radiative energy loss reduces in that case to the simple form,
\begin{equation}
\frac{dE_{\rm rad}}{dz}=\frac{\Delta E_{\rm rad}}{\lambda_f}\approx
\frac{C_2\alpha_s}{\pi}\langle q^2_{\perp}\rangle
\left[\ln\left(\xi+\sqrt{1+\xi^2}\right)+\xi
\ln\left(\frac{1}{\xi}+\sqrt{1+\frac{1}{\xi^2}}\right)\right],
\label{eq:dedz1}
\end{equation}
which depends on a dimensionless variable,
\begin{equation}
\xi=\frac{r_2E}{\mu_D^2\lambda_f}.
\end{equation}

We  see that the radiative energy loss $dE_{\rm rad}/dz$ thus obtained
interpolates between the factorization and Bethe-Heitler
limits as a function of the dimensionless variable $\xi$. In the factorization
limit, we fix $\mu_D\lambda_f\gg 1$ and let $E\rightarrow\infty$,
so that $\xi\gg 1$. In this case, we can neglect the second term
in Eq.~(\ref{eq:dedz1}) and have,
\begin{equation}
  \frac{dE_{\rm rad}}{dz}\approx \frac{C_2\alpha_s}{\pi}
 \langle q^2_{\perp}\rangle\ln\left(\frac{2r_2E}
 {\mu_D^2\lambda_f}\right);\;\; \xi\gg 1. \label{eq:dedzf}
\end{equation}
Thus, the radiative energy loss in the factorization limit has only
a logarithmic energy dependence (in addition to the energy dependence
of $\langle q_{\perp}^2\rangle$). Due to the non-Abelian nature of the
color interference, the resultant energy loss for a gluon ($C_2=C_A)$
is 9/4 times larger than that for a quark ($C_2=C_F$). In the other
extreme limit, we fix $E$ and let $\mu_D\lambda_f\rightarrow\infty$, so that
$\xi\ll 1$. In this case, the mean-free-path exceeds the effective 
formation time. The radiation from each scattering adds up. We
then recover the linear dependence of the energy loss $dE_{\rm rad}/dz$
on the incident energy $E$ (modulo logarithms),
\begin{equation}
\frac{dE_{\rm rad}}{dz}\approx  \frac{C_A\alpha_s}{2\pi\lambda_f}
\frac{\langle q^2_{\perp}\rangle}{\mu_D^2}
E\ln\left(\frac{2\mu_D^2\lambda_f}{r_2E}\right);\;\; \xi\ll 1, \label{eq:dedzb}
\end{equation}
as in the Bethe-Heitler formula. In both cases, the radiative energy 
loss is proportional to the average of the transverse momentum transfer, 
$\langle q_{\perp}^2\rangle$, which is controlled by
the color screening mass as in Eq.~(\ref{eq:avq}).

To see  more clearly how  the factorization limit
is approached, we now  estimate $\xi$ for a parton propagating 
inside a high temperature quark-gluon plasma. From 
Eq.~(\ref{eq:els}) and the quark and gluon densities in an ideal
system at a temperature $T$, the mean free path (for 3 quark flavors) is
\begin{equation}
1/\lambda_f^{(q)}=\sigma_{qq}\rho_q+\sigma_{qg}\rho_g\approx
\frac{2\pi\alpha^2_s}{\mu_D^2}4\times 7\zeta(3)
\frac{T^3}{\pi^2},\label{eq:lam1}
\end{equation}
\begin{equation}
1/\lambda_f^{(g)}=\sigma_{qg}\rho_q+\sigma_{gg}\rho_g\approx
\frac{2\pi\alpha^2_s}{\mu_D^2}9\times 7\zeta(3)
\frac{T^3}{\pi^2}, \label{eq:lam2}
\end{equation}
where $\zeta(3)\approx 1.2$.

Using Eqs. (\ref{eq:lam1}) and (\ref{eq:lam2}) and the 
perturbative color electric screening mass, $\mu_D^2=4\pi\alpha_sT^2$, 
we see that  $\xi$ appearing in the logarithms has a 
common energy and temperature dependence for both quarks and gluons,
\begin{equation}
\xi=\frac{r_2E}{\lambda_f\mu_D^2}=\frac{63\zeta(3)}{16\pi^3}\frac{E}{T}
\approx\frac{9}{2\pi^3}\frac{E}{T}, 
\;\;\mbox{for both $q$ and $g$}. \label{eq:xi}
\end{equation}

\begin{figure}
\centerline{\psfig{figure=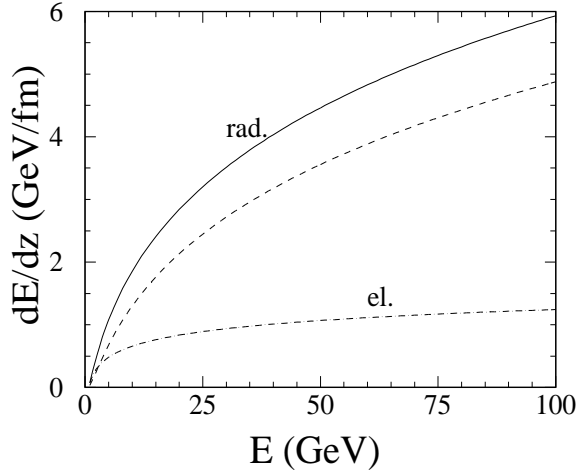,width=3in,height=2.5in}}
\caption{The energy dependence of energy loss, $dE/dz$, of a quark with 
        energy $E$ inside a quark-gluon plasma at temperature $T=300$ MeV.
        A weak coupling $\alpha_s=0.3$ is used. The solid line is the full
        expression and the dashed line is the factorization limit of the
        radiative energy loss. The dot-dashed line is the elastic energy
        loss.}
\label{fig32}
\end{figure}

With the above expression for $\xi$, we plot the radiative energy 
loss in Fig.~\ref{fig32} as a function of the jet energy inside 
a plasma at temperature $T=300$ MeV with $\alpha_s=0.3$.
The solid line is the full expression in Eq.~(\ref{eq:dedz1}) while
the dashed line is the factorization limit corresponding to the first
term in Eq.~(\ref{eq:dedz1}). We see that Eq. (\ref{eq:dedzf}) 
approximates Eq. (\ref{eq:dedz1}) quite well in this parameter range.
The energy dependence of the radiative energy loss
is due to the double logarithmic function in the formula one of which 
comes from the energy dependence of the average transverse momentum
$\langle q^2_{\perp}\rangle$ in Eq.~(\ref{eq:avq}).

The energy loss of a quark in dense matter due to elastic scattering 
was first estimated by Bjorken \cite{BJ82} and later was studied in detail
\cite{THOMA} in terms of finite temperature QCD.  For our purpose, a 
simple estimate taking into account both the thermal average and
color screening will suffice. In terms of elastic cross sections
and the density distributions for quarks and gluons in a plasma,
we have,
\begin{equation}
\frac{dE_{\rm el}}{dz}=\int_{\mu_D^2}^{s/4}dq^2_{\perp}
\frac{d\sigma_i}{dq^2_{\perp}}\rho_i\nu=\langle q^2_{\perp}\rangle
\sigma_i\langle \frac{\rho_i}{2\omega}\rangle,
\end{equation}
where $\nu\approx q^2_{\perp}/2\omega$ is the energy transfer of the
jet parton to a thermal parton with energy $\omega$ during an elastic
scattering, $\langle q^2_{\perp}\rangle$ is the average transverse
momentum transfer given by Eq.~(\ref{eq:avq}). Similar to 
Eqs.~(\ref{eq:lam1}) and (\ref{eq:lam2}), we have,
\begin{equation}
\sigma_{qq}\langle\frac{\rho_q}{2\omega}\rangle+
\sigma_{qg}\langle\frac{\rho_g}{2\omega}\rangle=
\frac{2\pi\alpha^2_s}{\mu_D^2}T^2,
\end{equation}
\begin{equation}
\sigma_{gq}\langle\frac{\rho_q}{2\omega}\rangle+
\sigma_{gg}\langle\frac{\rho_g}{2\omega}\rangle=
\frac{9}{4}\frac{2\pi\alpha^2_s}{\mu_D^2}T^2.
\end{equation}
The elastic energy loss of a fast parton inside a quark gluon 
plasma at a temperature $T$ is then given by,
\begin{equation}
\frac{dE_{\rm el}}{dz}=C_2\frac{3\pi\alpha_s^2}{2\mu_D^2}T^2
\langle q^2_{\perp}\rangle. \label{eq:dedz2}
\end{equation}
For comparison, we plot this elastic energy loss in Fig.~\ref{fig32}. 
In general, it is much smaller than the radiative energy loss 
and has a weaker energy dependence (single logarithmic).
 
Using Eqs.~(\ref{eq:avq}), (\ref{eq:dedzf}) and (\ref{eq:xi}), the 
total energy loss can be expressed as,
\begin{equation}
\frac{dE}{dz}=\frac{dE_{\rm el}}{dz}+\frac{dE_{\rm rad}}{dz}\approx
\frac{C_2\alpha_s}{\pi}\mu_D^2\ln\frac{3ET}{2\mu_D^2}\left(
\ln\frac{9E}{\pi^3T}+\frac{3\pi^2\alpha_s}{2\mu_D^2}T^2\right). 
\label{eq:dedzt}
\end{equation}
It is interesting to note that both the elastic and radiative energy
loss have the same color coefficient $C_2$. For high energy partons,
the radiative energy loss dominates over the elastic one. For 
$E=30$ GeV, $T=300$ MeV, and  $\alpha_s\approx 0.3$, 
the total energy loss for a propagating quark is $dE/dz\approx 3.6$ GeV/fm. 
Only about 25\% of this amount comes from elastic energy loss.

\begin{figure}
\centerline{\psfig{figure=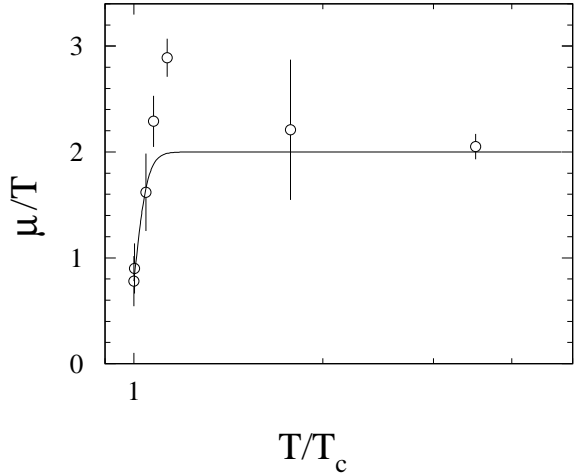,width=3in,height=2.5in}}
\caption{Lattice result of the Debye screen mass \protect\cite{GAO}, 
        $\mu_D$, as a 
        function of temperature. The curve is a parametrization used in
        the following calculation.}
\label{fig33}
\end{figure}

Comparing to the phenomenological string tension $\kappa\sim 1-2$ GeV/fm, 
which can be considered as the $dE/dz$ of quarks in the nonperturbative
QCD vacuum \cite{NIED}, one cannot really distinguish jet
quenching caused by energy loss in a quark gluon
plasma from that in a hadronic matter.  However, since the energy
loss is very sensitive to the infrared cutoff scale $\mu_D(T)$, the study
of jet quenching may be used as a signature of a quark gluon plasma
if the $\mu_D(T)$ dependence of $dE/dz$ can still be given by
Eq.~(\ref{eq:dedzt}) and if there is a significant variation
in $\mu_D(T)$ near the phase transition temperature $T_c$.
Up to now, little is known about $\mu_D(T)$ for physical QCD.
However, lattice simulations \cite{GAO} of a pure gauge QCD
indicate a rapid decrease of $\mu_D(T)$ near $T_c$ as
shown in Fig.~\ref{fig33}, where the weak coupling limit, 
$\mu(T)=gT$, is indeed recovered at high temperatures. Since a
sharp reduction of $\mu_D(T)$ would effectively suppress the 
energy loss according to our formula in Eq.~(\ref{eq:dedzt}), 
a long duration of the mixed phase in a near-first order
phase transition could lead to a decrease of jet quenching.

\begin{figure}
\centerline{\psfig{figure=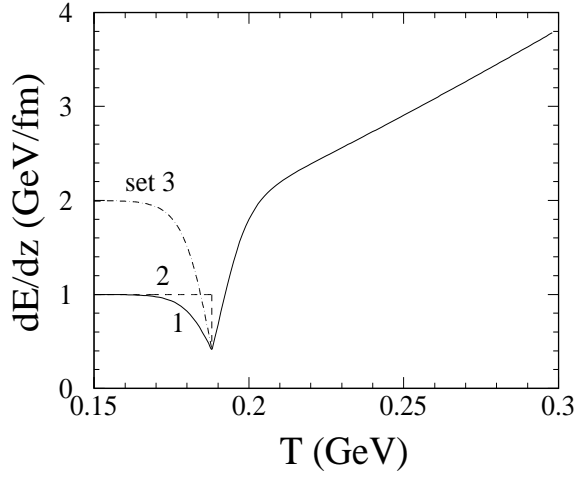,width=3in,height=2.5in}}
\caption{Three scenarios of energy loss as functions of temperature.}
\label{fig34}
\end{figure}

In the following, we will use a parametrization of the lattice
result as a phenomenological guide to describe the temperature 
dependence of the color screening mass $\mu_D(T)$ both near and 
above $T_c$. This will give us the temperature dependence of
the energy loss according to Eq.~(\ref{eq:dedzt}). The coupling 
constant which corresponds to $\mu_D(T)=gT$ in Fig.~\ref{fig33} 
at high temperatures is $\alpha_s\approx 0.3$. We shall freeze 
the coupling constant to this value at all temperatures in
calculating $dE/dz$ in the QGP phase. In hadronic phase 
at $T=0$, we assume a constant energy loss given by the string 
tension, $\kappa\sim 1$ GeV/fm, which is consistent with the 
deep inelastic lepton-nucleus scattering data \cite{GPSTR}.
To extrapolate from $T=0$ to $T_c$, one may choose a smooth
form, $dE/dz\rightarrow 1$ GeV/fm as $T\rightarrow 0$. This
scenario is plotted in Fig.~\ref{fig34} for a quark jet 
with $E=30$ GeV as set 1, in which there is a dip in $dE/dz$ 
during the phase transition due to the small value of $\mu_D$ 
at $T_c$. We have assumed a QCD phase transition temperature
$T_c=0.188$ GeV.  Alternatively, one may consider a 
scenario (set 2 in Fig.~\ref{fig34}) in which $dE/dz=$ cont.=1 GeV/fm 
in the hadronic phase, {\em i.e.}, there is a discontinuity in 
$dE/dz$ at $T=T_c$. For comparison, we also study the case that
$dE/dz$ in the $T=0$ limit is 2 GeV/fm (set 3).

\subsubsection{Suppression of dijets}

We now discuss the implications of the temperature dependence 
of $dE/dz$ on the suppression of di-jets and on the production 
rate of mono-jets in ultrarelativistic nuclear collisions.
Similarly to Eq.~(\ref{eq:sjet1}), the initial rate of back-to-back
parton jets of transverse momentum $p_T$ and rapidities $y_1$ and $y_2$
produced in hard scattering processes in central $AA$ collisions is,
\begin{equation}
\frac{dN^{(0)}_{\rm dijet}}{dp_T^2dy_1dy_2} = \int d^2b T_A^2(b)
K \sum_{a,b,c,d} x_1 f_{a/N}(x_1,p_T^2) x_2 f_{b/N}(x_2,p_T^2)\
\frac{d\sigma^{ab\rightarrow cd}}{d\hat{t}}(\hat{s},\hat{t},\hat{u})
\label{eq:noq}
\end{equation}
where we have neglected the nuclear shadowing effect for large
$p_T$ jet production, {\em i.e.}, $f_{a/A}(x,Q^2)\approx f_{a/N}(x,Q^2)T_A(b)$,
$T_A(b)$ is the nuclear thickness function as defined in Eq.~(\ref{eq:thick}).
The integration is performed over the plane transverse to the
beam direction, ${\bf b}$ is the transverse coordinate of the
production point in the transverse plane.

In the presence of dense matter, the high energy partons produced by a
hard scattering process can interact in the medium and thereby 
lose energy which will not show up in the final state of the jet
but contribute to the soft background. Thus, the number of
QCD hard jets to be observed in $AA$ collisions will be reduced
as compared to the expression Eq. (\ref{eq:noq}) (in Ref. \cite{MGMP90}
the term ``jet quenching'' has been introduced for this effect).
Taking into account interactions and energy loss in the medium, 
the rate of di-jets  of rapidities $y_1,y_2$ and transverse 
momenta $p_{T1}, p_{T2}$ reads
\begin{eqnarray}
\frac{dN_{\rm dijet}}{dp_{T1}dp_{T2}dy_1dy_2} & = &
\int d^2b\int dp_T^2 T_A^2(b) K \sum_{a,b,c,d} 
x_1 f_{a/N}(x_1,p_T^2)\ x_2 f_{b/N}(x_2,p_T^2) \nonumber\\
& & \cdot 
\frac{d\sigma^{ab\rightarrow cd}}{d\hat{t}}(\hat{s},\hat{t},\hat{u})
\delta(p_{T1}-p_T+\Delta E_{c,1}({\bf b},{\bf p_T},y_1))
\nonumber\\
& & \cdot \ \delta(p_{T2}-p_T+\Delta E_{d,2}({\bf b},{\bf p_T},y_2))
\label{eq:JQ}
\end{eqnarray}
The energy losses $\Delta E_{c,1}$ and $\Delta E_{d,2}$ depend not only
on the momenta and on the production point of the partons $c$ and $d$
but also on the space-time evolution of the dense matter. 
If we limit ourselves to jets in the central region, $y_1=y_2=0$,
the total energy loss of a massless parton inside the dense matter is
\begin{equation}
\Delta E \ =\ \int_{\tau_0}^{\tau_{f}} d\tau \ \frac{dE}{dz}(\tau)
\label{eq:totde}
\end{equation}
where $\tau_0$ is the formation time of the dense matter 
and $\tau_{f}$ is the proper time when the parton escapes from 
the dense medium, $\tau_f = (R^2-b^2\sin^2 \psi)^{1/2}-b \cos \psi$,
where $\psi$ is the angle between ${\bf b}$ and ${\bf p_T}$.

To evaluate the integrated energy loss suffered by a fast parton, 
we now need to know the the time dependence of the temperature. 
We shall assume a one-dimensional scaling expansion of the 
system \cite{BJOR}, thus the time dependence of the temperature
in both phases is given by $T\sim 1/\tau^{1/3}$. To determine
the duration of the mixed phase of a first order phase transition
in which the temperature is constant, $T_c$, we adopt a bag model 
equation of state (EOS) for the quark-gluon-plasma (QGP) phase:
\begin{equation}
\varepsilon(T) \ = \ \frac{37}{30} \pi^2 \ T^4\ + \ B \qquad
P(T) \ = \ \frac{1}{3} [\varepsilon(T)\ - \ 4B]
\label{eq:bag}
\end{equation}
where $\varepsilon$, $P$ and $B$ are the energy density, the pressure
and the bag constant, respectively. Define the energy densities at 
the critical temperature in the QGP phase and in the hadronic phase, 
$\varepsilon_Q$ and $\varepsilon_H$, respectively. The system then
enters the mixed phase at the proper time $\tau_Q=\tau_0 T_0^3/T_c^3$,
and goes from the mixed phase into the hadronic phase at
$\tau_H=(\varepsilon_Q+P_c)\tau_Q/(\varepsilon_H+P_c)$, where
$P_c=(\varepsilon_Q-4B)/3$ is the pressure at $T_c$. In the
following, we shall use $\varepsilon_Q=$2.5 GeV/fm$^3$,
$\varepsilon_H=$0.5 GeV/fm$^3$, and $B=$0.5 GeV/fm$^3$,
which implies a critical temperature $T_c=$0.188 GeV.

\begin{figure}
\centerline{\psfig{figure=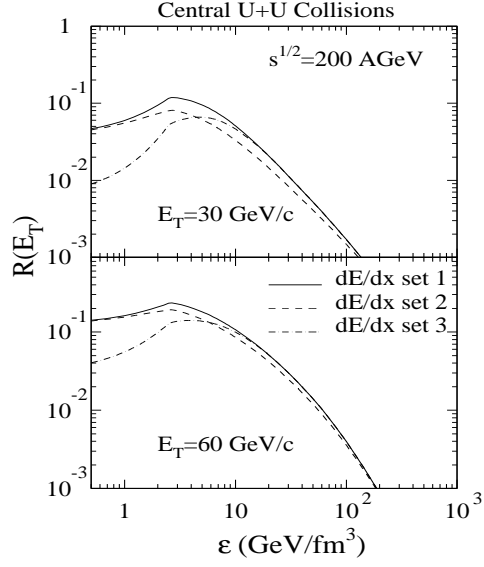,width=2.5in,height=3in}}
\caption{The suppression rate as a function of the initial energy density
         $\varepsilon_0$ for a dijet with the total energy
         $E_T=30$, 60 GeV in a central  $U+U$ collisions at 
         $\protect\sqrt{s}=200$ GeV.}
\label{fig35}
\end{figure}

\begin{figure}
\centerline{\psfig{figure=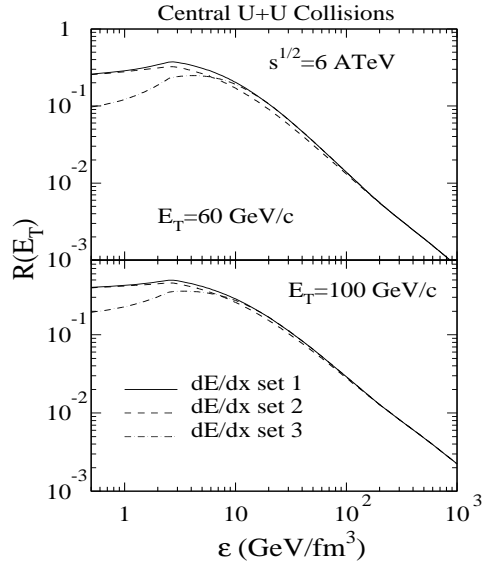,width=2.5in,height=3in}}
\caption{The same as Fig.~\protect\ref{fig35}, except at 
         $\protect\sqrt{s}=6$ TeV}
\label{fig36}
\end{figure}

In order to achieve maximum sensitivity to the quenching effect 
it is useful to consider the total transverse
energy $E_{\rm tot}=p_{T1}+p_{T2}$ for a pair of back-to-back
jets. The quenching ratio is defined as the number of dijets of
energy $E_{\rm tot}$ as measured in the experiment, divided
by the corresponding number expected in the absence of jet
interaction with the dense medium,
\begin{equation}
R_{AA}(E_{\rm tot}) \ =\
\frac{\displaystyle{
\int dp_{T1} \int dp_{T2} \ \delta(E_{\rm tot}-p_{T1}-p_{T2}) \
\left.\frac{dN_{\rm dijet}}{dp_{T1}dp_{T2}dy_1dy_2}\right|_{y_1=y_2=0}}}{
\displaystyle{\int dp_T^2 \ \delta(E_{\rm tot}-2p_T) \
\left.\frac{dN^{(0)}_{\rm dijet}}{dp_T^2dy_1dy_2}\right|_{y_1=y_2=0}}},
\label{eq:jqrate}
\end{equation}
for dijets at $y_1=y_2=0$ in the CM frame,

Figs.~\ref{fig35} and \ref{fig36} show the dependence of the 
quenching rate on the initial energy density $\varepsilon_0$
for central $U+U$ collisions at $\sqrt{s}=$200 AGeV
and $\sqrt{s}=$6 ATeV, respectively, for given
total dijet energies. The results were obtained 
by evaluating Eqs.~(\ref{eq:noq}) and (\ref{eq:JQ}) for
the Duke-Owens structure functions set 1 \cite{DUKE}
and the EOS discussed above. The initial proper time was 
taken to be $\tau_0=1/T_0$. The non-trivial dependence 
of $R_{AA}(E_{\rm tot})$ on $\varepsilon_0$
reflects the temperature dependence of $dE/dz$ as shown in 
{}Fig.~\ref{fig34}. For small initial energy densities, the 
high energy partons of the dijet suffer a temperature-independent 
energy loss per unit distance, $dE/dz= 1$ GeV/fm, in the hadronic 
phase of the dense matter at temperatures well below $T_c$.
As $\varepsilon_0$ increases the partons will spend more and
more time in the mixed phase. For all  scenarios , $dE/dz$  
takes its minimum value in the mixed phase and consequently,
the quenching rate increases with increasing initial energy 
density, {\em i.e.}, one observes less quenching. On the other hand, 
in the absence of such a local minimum, no increase of the dijet 
rate with $\varepsilon_0$ occurs.  
As $\varepsilon_0$ further increases, the leading partons of the 
jet probe the QGP phase at higher temperatures. Since $dE/dz$ grows 
with $T$ for $T>T_c$, $R_{AA}(E_{\rm tot})$
decreases with $\varepsilon_0$ in this region. All three curves 
come together asymptotically as they correspond
to the same $dE/dz$ in the QGP phase.
Thus a maximum of $R_{AA}(E_{\rm tot})$ at intermediate $\varepsilon_0$
reflects a minimum of $dE/dz$ at the critical temperature of the
phase transition.

\subsubsection{Monojet production}

Depending on the position of the initial hard parton scattering and
the direction of their momenta after the interaction, it can happen
that one of the high energy partons has to traverse a large distance
in the dense matter while the other one travels only a short
distance in the medium. If the first parton loses such a large
amount of energy that the corresponding jet can no longer be
distinguished from the soft and semi-hard background, one observes
a mono-jet. To be specific, let us define a mono-jet as a dijet where
one of the jets has a transverse momentum below a given value $p_{T{\rm cut}}$.
The number of mono-jets of transverse momentum $p_T$ and rapidity $y$
per central $AA$ collision at $y=0$ is given by [cf. Eq. (\ref{eq:JQ})]
\begin{equation}
 \left.\frac{dN_{\rm mono}(p_{T2}<p_{T{\rm cut}})}{dp_{T}dy}\right|_{y=0} \ = \
\int dy_2 \int dp_{T2} \ \Theta(p_{T{\rm cut}}-p_{T2}) \
\left.  \frac{dN_{\rm dijet}}{dp_{T}dp_{T2}dy dy_2}\right|_{y=0}
\label{eq:nmono}
\end{equation}
where one has to integrate over the (experimentally unknown) rapidity
and transverse momentum of the second parton. 

\begin{figure}
\centerline{\psfig{figure=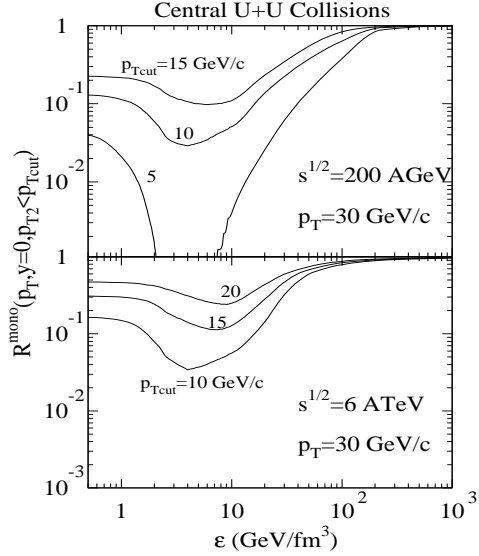,width=2.5in,height=3in}}
\caption{Monojet fraction as a function of the initial energy density
        $\varepsilon_0$ with a fixed $p_T=30$ GeV and different values
        of $p_{T{\rm cut}}$ for central $U+U$ collisions at 
        $\protect\sqrt{s}=200$ GeV, and 6 TeV.}
\label{fig37}
\end{figure}

For large $\varepsilon_0$, the number of mono-jets must decrease
because of the increase of $dE/dz$ with temperature in the QGP phase
which leads to the suppression of {\it all} jets. In this context it 
is useful to define the mono-jet fraction, {\em i.e.}, the mono-jet rate 
divided by the total jet rate, at a given transverse momentum $p_T$ 
and rapidity $y=0$,
\begin{equation}
R_{\rm mono}(p_T,p_{T2}<p_{T{\rm cut}},y=0) \ =\
\frac{\displaystyle{
\int dy_2 \int dp_{T2} \ \Theta(p_{T{\rm cut}}-p_{T2}) \
\left. \frac{dN_{\rm dijet}}{dp_{T}dp_{T2}dydy_2}\right|_{y=0} }}
{\displaystyle{ \int dy_2 \int dp_{T2} \
\left. \frac{dN_{\rm dijet}}{dp_{T}dp_{T2}dydy_2}\right|_{y=0} }}
\label{eq:rmono}
\end{equation}
In Fig.~\ref{fig37}, $R_{\rm mono}(p_T=30\ GeV/$c$,p_{T2}<p_{T{\rm cut}},y=0)$
is plotted as a function of $\varepsilon_0$ for scenario 1. Clearly,
the unquenching of jets associated with the minimum of $dE/dz$ at the 
critical temperature leads to a dip in the mono-jet rate which 
corresponds to the peak in the di-jet quenching rate
(cf. Figs.~\ref{fig35} and \ref{fig36}). In the limit of
large initial energy densities $R_{\rm mono}\rightarrow 1$, i.e.,
all jets are mono-jets at large $\varepsilon_0$. The reason
is that eventually $dE/dz$ becomes so large that, except for
negligible fraction of parton scatterings whose production points 
are close to the  nuclear surface in the transverse plane, only 
one of the two jets can survive.

Since the jet production rates depend on energy, $A$
which determine the transverse size of the dense matter, $R_A$, 
and the initial energy density $\varepsilon_0$, one would like to fixed 
the first two variables and isolate only the dependence on $\varepsilon_0$.
In experiments, there is no practical method to vary the initial
energy density $\varepsilon_0$ for both fixed $\sqrt{s}$ and $A$.
In order to best study jet quenching and monojet production as a
probe of the phase transition, one can vary the initial energy density
by either changing $\sqrt{s}$ for fixed beams or changing $A$ number 
for fixed energy \cite{PWG93}.

\section{Summary and Discussion}

In this report, I presented a pQCD-based picture of ultrarelativistic
heavy-ion collisions. In this framework, a nucleus in the infinite
momentum frame consists of many partons (quarks and gluons). The 
interactions among these partons can be divided into perturbative,
which can be described by pQCD calculation, and nonperturbative,
which can only be modeled phenomenologically. I have demonstrated
that pQCD processes dominate the underlying dynamics of heavy-ion
collisions at extremely high energies. I argued that the soft
component of a Pomeron exchange would be suppressed in the presence
of a dense and hot partonic plasma, though it is still important
in the initial parton scatterings in the early stage of the heavy
ion collisions and is responsible for the initial soft parton
production. It is then reasonable to assume that the evolution
of the initially produced partons can be described by pQCD
processes. Using the initial conditions estimated by the HIJING Monte
Carlo model, the following picture emerges:

(1) During the early stage in ultrarelativistic heavy-ion
collisions, hard or semihard parton scatterings, which happen
in a time scale of about 0.2 fm/$c$, produce a hot and undersaturated parton
gas. This parton gas is dominated by gluons and is far from
chemical equilibrium. Multiple hard scatterings suffered by a single
parton during this short period of time when the beam partons
pass through each other are suppressed due to the interference
embedded in the Glauber formula for multiple scatterings. This
leads to the predicted disapparence of the Cronin effect. 
Interference and parton fusion also lead to the depletion
of small $x$ partons in the effective parton distributions
inside a nucleus. This nuclear shadowing of parton distributions
reduces the initial parton production.

(2) After the two beams of partons pass through each other, the produced
parton gas in the central rapidity region starts its evolution
toward (kinetic) thermalization and (chemical) equilibration through 
elastic scatterings and induced radiations. The kinematic separation 
of partons in the central slab of about 1 fm 
through free-streaming gives an estimate of the time scale 
$\tau_{\rm iso}\sim 0.5 - 0.7$ fm/$c$, when local isotropy in momentum
distributions is reached. Further evolution of the parton gas toward
a fully equilibrated parton plasma is dictated by the parton
proliferation through induced radiation and gluon fusion. Though
the gluon equilibration rate is reduced by the inclusion of the
Landau-Pomeranchuk-Migdal effect, gluon fugacity still increases
rapidly toward its equilibrium value. Due to the consumption of
energy by the additional parton production, the effective temperature
of the parton plasma cools down considerably faster than the ideal
Bjorken's scaling solution. Therefore, the life time of the plasma
is reduced to 4 - 6 fm/$c$ before the temperature drops below the
QCD phase transition temperature.

(3) The evolution of the quark distribution always lags behind that
of gluons due to a smaller equilibration rate and the initial density.
For heavy quarks, the equilibration rate is even smaller. Take charm
quarks for example. The thermal production during the equilibration 
period is much smaller than the initial direct production, due to the 
small initial gluon fugacity and the short life time during which 
the temperature remains high enough to produce charm quarks.
For the same reason, dilepton and photon production during the 
evolution of the parton plasma is also small because of small
quark number density. Therefore, observation of large charm and dilepton
enhancement would imply high initial gluon and quark density and thus 
a longer life time of the parton plasma \cite{BMXW92,LMW94,LG94}. 

(4) Even though the initial parton system is not in a full
equilibrium, a study of color screening \cite{BMW92} shows
that the system is already in a deconfined state with large
average momentum (or effective temperature). Such a deconfined
parton system, though not in equilibrium, will dissociate
hadronic states like $J/\psi$. It was shown that $J/\psi$ can
be substantially suppressed during the evolution of the
parton plasma toward equilibrium. The measurements of this
suppression can reveal the initial conditions and the evolution
history of the parton plasma in high-energy heavy-ion collisions.

Readers should be reminded that the above picture is the
result of only a qualitative estimate based on a pQCD-based
model. Many uncertainties prevent us from making a more
quantitative calculation. All in all, these uncertainties
arise from our ignorance of the nonperturbative physics
and our inability to calculate the soft processes in
the framework of QCD. In the pQCD-based model reviewed
here, the uncertainties really lie in the cut-off, $p_0$,
which is supposed to separate nonperturbative soft interactions
from perturbative hard processes. Since soft and hard physics
do not have a definite boundary, the resultant parton production
from hard or semihard processes is very sensitive to the cut-off. The
accompanying soft parton production is not known in this
model and may only be estimated by effective models like the
color flux-tube model \cite{KEMG}. Recently, it has been
argued that the presence of larger $p_T$ partons can provide
screening to the parton production with smaller $p_T$ \cite{EBW95}.
The screening mass can then replace the {\it ad hoc} cut-off
and regulates the parton production cross section self-consistently.
There have been
recent developments in the field theory of particle production 
from mean fields \cite{KLUGER}. Such a field theoretical
approach to particle production could be the ultimate and 
consistent way to address the production and formation of a parton
plasma in heavy-ion collisions, once the collision terms
are included and the theory is generalized to the QCD case 
with rather random color fields.

The other nonperturbative uncertainty comes from the
parton distribution functions inside a nucleus. Though
quark distributions have been measured in deeply inelastic
lepton-nucleon (nucleus) collisions, the gluon distribution
is not known very well. Furthermore, nuclear modification
of the gluon distribution has not at all been measured
experimentally. Since gluon scatterings are the dominant
QCD processes in parton production, our results are very 
sensitive to the gluon distribution. On top of that, nuclear 
shadowing will cause additional 20-50\% uncertainty in
the initial parton density. One way to reduce this
uncertainty is to measure $pp$ and $pA$ collisions
systematically as proposed in Ref.~\cite{WG92,LIGY95}.

For a thermalized parton plasma, the uncertainties related to
nonperturbative physics will be reduced, since the soft
processes (long distance interactions) will be screened
by the interactions with the hot medium. For example, the
original divergent cross sections are regularized  by the
color screening mass $\mu_D$ and the infrared and collinear 
singularities associated with gluon radiations can be 
regularized by the LPM suppression. Recent developments
in QCD transport theory \cite{BLIA} have systematically studied 
these problems and formally included the resummation of hard
thermal loops, first proposed by Braaten and Pisarski \cite{BP90} 
in finite temperature field theory. However, one finds
that, even though the electric sector of the strong 
interaction is regularized by an effective Debye screening
mass, the magnetic interaction is still not regularized.
For most of the transport coefficients, this is not a problem
since the cancellation between in and out-state scatterings
makes the effective cross section less divergent and the
imaginary part of the gluon self-energy is enough to
regulate the infrared divergency \cite{HEIS,VISC,HHXW}. This is
so-called dynamic screening by Landau damping. For 
some other quantities, like gluon damping rate \cite{damp}
and color diffusion rate \cite{ASMG,HEIS}, one still has to 
invoke a magnetic screening mass which will always be a
nonperturbative quantity. Higher order corrections to
the Debye screening mass also depend on the non-perturbative
magnetic screening mass \cite{AY}. Therefore, some aspect of
the parton thermalization and equilibration
are still influenced by nonperturbative physics.

Throughout the production and evolution of the partonic system,
interference effects play an important role in multiple
collision processes, the correct implementation of which will be 
a great challenge to any transport model based on {\em classical} 
parton cascades. First, the interference between different amplitudes
in multiple parton scatterings leads to the Glauber formula which
suppresses multiple hard scatterings of beam partons.
Second, the LPM effect, {\em i.e.}, the destructive interference among 
different amplitudes of gluon radiation induced by multiple scatterings,
suppresses soft gluon radiation whose effective formation time,
$\tau_{\rm QCD}$, is larger than the mean free path, $\lambda_f$,
of the parton scatterings. This leads both to reduced 
gluon equilibration rates and reduced radiative energy loss.
A detailed analysis of the radiation amplitude has revealed the
underlying physics which contradicts the intuitive picture of the
classical parton cascade model. 
A propagating parton can no longer
be considered as time-like with a degrading virtuality as it
radiates between multiple scatterings, since the parton can be
space-like ($q^2<0$ for initial state radiation) in one amplitude 
and time-like ($q^2>0$ for final state radiation) in another. It
is the cancellation between these amplitudes that suppresses the
soft radiation with long formation time. One way to incorporate
the LPM effect in the parton interaction simulation, as I have 
illustrated, is to consider both the initial and final state 
radiation together for each scattering and impose the formation
time requirement, $\tau_{\rm QCD}<\lambda_f$, for the integration
over the phase-space of the radiated gluons. Shuryak and 
Xiong \cite{XS93} have used multiple gluon production
amplitudes given by Parke and Taylor \cite{PT86}. 
This approach has improved the gluon radiation
calculation beyond the leading logarithm approximation.
However, the space-time structure of the multiple gluon
amplitudes is not clear and a similar LPM analysis cannot be
performed. In addition, virtual corrections have to be 
included to satisfy the unitarity requirement.

In conclusion, pQCD-based models predict the formation of a hot
and dense parton plasma in ultrarelativistic heavy-ion collisions,
in which color deconfinement can be achieved and gluons dominate
the number of degrees of freedom. The answer to the question 
whether such a system can finally evolve into a fully thermalized 
and equilibrated parton plasma still depends crucially on the
reduction of the uncertainties involved and the understanding of
the underlying nonperturbative physics. Although theoretical progress
can and must be made in these directions, the ultimate quantitative
solution still awaits the experimental measurements at RHIC and LHC.

\section*{Acknowledgements}

I am deeply indebted to M.~Gyulassy for his encouragement and 
constant flow of ideas which have inspired most of the work I 
discussed here, and his enthusiastic collaboration in the 
development of the HIJING model. I thank K.~J.~Eskola for his 
collaboration which I have enjoyed very much.
I am grateful to B.~M\"uller for his inspiring collaboration 
through the last few years. I would also like to thank some of
my collaborators, T.~S.~Bir\'o, D. Kharzeev, P. L\'evai, J.~Qiu,
H. Satz, M.~H.~Thoma and X.-M. Xu who made 
significant contributions to the work I reviewed in this paper. 
Stimulating discussions with K.~Geiger, H.~Heiselberg, J.~I.~Kapusta, 
L.~McLerran, E.~Shuryak, and  L.~Xiong are also
gratefully acknowledged. Finally, I would like to thank Ulrich Heinz
for his comments and critical reading of this review.

\end{document}